\documentclass[twocolumn,aps,prl,superscriptaddress,tightenlines]{revtex4-1}
\usepackage{graphicx}
\usepackage{dcolumn}
\usepackage{soul}
\usepackage{empheq}
\usepackage{epsfig}
\usepackage{amsmath}
\usepackage{amsfonts}
\usepackage{dsfont}
\usepackage{tikz}
\usetikzlibrary{matrix}
\usepackage{amsmath,bm}
\usepackage{commath}
\usepackage{amssymb}
\usepackage{amssymb}
\usepackage{empheq}
\usepackage{xcolor}
\usepackage{bbm}
\usepackage[caption=false]{subfig}
\usepackage{placeins}
\usepackage{soul}
\usepackage{physics}
\usepackage{times} 
\usepackage{CJKutf8}
\usepackage{setspace}
\usepackage{mathtools}
\usepackage{cases}
\usepackage{esint}
\usepackage{extarrows}
\usepackage{slashed}
\usepackage{pifont}
\usepackage{mathrsfs}
\usepackage{changes} 

\definecolor{myblue}{rgb}{0.18039,0.1882353,0.57255}
\definecolor{myred}{rgb}{1,0.,0.3}

\usepackage{mathrsfs}
\usepackage{changes} 
\usepackage{hyperref}
\definechangesauthor[name={Per cusse}, color=red]{per} 
\hypersetup{
    colorlinks=true,
    citecolor=myblue,
    linkcolor=myblue,
    urlcolor=myblue,
}

\usepackage{orcidlink}

\makeatletter
\newcommand*\bigcdot{\mathpalette\bigcdot@{.8}}
\newcommand*\bigcdot@[2]{\mathbin{\vcenter{\hbox{\scalebox{#2}{$\m@th#1\bullet$}}}}}
\makeatother
\begin{document}
\newcommand{\A}{\text{A}}
\newcommand{\B}{\text{B}}
\title{Decoherence-free Behaviors of Quantum Emitters in Dissipative Photonic Graphene}

\author{Qing-Yang Qiu\orcidlink{0009-0007-3214-4892}}
\affiliation{School of Physics and Institute for Quantum Science and Engineering, Huazhong University of Science and Technology, and Wuhan institute of quantum technology, Wuhan, 430074, China}

\author{Guoqing Tian\orcidlink{0009-0000-6801-1361}}
\affiliation{School of Physics and Institute for Quantum Science and Engineering, Huazhong University of Science and Technology, and Wuhan institute of quantum technology, Wuhan, 430074, China}

\author{Zhi-Guang Lu\orcidlink{0009-0007-4729-691X}}
\affiliation{School of Physics and Institute for Quantum Science and Engineering, Huazhong University of Science and Technology, and Wuhan institute of quantum technology, Wuhan, 430074, China}

\author{Franco Nori\orcidlink{0000-0003-3682-7432}}
\affiliation{Quantum Computing Center, RIKEN, Wakoshi, Saitama 351-0198, Japan}
\affiliation{Department of Physics, The University of Michigan, Ann Arbor, Michigan 48109-1040, USA}

\author{Xin-You L\"{u}\orcidlink{0000-0003-3561-6684}}\email{xinyoulu@hust.edu.cn}
\affiliation{School of Physics and Institute for Quantum Science and Engineering, Huazhong University of Science and Technology, and Wuhan institute of quantum technology, Wuhan, 430074, China}

\date{\today}
\begin{abstract}
  Achieving decoherence-free quantum state manipulation is a paramount goal in modern quantum technologies. To this end, we demonstrate its implementation in a two-dimensional dissipative photonic graphene featuring exceptional rings. Employing the resolvent method, we analytically explore the quantum dynamics of emitters coupled to photonic graphene. In the thermodynamic limit, our analysis predicts a dissipation-dependent logarithmic relaxation for a single quantum emitter, alongside a pronounced quantum Zeno effect that slows the decay with increased dissipation. Notably, within a finite lattice, the excitation of single quantum emitter stabilizes in a decoherence-protected quantum state, which is identified as a dissipation-robust quasilocalized state. Interestingly, this state, together with a dark state, facilitates decoherence-free interactions between quantum emitters. This capability can be extended to topological graphenic platforms, where edge states mediate analogous protected interactions among giant atoms. Our findings highlight a promising path toward protecting quantum coherence in practical, high-dimensional photonic environment through dissipation engineering.
\end{abstract}

\maketitle
\emph{Introduction.}---Engineering decoherence-free interactions stands as a central challenge in quantum technologies and many-body physics\,\cite{PhysRevA.63.042307,Douglas2015,Friesen}. In particular, operating in dissipative environments hinders the scalability of quantum networks, as it forces a fundamental trade-off between achieving long-range interactions and maintaining protection from decoherence. This trade-off can be mitigated either by leveraging environmental nonlinearities\,\cite{PhysRevResearch.7.L012014,mck6-2rqd} or by encoding quantum information into decoherence-free subspaces\,\cite{PhysRevLett.85.1758,PhysRevLett.109.170501,Paulisch_2016,PhysRevA.111.033718}, which selectively shields particular Hilbert subspaces of the quantum system from environmental-induced decoherence.

Condensed matter systems have recently witnessed intense focus on graphene\,\cite{PhysRevLett.97.187401,RevModPhys.81.109,RevModPhys.83.407}, a carbon-based monolayer material, due to its profound fundamental interest and its broad, exploitable application potential. It serves not only as a versatile platform for observing quantum Hall effects\,\cite{Zhang,Novoselov2005,YatingSha}, but also hosts unconventional superconductivity in its magic-angle twisted bilayer configuration\,\cite{Cao1038,Cao26160,PhysRevLett.122.106405}. Apart from the real electronic materials, the designs of artificial graphene have flourished, especially in photonic realizations\,\cite{Rechtsman,Bellec,PhysRevLett.131.013804,PhysRevA.111.043526}, offering the key advantage of faithfully emulating graphenic pivotal features while circumventing the intrinsic limitations and structural instabilities of atomically real materials.

At the heart of such photonic graphene platforms lies the tailored design of their energy dispersion relations around characteristic Dirac cones.  In particular, controlled distortions of these cones, achieved through pathways such as strain\,\cite{Polini,PNigge,PhysRevB.111.235407,Shifang137} or anisotropy\,\cite{PhysRevLett.119.016401,PhysRevX.9.031010,PhysRevB.101.045130}, enable unusual transmission properties\,\cite{PhysRevLett.100.013904,PhysRevB.78.045122}. Beyond these artificial modifications, the inevitable and ubiquitous particle dissipation in realistic settings generically reshapes Dirac points into rings of exceptional points (EPs), a regime extensively studied both in theory\,\cite{PhysRevLett.118.045701,PhysRevB.100.245205,PhysRevLett.127.196801} and experimental implementation\,\cite{Zhen525,Cerjan,Lei2025}. Crucially, given the remarkable tunability of photonic graphene, a pivotal question follows: can it be engineered to demonstrate not only robustness against imperfections like loss and disorder, but also universal decoherence protection across the entire Hilbert space?

In this Letter, we address this question and point out the possibility of decoherence-free behaviors that emerge for quantum emitters (QEs) coupled to a dissipative photonic graphene. Within this bath, dissipation deforms the Dirac cones into EP rings. Employing the resolvent operator approach\,\cite{Cohen_book,GonzalezTudela2018,sciadv0297,yygz-71tr}, we drill down into the atom decay dynamics of this quantum electrodynamics (QED) setting. For a single QE, a nonperturbative treatment in the thermodynamic limit yields a logarithmic relaxation, whereas the Markov approximation predicts no decay. This relaxation is dependent on the dissipation strength, exhibiting a pronounced quantum Zeno effect where stronger dissipation leads to slower decay. In a finite bath, atomic excitation undergoes fractional decay, and its population ultimately oscillates around a dissipation-independent value. This decoherence-protected behavior originates from the quasilocalized state (QLS). For two QEs simultaneously coupled to a common cavity mode in photonic graphene, we identify purely coherent interactions that are immune to single-sublattice dissipation and robust against off-digonal disorder. The mechanism stems from the interference between a dark state and a QLS. We subsequently extend the graphenic QED to topological platforms, enabling the discovery of decoherence-free interactions between giant atoms to be mediated by edge states.

\begin{figure}
  \centering
  \includegraphics[width=8.7cm]{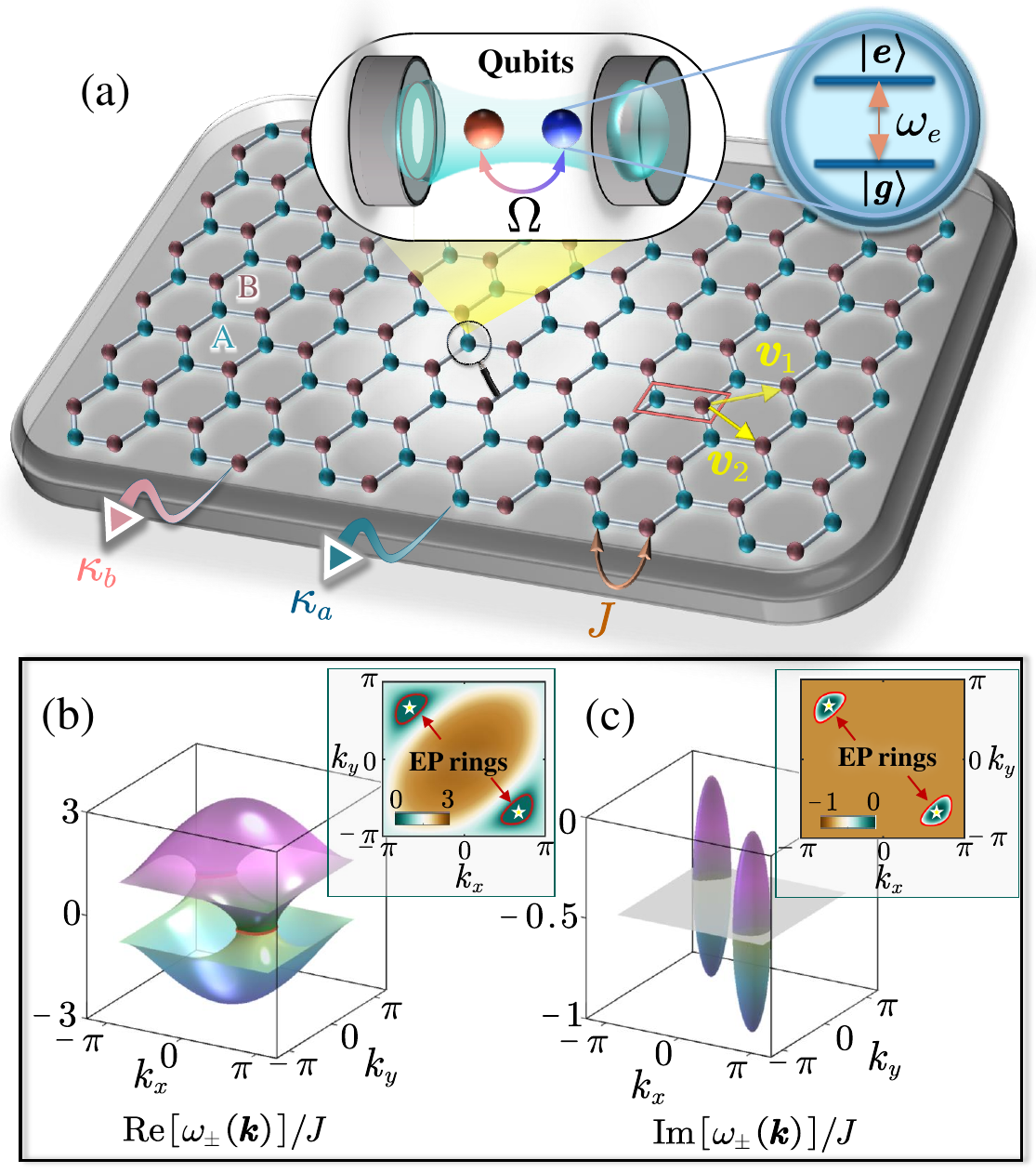}
  \caption{(a) Schematic of QEs coupled to a photonic graphene. The dissipative photonic environment is composed of two interspersed triangular sublattices, denoted as A (dark blue) and B (dark red). (b) Real and (c) imaginary components of the complex energy spectrum, $\mathrm{Re}[\omega_{\pm}(\boldsymbol{k})]/J$ and $\mathrm{Im}[\omega_{\pm}(\boldsymbol{k})]/J$, for the bath with single-sublattice dissipation, i.e., $\kappa_{a}=2J,\kappa_{b}=0$. The insets in (b) and (c) show the spectra $\mathrm{Re}[\omega_{+}(\boldsymbol{k})]/J$ and $\mathrm{Im}[\omega_{+}(\boldsymbol{k})]/J$ for the upper band. The dissipation-free photon modes are marked by pentagrams.}\label{fig1}
\end{figure}
\textit{Model and Hamiltonian.}---As depicted in Fig.\,\ref{fig1}(a), the system consists of $N_{e}$ QEs, each modeled as a two-level system with transition frequency $\omega_{e}$, coupled to a $N\times N$ photonic graphene with a uniform strength $\mathrm{g}$. The bath can be viewed as two interspersed triangular lattices\,\cite{PhysRevA.97.043831,Leonforte_2025,dibenedetto2025}, designated as A and B sublattices. This tight-binding model resides on a Bravais lattice specified by the primitive lattice vectors $\boldsymbol{v}_{1}=(3,\sqrt{3})/2$ and $\boldsymbol{v}_{2}=(3,-\sqrt{3})/2$, with the corresponding unit cell highlighted by a red box in Fig.\,\ref{fig1}(a). The location of any lattice site is given by $\boldsymbol{n} = n_1 \boldsymbol{v}_1 + n_2 \boldsymbol{v}_2$ with the indices $n_{1,2}=1,2,...,N$, thereby allowing for the definition of annihilation operators $a_{\boldsymbol{n}}$ and $b_{\boldsymbol{n}}$ on the A and B sublattices. Our discussion focuses on isotropic photonic graphene with a nearest-neighbor coupling strength $J$, while the detailed analysis for the anisotropic case is provided in\,\cite{SM}. Notably, the loss of photonic graphene is engineered \,\cite{PhysRevLett.118.200401,PhysRevLett.129.223601,Roccati,PhysRevLett.134.180401} with photon loss rates for sublattices A and B controlled by $\kappa_a$ and $\kappa_b$, respectively. Within the Markovian and rotating-wave approximations\,\cite{PhysRevA.4.1778,PhysRevA.40.4077,PhysRevA.91.042116}, the dynamics for the system plus the field is given by a standard Lindblad master equation ($\hbar=1$)
\begin{align}\label{eq1}	
\dot{\rho}_t=-i[H_{\text{atom}}+H_{\text{pg}}+H_{\text{int}},\rho_t]+{\mathcal{L}_a}{\rho _t}+{\mathcal{L}_b}{\rho _t}.
\end{align}
The considered atom Hamiltonian is $H_{\text{atom}}=\omega_{e}\sigma^{\dagger}\sigma$ for a single QE, and $H_{\text{atom}}\!=\!\omega_{e}\sum_{\ell=1}^{2}\sigma_{\ell}^{\dagger}\sigma_{\ell}^{}+\Omega (\sigma_{1}^{\dagger}\sigma_{2}^{}+{\rm{H.c.}})$ for two QEs, respectively. Here, $\omega_{e}$ denotes the atomic transition frequency and $\Omega$ the strength of direct inter-QE coupling.
The bath Hamiltonian for photonic graphene in the real space reads
\begin{align}\label{eq2}	
\!\!\!\!H_{\text{pg}}=\sum\limits_{\boldsymbol{n}}\omega_{c}(a^{\dagger}_{\boldsymbol{n}}a^{}_{\boldsymbol{n}}\!+\!b^{\dagger}_{\boldsymbol{n}}b^{}_{\boldsymbol{n}})\!+\!J\!\!\!\!\!\!\sum\limits_{\boldsymbol{v}=0,\boldsymbol{v}_{1},\boldsymbol{v}_{2}}\!\!\!\!\!\left(a^{\dagger}_{\boldsymbol{n}+\boldsymbol{v}}b^{}_{\boldsymbol{n}}
\!+\!\mathrm{H.c.}\right),
\end{align}\\[-0.8em]
where $\omega_{c}$ is the cavity resonant frequency. The local emitter-field interaction is described by $H_{{\rm int}}=\mathrm{g}\sum_{\ell=1}^{N_{e}}(a_{\boldsymbol{n}_{\ell}}^{\dagger}\sigma_{\ell}^{}+\mathrm{H.c.})$, where $\boldsymbol{n}_{\ell}$ indexes the unit cell to which the $\ell$th QE is coupled, with $\sigma_{\ell}^{}=\ket{g}_{\ell}\bra{e}$ being the corresponding atomic coherent operator. Here, the emitters are assumed to be coupled exclusively to the A sublattice. The sublattice dissipators are described by $\mathcal{L}_{a(b)}=\kappa_{a(b)}\sum_{\boldsymbol{n}}\mathcal{D}[a(b)_{\boldsymbol{n}}]$ with the Lindblad superoperator $\mathcal{D}[L]\rho \equiv L\rho L^\dagger - \{L^\dagger L,\rho\}/2$. By restricting our analysis to the single-excitation sector, the overall dynamics is fully governed by an effective non-Hermitian Hamiltonian $H_{\text{eff}}=H_{\text{atom}}+H^{\text{eff}}_{\text{pg}}+H_{\text{int}}$, where $H^{\text{eff}}_{\text{pg}} = H_{\text{pg}} - i\sum_{\boldsymbol{n}} (\kappa_a a_{\boldsymbol{n}}^\dagger a^{}_{\boldsymbol{n}} + \kappa_b b_{\boldsymbol{n}}^\dagger b^{}_{\boldsymbol{n}})/2$.

We proceed by illustrating the exotic spectral properties of this dissipative photonic graphene. Under periodic boundary conditions, we apply discrete Fourier transforms $a_{\boldsymbol{n}}=\sum_{\boldsymbol{k}}e^{i\boldsymbol{k}\cdot\boldsymbol{n}}a_{\boldsymbol{k}}/N$ and $b_{\boldsymbol{n}}=\sum_{\boldsymbol{k}}e^{i\boldsymbol{k}\cdot\boldsymbol{n}}b_{\boldsymbol{k}}/N$, which allows us to rewrite effective bath Hamiltonian as $H^{\text{eff}}_{\text{pg}}=\sum_{\boldsymbol{k}}\boldsymbol{{\rm{o}}}_{\boldsymbol{k}}^\dagger \boldsymbol{{\rm{h}}}^{}_{\boldsymbol{k}}\boldsymbol{{\rm{o}}}^{}_{\boldsymbol{k}}$ with $\boldsymbol{{\rm{o}}}_{\boldsymbol{k}}=[a_{\boldsymbol{k}},b_{\boldsymbol{k}}]^T$. The corresponding Bloch Hamiltonian in the rotating frame at the cavity frequency $\omega_{c}$ is given by
\begin{align}\label{eq3}
\boldsymbol{{\rm{h}}}_{\boldsymbol{k}}=\Re[f(\boldsymbol{k})]\sigma_x-\Im[f(\boldsymbol{k})]\sigma_y-i\kappa_{-}\sigma_{z}-i\kappa_{+}\sigma_{0},
\end{align}
where $f(\boldsymbol{k})=1+e^{-i\boldsymbol{k}\cdot\boldsymbol{v}_{1}}+e^{-i\boldsymbol{k}\cdot\boldsymbol{v}_{2}}$ and $\kappa_{\pm}=(\kappa_{a}\pm\kappa_{b})/4$ have been introduced\,\cite{SM}. The kernel Hamiltonian $\boldsymbol{{\rm{h}}}_{\boldsymbol{k}}$ is endowed with passive PT symmetry\,\cite{PhysRevLett.80.5243,PhysRevLett.103.093902,Joglekar18}, satisfying $\sigma_{x}(\boldsymbol{{\rm{h}}}_{\boldsymbol{k}}+i\kappa_{+}\sigma_{0})^{*}\sigma_{x}=\boldsymbol{{\rm{h}}}_{\boldsymbol{k}}+i\kappa_{+}\sigma_{0}$, and exhibits an exact PT symmetry once the background loss $-i\kappa_{+}\sigma_{0}$ is neglected. The complex dispersion relation follows from diagonalizing $\boldsymbol{\mathrm{h}}_{\boldsymbol{k}}$ and is given by $\omega_{\pm}(\boldsymbol{k})=-i\kappa_{+}\pm\sqrt{\left|f(\boldsymbol{k})\right|^{2}-\kappa_{-}^{2}}$. Throughout this work, we focus on the case of single-sublattice dissipation by setting $\kappa_a \neq 0$ and $\kappa_b = 0$ as our default, relegating the case of two-sublattice dissipation to\,\cite{SM} for completeness.

For a weak dissipation ($\kappa_{a}<4J$), the eigenvalues $\omega_{\pm}(\boldsymbol{k})$ become complex, and the original Dirac points $\boldsymbol{{\rm{K}}}_{\pm} = 2\pi(\pm 1, \mp 1)/3$ morph into a pair of EP rings [cf.\,the red curves in Figs.\,\ref{fig1}(b) and \ref{fig1}(c)] defined by $\left|f(\boldsymbol{k_{\text{ring}}})\right|=\kappa_{-}$. At moderate dissipation [$\kappa_{a}\in(4,12)J$],  this pair of rings merges into a single ring, which then vanishes entirely when $\kappa_{a}>12J$.  It is noteworthy that there always exist certain dissipation-free photon modes across all three dissipative regimes\,\cite{SM}. For the subsequent analysis, we focus on the weak dissipation regime to enable an analytical dynamical description. In this regime, the eigenvalues on the EP rings become purely imaginary, i.e., $\omega_{\pm}(\boldsymbol{k}_{\text{ring}})/J=-i\kappa_{a}/4$, and the two eigenstates coalesce. Of particular importance is the existence of zero-energy modes within the EP rings in the upper band [cf.\,the pentagram marked in Fig.\,\ref{fig1}(c)], characterized by $\Re[\omega_{+}(\mathbf{K}_{\pm})]=\Im[\omega_{+}(\mathbf{K}_{\pm})]=0$, which inspires the exploration of dissipation-immune behaviors.

\begin{figure}
  \centering
  \includegraphics[width=8.7cm]{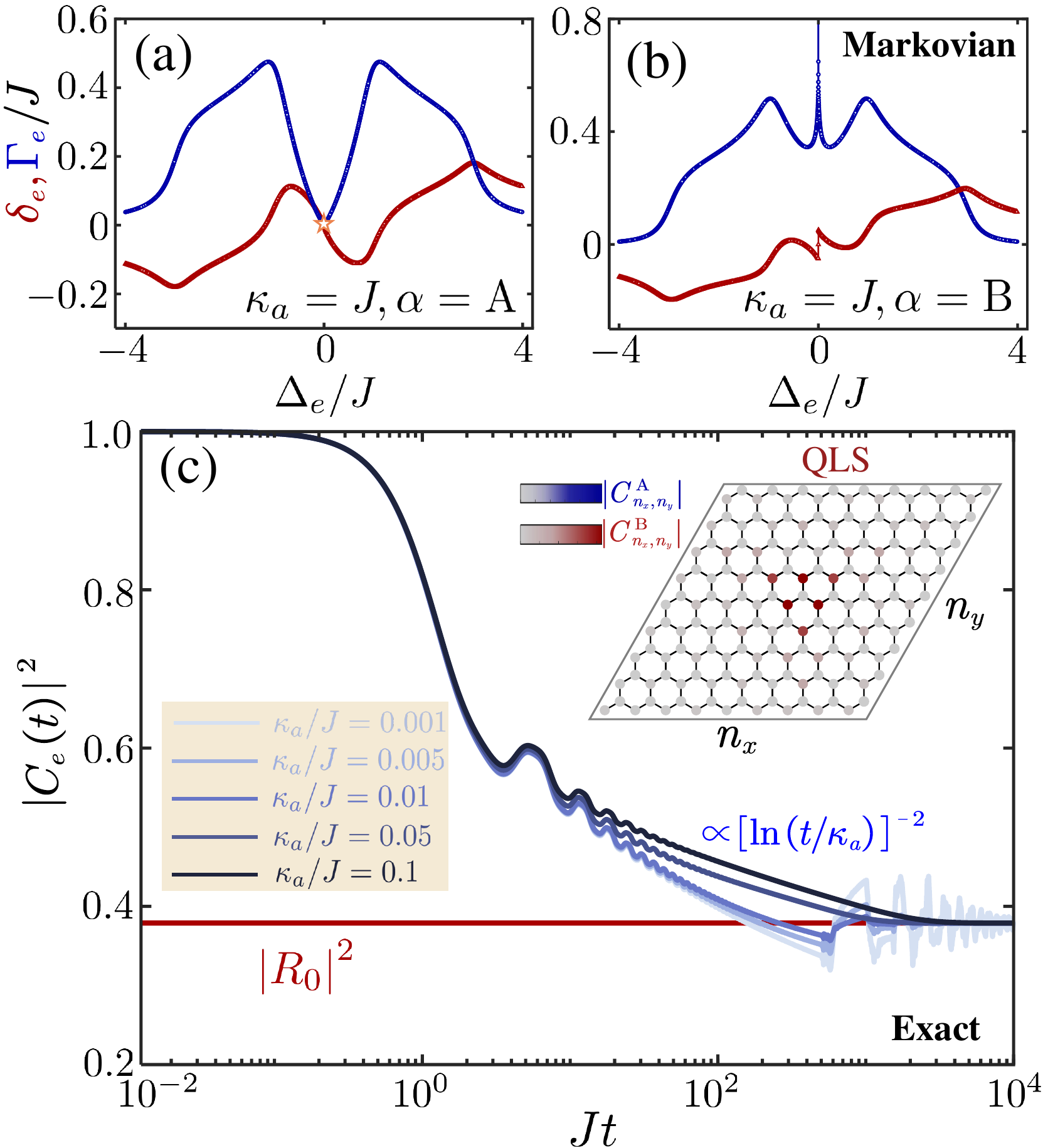}
  \caption{The Markovian decay rates $\Gamma_{e}$ (blue lines) and frequency shifts $\delta_{e}$ (red lines) for a single QE coupled to sublattice A (a) and B (b), plotted against the detuning $\Delta_e/J$ with $\kappa_a=J,\kappa_b=0$. (c) Excited-state population $|C_e(t)|^2$ of a resonant QE ($\Delta_e=0$) located at the center of sublattice A in a photonic graphene of size $N=512$, plotted as a function of scaled time $Jt$ with $\mathrm{g}=0.5J$. The inset of (c) depicts the photonic spatial density distribution $|C_{n_x,n_y}^{\A/\B}|^2$ corresponding to the QLS.}\label{fig2}
\end{figure}

\textit{Dissipation-immune single QE dynamics.}---We are now in a position to explore the dissipation-immune dynamics of a single QE that is coupled to a photonic graphene subject to single-sublattice dissipation. To be specific, we concentrate on the nonunitary evolution described by $\ket{\Psi_t}=e^{-iH_{\text{eff}}t}\ket{\Psi_0}$ \,\cite{PhysRevX.7.031024,PhysRevA.106.053517,Holzinger}, which initiates from an excited QE $\ket{\Psi_0}=\ket{e;\mathrm{vac}}$. The time-evolved state in the single-excitation manifold can be expanded as $\ket{\Psi_t}=\left[C_{e}(t)\sigma^{\dagger}+\sum_{\boldsymbol{n}}\left(C^{\text{A}}_{\boldsymbol{n}}(t)a_{\boldsymbol{n}}^{\dagger}+C^{\text{B}}_{\boldsymbol{n}}(t)b_{\boldsymbol{n}}^{\dagger}\right)\right]|g;\text{vac}\rangle$, where $C_{e}(t)$ and $C^{\text{A}({\text{B}})}_{\boldsymbol{n}}(t)$ are the probability amplitudes at time $t$ to find an excitation in the QE and the $\boldsymbol{n}$th cavity mode in sublattice A (B), respectively. Relying on the resolvent method, the coefficient $C_{e}(t)$ is given by a Fourier integral
\begin{align}\label{eq4}
\small
C_{e}(t)=\,-\frac{1}{2\pi i}\int_{-\infty}^{\infty}\dd{E}\frac{e^{-iEt}}{E-\Delta_{e}-\Sigma_{e}(E+i0^{+})},
\end{align}
where $\Sigma_{e}(z)$ and $\Delta_{e}=\omega_{e}-\omega_{c}$ are the single-QE self-energy and detuning, respectively. For QE coupled to a dissipative (lossless) sublattice A (B), the corresponding self-energy, denoted as $\Sigma_{e}^{\text{A(B)}}(z)$, can be calculated analytically in the thermodynamic limit, obtaining\,\cite{SM}
\begin{align}\label{eq5}
\!\!\!\Sigma_{e}^{\text{A(B)}}(z)=\frac{\mathrm{g}^{2}\left(2z+i\kappa_{b(a)}\right)}{\pi(\sqrt{z_{{\rm nh}}^{2}}+3J)^{\frac{1}{2}}(\sqrt{z_{{\rm nh}}^{2}}-J)^{\frac{3}{2}}}K\left[m(z_{\text{nh}})\right],
\end{align}
where $z_{{\rm nh}}=\sqrt{(z+i\kappa_{+})^{2}+\kappa_{-}^{2}}$, $K(m)$ is the complete elliptical integral of the first kind, and $m(z_{\text{nh}})$ is defined in\,\cite{SM}.

To elucidate the underlying dynamics, it is instructive to begin by examining the Markovian picture. This is implemented by approximating the self-energy in Eq.~(\ref{eq4}) as $\Sigma_{e}(E+i0^{+}) \approx \Sigma_{e}(\Delta_{e}+i0^{+})=\delta_e-i\Gamma_e/2$, where $\delta_e$ and $\Gamma_e$ are the corresponding Lamb shift and decay rate. The resulting dynamics of $|C_e(t)|^2$ thus predicts a perfect exponential decay with rate $\Gamma_e $ given by Fermi's Golden Rule (FGR). We present the perturbative results for $\delta_e$ (red lines) and $\Gamma_e$ (blue lines) in Figs.\,\ref{fig2}(a) and \ref{fig2}(b), for the cases of the QE coupled to sublattices A and B, respectively. Close to the band center ($\Delta_e = 0$), it yields a vanishing decay rate for a QE coupled to sublattice A, but predicts a divergent rate and a discontinuous $\delta_e$ for sublattice B, suggesting rapid de-excitation. The freezing of atomic excitation for the former scenario can be understood by expanding the self-energy $\Sigma_{e}^{\text{A}}(E+i0^{+})$ around the middle of the band ($|E|\ll J$), which yields\,\cite{SM}
\begin{align}\label{eq6}
\Sigma_{e}^{\text{A}}(E+i0^{+})\approx\frac{\mathrm{g}^{2}}{\sqrt{3}J^{2}}\left[-\frac{|E|}{2}i+\frac{E}{\pi}\ln\left(\frac{|E|\kappa_a}{18J^{2}}\right)\right],
\end{align}
where the linear scaling of $\Gamma_e$ indicates a vanishing decay at the band center.

We then turn to a nonperturbative dynamical description. A marked discrepancy between perturbative and nonperturbative results emerges for both infinite and finite bathes near the band center\,\cite{SM}, motivating our focus on the resonant case $\Delta_e = 0$ [cf.\,the pentagram marked in Fig.\,\ref{fig2}(a)]. Under the thermodynamic limit, the atomic dynamics features dissipation-dependent logarithmic relaxation, thereby violating FGR. This behavior is captured by its asymptotic expression in the long-time limit, which yields\,\cite{SM}
\begin{align}\label{eq7}
\!\!\!\lim_{t\rightarrow\infty}C_{e}(t)\approx\underset{t\rightarrow\infty}{\mathrm{lim}}\int_{0}^{\infty}\!\!\dd{y}\frac{\mathcal{E}_{e}e^{-yt}}{y\left[\ln(\frac{\kappa_a y}{18J^{2}})\right]^{2}}=\frac{\mathcal{E}_{e}}{\ln(\frac{18J^{2}}{\kappa_{a}}t)},
\end{align}
where $\mathcal{E}_{e}=5\sqrt{3}J^{2}\pi/ \mathrm{g}^{2}$. Following Eq.~(\ref{eq7}), a stronger dissipation indicates a longer lifetime of the excitation, which coincides with a signature of the quantum Zeno effect\,\cite{PhysRevA.95.042132,PhysRevA.98.052110,PhysRevLett.126.190402,PhysRevResearch.6.033243}.

For finite photonic graphene, we plot the exact excited-state population $|C_e(t)|^2$ for a QE resonantly coupled to sublattice A, obtained from a parameter scan of $\kappa_a/J$ across the range $10^{-3}$ to $10^{-1}$. The asymptotic behavior observed in the early-time dynamics of Fig.\,\ref{fig2}(c) further provides direct evidence for the quantum Zeno effect. Moreover, as boundary effects set in\,\cite{PhysRevA.97.043831,Redondo-Yuste_2021}, $|C_e(t)|^2$ settles into persistent oscillations around an asymptotic value, with all numerical predictions for different $\kappa_a$ converging to a common dissipation-independent constant $|R_0|^{2}$ [c.f. the red line in Fig.\,\ref{fig2}(c)]. Here, $R_0$ denotes the overlap between the initial wavefunction and QLS, which takes the form of $\ket{\Psi_{\text{QLS}}}=\left[R_0\sigma^{\dagger}+\sum_{\boldsymbol{n}}C^{\text{B}}_{\boldsymbol{n}}(t)b_{\boldsymbol{n}}^{\dagger}\right]|g;\text{vac}\rangle$. This overlap can be computed as $R_{0}=\left(1+\mathrm{g}^{2}\mathcal{G}(N)/J^{2}\right)^{-1}$,
where $\mathcal{G}(N)\approx 0.2057+2\ln N/(\sqrt{3}\pi)$ is size-dependent and also related to lattice isotropy\,\cite{SM}. The photonic component in $\ket{\Psi_{\text{QLS}}}$ is analytically determined\,\cite{SM}, as illustrated in the inset of Fig.\,\ref{fig2}(c). Defined by its non-square-integrability in the thermodynamic limit (where $N \rightarrow \infty$ implies $R_0 \rightarrow 0$), the QLS conceptually aligns with a vacancy-like dressed state\,\cite{PhysRevLett.126.063601,PhysRevA.104.053522,PhysRevB.107.054301}, and exhibits photonic components localized exclusively on sublattice B.

\begin{figure}
  \centering
  \includegraphics[width=8.7cm]{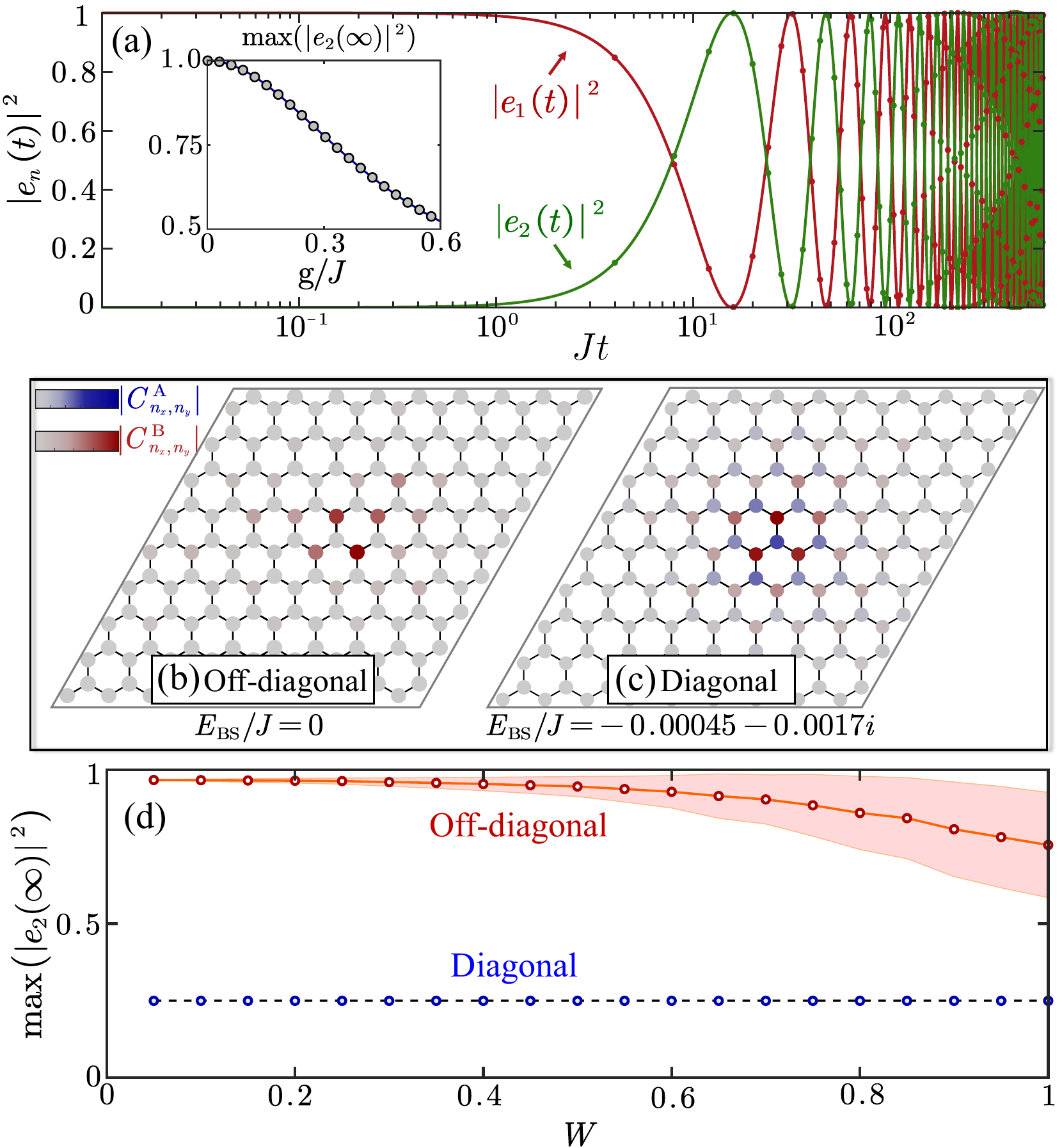}
  \caption{ (a) Time evolution of $|e_1(t)|^2$ (red) and $|e_2(t)|^2$ (green) with coupling strength $\mathrm{g} = 0.01J$. Solid lines and markers denote numerical and analytical results, respectively. (b) and (c) are the spatial photonic wave function profiles $|C_{n_x,n_y}^{\A}|$ (blue) and $|C_{n_x,n_y}^{\B}|$ (red) corresponding to off-diagonal and diagonal disorder with a strength of $W=0.5$. (d) Maximum transferred population, $\max(|e_2(\infty)|^2)$, versus disorder strength $W$ for both disorder types with $\mathrm{g} = 0.1J$. The red (blue) dots and shaded region denote the average and standard deviation over $10^3$ realizations for off-diagonal (diagonal) disorder. Other parameters implemented here are $\Delta_{e} = -\Omega = 0.1J$, $N = 2^6$, $\kappa_a = 10J$, and $\kappa_b = 0$.}\label{fig3}
\end{figure}

\textit{Decoherence-free interactions between QEs.}---Next, we aim to engineer purely coherent atomic interactions even within this dissipative photonic platform. To this end, we consider a configuration where both QEs are co-coupled to a same cavity in dissipative sublattice A. Notably, our model incorporates a direct inter-QE coupling of strength $\Omega$. And the relevant evolved state is given by $\ket{\Psi_t} =\left[\sum_{\ell=1}^{2}e^{}_{\ell}(t)\sigma_{\ell}^{\dagger}+\sum_{\boldsymbol{n}}\left(C^{\text{A}}_{\boldsymbol{n}}(t)a_{\boldsymbol{n}}^{\dagger}
+C^{\text{B}}_{\boldsymbol{n}}(t)b_{\boldsymbol{n}}^{\dagger}\right)\right]|g,g;\text{vac}\rangle$ in the single-excitation sector. Within the resolvent formalism and for the initial state  $\ket{\Psi_0}=\ket{e,g;\text{vac}}$, the explicit probability amplitudes $e_{1,2}(t)$ are given by\,\cite{SM}
\begin{align}
\!\!e_{1}(t)&=\!\!\int_{\mathcal{C}}\frac{\dd{z}}{2\pi i}\frac{-[z-\Delta_{e}-\Sigma_{22}^{\A\A}(z)]e^{-izt}}{[z-\Delta_{e}+\Omega][z-\Delta_{e}-\Omega-2\Sigma_{11}^{\A\A}(z)]},\label{eq8}\\
\!\!e_{2}(t)&=\!\!\int_{\mathcal{C}}\frac{\dd{z}}{2\pi i}\frac{-[\Omega+\Sigma_{12}^{\A\A}(z)]e^{-izt}}{[z-\Delta_{e}+\Omega][z-\Delta_{e}-\Omega-2\Sigma_{11}^{\A\A}(z)]},\label{eq9}
\end{align}
where $\Sigma_{\ell\ell^{\prime}}^{\alpha\beta}(z)=\mathrm{g}^{2}\langle{\rm vac}|c_{\boldsymbol{n}_{\ell},\alpha}(z-H^{\text{eff}}_{\text{gp}})^{-1}c_{\boldsymbol{n}_{\ell^{\prime}},\beta}^{\dagger}|{\rm vac}\rangle$ denotes the self-energy of QEs. Conditioned on $\Delta_e + \Omega = 0$, the long-time dynamics of $e_{1,2}(t)$ is dictated by two physical poles: $z_1 = \Delta_e - \Omega$ and $z_2 = 0$, which respectively correspond to the eigenenergies of dark state $\ket{\Psi_\text{Dark}}=(\sigma_1^\dagger-\sigma_2^\dagger)|g,g;\text{vac}\rangle/\sqrt{2}$ and two-emitter QLS whose form parallels the single-emitter case\,\cite{SM}, respectively. By strictly solving Eqs.\,(\ref{eq8}) and (\ref{eq9}), the analytical amplitudes $e_{1,2}(t)$ in the long-time limit can be computed as\,\cite{SM} $e_{1/2}(t) = \left( [1 + 2\mathrm{g}^{2}\mathcal{G}(N)/J^{2}]^{-1} \pm e^{2i\Omega t} \right)/2$. Consequently, this result reveals a purely and dissipation-immune coherent interaction between the QEs, as confirmed numerically in Fig.\,\ref{fig3}(a) (solid lines) and analytically (dotted lines). Moreover, it also predicts a larger transferred population $\max(|e_2(\infty)|^2)$ for a smaller $\mathrm{g}/J$  [cf. the inset of Fig.\,\ref{fig3}(a)].

We proceed by delving into the robustness of dissipation-free interactions against disorder. Since the dark state is unaffected by disorder due to its environmental decoupling, we therefore focus our analysis of the robustness solely on the disorder-induced modification to the QLS. For the purpose of illustration, we examine off-diagonal and diagonal disorder. The former introduces random tunneling amplitudes via $H^{\text{eff}}_{\text{pg}}\to H^{\text{eff}}_{\text{pg}}+\sum_{\boldsymbol{n}}(\epsilon_{1,\boldsymbol{n}}a_{\boldsymbol{n}}^{\dagger} b_{\boldsymbol{n}}^{}+\epsilon_{2,\boldsymbol{n}}a_{\boldsymbol{n}}^{\dagger} b_{\boldsymbol{n}+\boldsymbol{v}_{1}}^{}+\epsilon_{3,\boldsymbol{n}}a_{\boldsymbol{n}}^{\dagger} b_{\boldsymbol{n}+\boldsymbol{v}_{2}}^{}+\text{H.c.})$, preserving the chiral symmetry. In contrast, the latter adds random on-site energies into the bath Hamiltonian as $H^{\text{eff}}_{\text{pg}}\to H^{\text{eff}}_{\text{pg}}+\sum_{\boldsymbol{n}}(\epsilon_{a,\boldsymbol{n}}a_{\boldsymbol{n}}^{\dagger} a_{\boldsymbol{n}}^{}+\epsilon_{b,\boldsymbol{n}}b_{\boldsymbol{n}}^{\dagger} b_{\boldsymbol{n}}^{})$, thereby breaking the chiral symmetry of the model. All disorder parameters $\epsilon_{\mu,\boldsymbol{n}}/J\ (\mu=a,b,1,2,3)$ are independently drawn from a uniform distribution $[-W, W]$ per unit cell, with $W$ defining the disorder strength.

In Figs.\,\ref{fig3}(b) and \ref{fig3}(c), we compare the real-space profiles of the photonic bound states modified by off-diagonal and diagonal disorder. Notably, we observe that the off-diagonal disorder slightly alters the QLS's spatial photon profile relative to the clean system [c.f. the inset of Fig.\,\ref{fig2}(c)]. It's photonic component still localizes on sublattice B and the bound-state energy remains at zero. However, the diagonal disorder drastically modifies the bound state, spreading its photonic components across both sublattices A and B. It also imparts a negative imaginary part to the energy [cf.\,the annotations below Figs.\,\ref{fig3}(b) and \ref{fig3}(c)], implying the disappearance of QLS. Furthermore, as shown in Fig.\,\ref{fig3}(d), the symmetry-preserving case confers strong robustness against disorder. In contrast, breaking chiral symmetry fixes the maximal excitation transfer $\max(|e_2(\infty)|^2)$ at a constant of $1/4$, contributed solely from the dark state. Consequently, we conclude that the coexistence of a dark state and a QLS mediates the perfect dissipation-immune interactions.

\begin{figure}
  \centering
  \includegraphics[width=8.7cm]{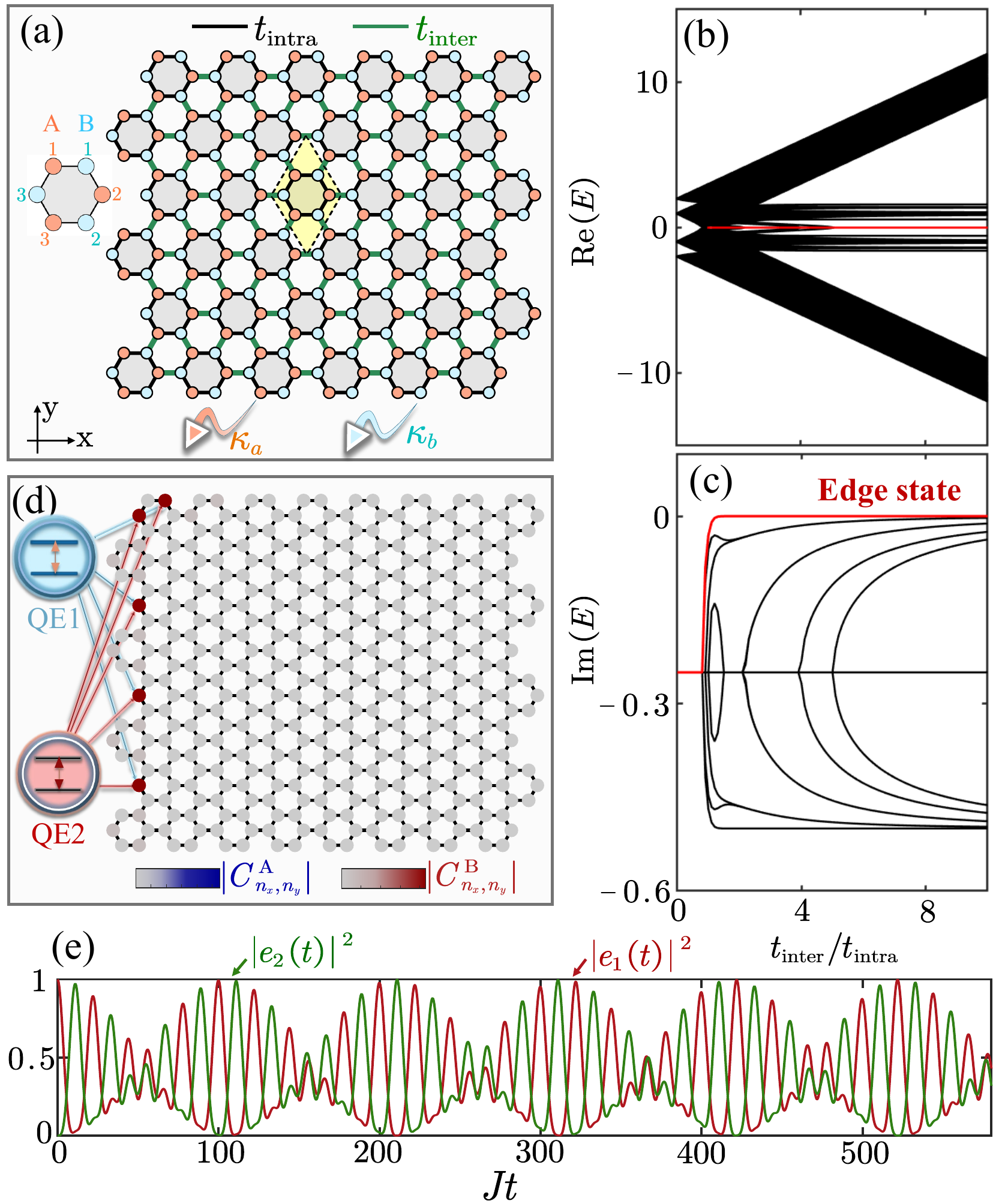}
  \caption{(a) Schematic of a $6\times6$ photonic graphene with SSH-like hoppings $t_{\mathrm{intra}}$ (black links) and $t_{\mathrm{inter}}$ (green links). The unit cell comprises six optical cavities, with three from sublattice A (orange) and three from sublattice B (blue) assigned dissipation strengths $\kappa_a$ and $\kappa_b$, respectively. (b) and (c) are the real and imaginary components of the complex energy spectrum for a lattice of size $L=10$. (d) Spatial profiles of the photonic wavefunctions $|C^{\mathrm{A}}_{n_x,n_y}|$ (blue) and $|C^{\mathrm{B}}_{n_x,n_y}|$ (red) of the edge state for $L=8$ and $t_{\mathrm{inter}}/t_{\mathrm{intra}}=15$. (e) Excited-state population $|e_{1,2}(t)|^2$ of giant atoms coupled to the edge state, where the performed parameters are $L=8$, $\Delta_e=0$, $\Omega=0.02J$, $t_{\mathrm{inter}}/t_{\mathrm{intra}}=15$, and ${\rm{g}}=0.2J$. All panels share the same dissipation strengths $\kappa_a=J$ and $\kappa_b=0$.}\label{fig4}
\end{figure}

\textit{Decoherence-free giant-atom interactions in topological graphene}---We now demonstrate that decoherence-free interactions can be engineered between giant artificial atoms\,\cite{PhysRevLett.120.140404,PhysRevLett.122.203603,Kannan2020,PhysRevA.101.053855,PhysRevLett.128.223602,Qiulaser,Jouanny2025} that are coupled to topological dissipative photonic graphene. Similar to the Su-Schrieffer-Heeger (SSH) model, we consider a topological photonic graphene\,\cite{PhysRevLett.122.086804,PhysRevLett.125.255502,PhysRevResearch.3.023121,ACSXie} with a Kekulé-type hopping texture, as illustrated in Fig.\,\ref{fig4}(a). This $L\times L$ modified graphene is characterized by intracell $t_{\mathrm{intra}}$ and intercell $t_{\mathrm{inter}}$ hoppings. Within a unit cell [cf. the yellow diamond marked in Fig.\,\ref{fig4}(a)], there are six optical sites. In particular, this finite lattice is bounded by armchair edges along the $x$-axis and zigzag edges along the $y$-axis.

In Figs.\,\ref{fig4}(b) and \ref{fig4}(c), we plot respectively the real $\Re(E)$ and imaginary $\Im(E)$ components of the complex spectrum for this engineered bath versus $t_{\mathrm{inter}}/t_{\mathrm{intra}}$. The real part $\Re(E)$ is symmetric about zero, while the imaginary part $\Im(E)$ is symmetric about $-\kappa_a/4$. Specifically, we identify zero-energy edge states (dissipationless) for $t_{\mathrm{inter}} > t_{\mathrm{intra}}$, indicated by the red curves in Fig.\,\ref{fig4}(b) and \ref{fig4}(c). The edge state is fully localized on sublattice B, as characterized by its photonic wavefunction [see Fig.\,\ref{fig4}(d)]. In this scenario, coupling two directly-interacting giant QEs to this edge state yields a modulated, non-decaying envelope in their coherent dynamics, arising from complex interference, as depicted in Fig.\,\ref{fig4}(e). Physically, this dissipation-immune interaction is mediated by the cooperation between a dark state and a edge state. The mediating mechanism for these coherent interactions can be extended to other topological graphene platforms as exemplified in\,\cite{SM}, facilitated by engineering the edges of photonic graphene.

\textit{Summary and outlook.}---To sum up, we have established a nonperturbative description for decoherence-free interactions between coupled QEs in dissipative photonic graphene. We find that the graphenic QED exhibits strong robustness, featuring immunity to lattice dissipation and stability against energetic disorder. Furthermore, we demonstrate a dissipation-dependent logarithmic relaxation, which violates the FGR, alongside a pronounced quantum Zeno effect. Notably, our results reveal that the decoherence-free interactions are enabled for both natural and artificial giant QEs. Several interesting perspectives can be inspired by our current study. A major direction is to extend our results to multi-emitter or multi-excitation regimes\,\cite{PhysRevLett.131.033605,PhysRevX.14.011020,hmbj-lg7p}, where collective effects and equilibrium properties can be explored. It may also be interesting to study non-Hermitian phenomena in superstructures of multilayer graphene or in curved geometries like nanotubes and nanospheres\,\cite{PhysRevB.68.035430,PhysRevB.72.195426,PhysRevB.91.125414,Xiaoyu}, which could offer versatile interfaces for engineering fascinating bath-mediated interactions.

\textit{Acknowledgments.}---This work is supported by the National Science Fund for Distinguished Young Scholars of China (Grant No.\,12425502), the Innovation Program for Quantum
 Science and Technology (Grant No.\,2024ZD0301000), the National Key Research and Development Program of China\,(Grant No.\,2021YFA1400700),\,and the Fundamental Research Funds for the Central Universities (Grant No. 2024BRA001). F.N. is supported in part by the Japan Science and Technology Agency (JST) [via the CREST Quantum Frontiers program Grant No. JPMJCR24I2, the Quantum Leap Flagship Program (Q-LEAP), and the Moonshot R\&D Grant No. JPMJMS2061], and the Office of Naval Research (ONR) Global (via Grant No. N62909-23-1-2074).

\textit{Data availability.}---The data that support the findings of this Letter are not publicly available. The data are available from the authors upon reasonable request.
\bibliographystyle{myapsrev4-1}
\clearpage
\setcounter{secnumdepth}{2}
\onecolumngrid
\newcommand{\note}[1]{\aucmnt{\textcolor{red}{#1}}}
\renewcommand{\Re}{\mathop{\rm Re}}		
\renewcommand{\Im}{\mathop{\rm Im}}		
\newcommand\specialsectioning{\setcounter{secnumdepth}{-2}}
\setcounter{equation}{0} \setcounter{figure}{0}
\setcounter{table}{0}
\makeatletter
\renewcommand{\theequation}{S\arabic{equation}}
\renewcommand{\thefigure}{S\arabic{figure}}
\renewcommand{\bibnumfmt}[1]{[S#1]}
\renewcommand{\citenumfont}[1]{S#1}
\begin{center}
    {\Large \textbf{ Supplemental Material for\\
            ``Decoherence-free Behaviors of Quantum Emitters in Dissipative Photonic Graphene"}}
\end{center}

\begin{center}
Qing-Yang Qiu$^{1}$, Guoqing Tian$^{1}$, Zhi-Guang Lu$^{1}$, Franco Nori$^{2,3,4}$, Xin-You L\"{u}$^{1, *}$
\end{center}
\begin{minipage}{17cm}
\centering 
\small\itshape
$^{1}$School of Physics and Institute for Quantum Science and Engineering, Huazhong University of Science and Technology, and Wuhan Institute of Quantum Technology, Wuhan 430074, China \\
\small\itshape
$^{2}$Theoretical Quantum Physics Laboratory, Cluster for Pioneering Research, RIKEN, Wakoshi, Saitama 351-0198, Japan \\
\small\itshape
$^{3}$Quantum Information Physics Theory Research Team, Quantum Computing Center, RIKEN, Wakoshi, Saitama 351-0198, Japan \\
\small\itshape
$^{4}$Physics Department, The University of Michigan, Ann Arbor, Michigan 48109-1040, USA
\end{minipage}
\vspace{8mm}

This supplement material contains six parts: I. Exact dynamics of quantum system coupled to Dirac conelike lattices; II. Light-matter interactions for a single quantum emitter with homogeneous dissipation; III. Light-matter interactions in Dirac conelike bath with single-sublattice dissipation. IV. Quantum dynamics in anisotropic Dirac photonic bath with single-sublattice dissipation. V. Robustness of dissipation-free interactions against disorder. VI. Dissipation-free interactions in topological platforms with mixed boundaries.

\tableofcontents
\newpage

\section{Exact Dynamics of Quantum System Coupled to Dirac Conelike Lattices}\label{I}
\setcounter{equation}{0}
\renewcommand\theequation{S\arabic{equation}}
\makeatletter
\renewcommand{\thefigure}{S\@arabic\c@figure}
\makeatother
In this section, we present the detailed dynamical descriptions of quantum emitters (QEs) coupled to an isotropic photonic graphene which, due to its openness and energy leakage, is described by a non-Hermitian bath Hamiltonian. The bath is modeled by a nearest-neighbor tight-binding hexagonal lattice, as schematically depicted in Fig.~\ref{figS1}. We commence with a description of this photonic environment. The spectral properties of the photonic graphene are examined in subsection~\ref{IA} for the dissipationless case and subsequently for the dissipative case in subsection~\ref{IB}. We then detail the dynamics of a single QE coupled to this photonic environment in subsection~\ref{IC}, and extend the analysis to a pair of QEs in subsection~\ref{ID}. These dynamical treatments lay the theoretical framework for the rich physical phenomena explored in the later discussions.

\subsection{The spectrum of photonic graphene without dissipation}\label{IA}
\begin{figure}
  \centering
  \includegraphics[width=10cm]{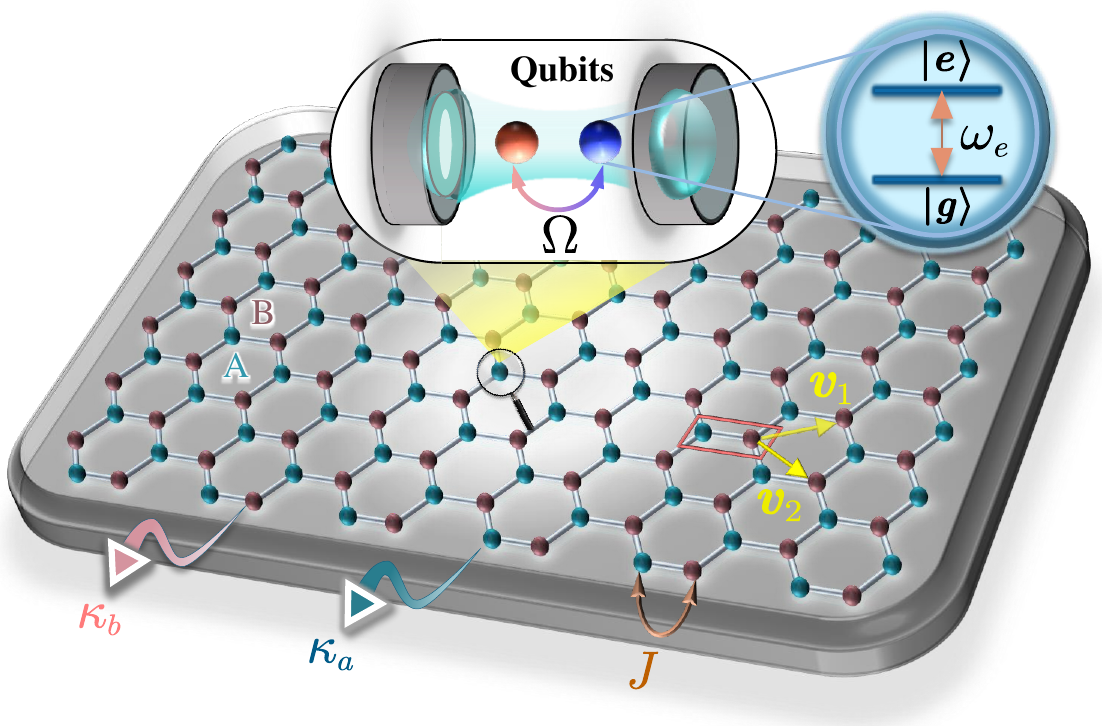}
  \caption{Schematic of the quantum emitters coupled to a photonic tight-binding lattice featuring Dirac cones. The photonic environment consists of two interspersed triangular lattices (marked by dark blue and dark red balls for sites A and B, respectively), where each site represents a bosonic or cavity mode. The primitive vectors $\boldsymbol{v}_{1},\boldsymbol{v}_{2}$ defining the Bravais lattice are plotted in cyan. Each A (B) site is coupled to its three nearest-neighbor B (A) sites via the hopping interactions indicated by brown double-headed arrow. The optical environment features engineered dissipation, manifesting as independently controlled loss rates $\kappa_a$ and $\kappa_b$ for sublattices A and B, respectively. In this scheme, a pair of qubits (represented by the red and blue balls) couples to a single cavity mode of the A sublattice, with a direct coupling $\Omega$ between them. Each quantum emitter is modeled as a two-level system with ground ($|g\rangle$) and excited ($|e\rangle$) states.}\label{figS1}
\end{figure}

The photonic graphene can be viewed as two interspersed triangular lattices, designated as A and B sublattices and depicted by the blue and red balls in Fig.~\ref{figS1}, respectively. Each B site couples to three A sites: one intracell coupling (within the same unit cell, white box in Fig.\,\ref{figS1}) and two intercell tunnelings to neighboring A sites. This tight-binding model resides on a Bravais lattice specified by the primitive lattice vectors $\boldsymbol{v}_{1}=(3,\sqrt{3})/2$ and $\boldsymbol{v}_{2}=(3,-\sqrt{3})/2$. The three coupling coefficients are uniform and share the same magnitude $J$, a condition we will later relax to study graphenic quantum electrodynamics (QED) in an anisotropic non-Hermitian lattice. Based on these primitive lattice vectors, the position of any lattice site can be expressed as $\boldsymbol{n}=n_{1}\boldsymbol{v}_{1}+n_{2}\boldsymbol{v}_{2}$, where $n_{1}$ and $n_{2}$ are integer numbers $n_{1,2}=1,2,\cdots,N$ with $N$ the bath size. The bath Hamiltonian for photonic graphene without external dissipation reads
\begin{align}
H_{\text{pg}}=\omega_{c}\sum\limits_{\boldsymbol{n}}(a^{\dagger}_{\boldsymbol{n}}a^{}_{\boldsymbol{n}}+b^{\dagger}_{\boldsymbol{n}}b^{}_{\boldsymbol{n}})+J\sum\limits_{\boldsymbol{n}}\left(a^{\dagger}_{\boldsymbol{n}}b^{}_{\boldsymbol{n}}+a^{\dagger}_{\boldsymbol{n}+\boldsymbol{v}_{1}}b^{}_{\boldsymbol{n}}
+a^{\dagger}_{\boldsymbol{n}+\boldsymbol{v}_{2}}b^{}_{\boldsymbol{n}}+\mathrm{H.c.}\right),\label{S1}
\end{align}
where the operators $a_{\boldsymbol{n}}$ ($a_{\boldsymbol{n}}^{\dagger}$) and $b_{\boldsymbol{n}}$ ($b_{\boldsymbol{n}}^{\dagger}$) annihilate (create) a bosonic excitation on the A and B sites at position $\boldsymbol{n}$, respectively, and $\omega_{c}$ is the resonant frequency of these bosonic modes or cavity modes.

The aforementioned model, with its A and B sublattices, yields a two-band structure, which can be directly obtained via a discrete Fourier transform under periodic boundary conditions. We begin by rewriting the bath Hamiltonian \ref{S1} in a frame rotating with $\omega_{c}$, i.e., taking $\omega_{c}$ as an energy reference. Subsequently, we apply the following transformations
\begin{align}
a_{\boldsymbol{n}}=\frac{1}{N}\sum\limits_{\boldsymbol{k}}e^{i\boldsymbol{k}\cdot\boldsymbol{n}}a_{\boldsymbol{k}},\,\,\,b_{\boldsymbol{n}}=\frac{1}{N}\sum\limits_{\boldsymbol{k}}e^{i\boldsymbol{k}\cdot\boldsymbol{n}}b_{\boldsymbol{k}},
\,\,\, \boldsymbol{k}=k_{1}\boldsymbol{d}_{1}+k_{2}\boldsymbol{d}_{2}\label{S2}
\end{align}
to the bosonic modes with $k_{1},k_{2}=\frac{2\pi}{N}\left(-\frac{N}{2},\cdots,\frac{N}{2}-1\right)$, which yields
\begin{align}
H_{\text{pg}}= &\,J\sum\limits_{\boldsymbol{k}}(1+e^{-i\boldsymbol{k}\cdot\boldsymbol{v}_{1}}+e^{-i\boldsymbol{k}\cdot\boldsymbol{v}_{2}})a^{\dagger}_{\boldsymbol{k}}b^{}_{\boldsymbol{k}}+\mathrm{H.c.}=\sum_{\boldsymbol{k}}\left[a_{\boldsymbol{k}}^{\dagger},b_{\boldsymbol{k}}^{\dagger}\right]\left[\begin{array}{cc}
0 & f(\boldsymbol{k})\\
f^{*}(\boldsymbol{k}) & 0
\end{array}\right]\left[\begin{array}{c}
a_{\boldsymbol{k}}\\
b_{\boldsymbol{k}}
\end{array}\right],\label{S3}
\end{align}
where $\boldsymbol{d}_{1}$ and $\boldsymbol{d}_{2}$ are the primitive basis vectors of reciprocal space in the first Brillouin zone, satisfying $\boldsymbol{v}_{i}\cdot\boldsymbol{d}_{j}=\delta_{ij}$, and the natation $f(\boldsymbol{k})=1+e^{-i\boldsymbol{k}\cdot\boldsymbol{v}_{1}}+e^{-i\boldsymbol{k}\cdot\boldsymbol{v}_{2}}$ has been introduced for conciseness. Notice, the bath Hamiltonian can be rewritten as $H_{\text{pg}}=\sum_{\boldsymbol{k}}\boldsymbol{\mathrm{o}}_{\boldsymbol{k}}^\dagger \boldsymbol{\mathrm{h}}^{(0)}_{\boldsymbol{k}}\boldsymbol{\mathrm{o}}^{}_{\boldsymbol{k}}$ with $\boldsymbol{\mathrm{o}}_{\boldsymbol{k}}=[a_{\boldsymbol{k}},b_{\boldsymbol{k}}]^T$, and the corresponding Bloch Hamiltonian or kernel Hamiltonian is given by
\begin{align}
\boldsymbol{\mathrm{h}}^{(0)}_{\boldsymbol{k}}=\left[\begin{array}{cc}
0 & f(\boldsymbol{k})\\
f^{*}(\boldsymbol{k}) & 0
\end{array}\right]=\Re[f(\boldsymbol{k})]\sigma_x-\Im[f(\boldsymbol{k})]\sigma_y,\label{S4}
\end{align}
where $\sigma_x$ and $\sigma_y$ are the Pauli matrixes.

At first glance, the Hamiltonian $\boldsymbol{\mathrm{h}}^{(0)}_{\boldsymbol{k}}$ is noted to possess chiral (sublattice) symmetry, guaranteeing that all eigenmodes group into chiral pairs with opposite energies. This is directly revealed by simply diagonalizing $\boldsymbol{\mathrm{h}}^{(0)}_{\boldsymbol{k}}$, which yields
\begin{align}
H_{\text{pg}}=\sum_{\boldsymbol{k}}\left[u_{\boldsymbol{k}}^{\dagger},l_{\boldsymbol{k}}^{\dagger}\right]\left[\begin{array}{cc}
\omega_{\text{pg}}(\boldsymbol{k}) & 0\\
0 & -\omega_{\text{pg}}(\boldsymbol{k})
\end{array}\right]\left[\begin{array}{c}
u_{\boldsymbol{k}}\\
l_{\boldsymbol{k}}
\end{array}\right]=\sum_{\boldsymbol{k}}\omega_{\text{pg}}(\boldsymbol{k})(u_{\boldsymbol{k}}^{\dagger}u_{\boldsymbol{k}}-l_{\boldsymbol{k}}^{\dagger}l_{\boldsymbol{k}}),\label{S5}
\end{align}
where the eigenoperators $u_{\boldsymbol{k}}=(a_{\boldsymbol{k}}+ b_{\boldsymbol{k}}e^{i\phi(\boldsymbol{k})})/\sqrt{2}$ and $l_{\boldsymbol{k}}=(a_{\boldsymbol{k}}- b_{\boldsymbol{k}}e^{i\phi(\boldsymbol{k})})/\sqrt{2}$ represent the annihilation operators for the upper and lower band modes, respectively, and the phase factor is defined by $\phi(\boldsymbol{k})\equiv\arctan\left(\Im[f(\boldsymbol{k})]/\Re[f(\boldsymbol{k})]\right)$. The dispersion relation $\omega_{\text{pg}}(\boldsymbol{k})$ for this photonic graphene without dissipation is given by\,\cite{SMRevModPhys.81.109}
\begin{align}
\omega_{\text{pg}}(\boldsymbol{k})=J\sqrt{3+2\cos(k_{1}-k_{2})+2\cos k_{1}+2\cos k_{2}}.\label{S6}
\end{align}
We now summarize the key spectral properties of the photonic graphene in the absence of dissipation: (i) The upper and lower bands touch at two seperate momenta points, $\boldsymbol{\mathrm{K}}_{+} = 2\pi(1, -1)/3$ and $\boldsymbol{\mathrm{K}}_{-} = 2\pi(-1, 1)/3$, known as a pair of inequivalent Dirac points, where the frequency $\omega_{\text{ph}}(\boldsymbol{\mathrm{K}}_{\pm})$ is exactly zero. In the vicinity of these points, the energy dispersion becomes linear. To first order, we have $f(\boldsymbol{\mathrm{K}}_{\pm} + \boldsymbol{q}) \approx J (\boldsymbol{w}_{\pm} \cdot \boldsymbol{q})$, where $\boldsymbol{w}_{\pm} = i(e^{\pm 2\pi i/3}, e^{\mp 2\pi i/3})$. (ii) The energy band extends from $-3J$ to $3J$ and is symmetric about the reference energy, i.e., $E=0$. A key feature emerges in the thermodynamic limit: the density of state (DOS) vanishes exactly at the band center (the Dirac points) yet diverges around $|E| = J$. It suggests that an initially excited QE coupled to this photonic graphene, when its energy is resonant with bosonic modes, becomes protected from radiating energy into the lattice. In contrast, a QE with energy $|E|=J$, falling within the divergent DOS region, would dissipate its energy almost instantaneously.

\subsection{The spectrum of photonic graphene with dissipation}\label{IB}
Having elucidated the ideal case, we now introduce a more realistic scenario: a non-Hermitian photonic graphene with engineered photon loss\,\cite{SMPhysRevLett.129.223601}, manifesting as independently controlled loss rates $\kappa_a$ and $\kappa_b$ for sublattices A and B, respectively. Following the standard treatment for open quantum systems, the related dynamics is described by a quantum master equation derived under the Born-Markov and rotating-wave approximations\,\cite{SMPhysRevA.91.042116,SMPhysRevResearch.2.013369}, which yields
\begin{align}\label{S7}
	\dot{\rho}_t=-i[H_{\text{atom}}+H_{\text{pg}}+H_{\text{int}},\rho_t]+\kappa_{a}\sum_{\boldsymbol{n}}\mathcal{D}[a_{\boldsymbol{n}}]\rho_t
+\kappa_{b}\sum_{\boldsymbol{n}}\mathcal{D}[b_{\boldsymbol{n}}]\rho_t,
\end{align}
where $H_{\text{atom}}$ and $H_{\text{int}}$ are the system Hamiltonian containing only the QEs and the light-matter interaction Hamiltonian, respectively. The bath Hamiltonian $H_{\text{pg}}$ is defined by Eq.~(\ref{S1}). The photon dissipators for the two sublattices are captured by the second and third terms in Eq.~(\ref{S7}), with the loss rates for sublattices A and B controlled by $\kappa_{a}$ and $\kappa_{b}$, respectively. The dissipation is described by the Lindblad superoperator $\mathcal{D}[L]$, defined as $\mathcal{D}[L]\rho \equiv L\rho L^\dagger - \{L^\dagger L,\rho\}/2$ with $\{\mathcal{O}_{1},\mathcal{O}_{2}\}=\mathcal{O}_{1}\mathcal{O}_{2}+\mathcal{O}_{2}\mathcal{O}_{1}$ the anticommutator. To solve the equation of motion described by Eq.~(\ref{S7}) in the single-excitation subspace, we recast the master equation into a convenient form
\begin{align}\label{S8}
\dot{\rho}_t = -i(H_{\mathrm{eff}} \rho_t - \rho_t H_{\mathrm{eff}}^\dagger) + \kappa_a \sum_{\boldsymbol{n}}a_{\boldsymbol{n}}\rho_t a_{\boldsymbol{n}}^\dagger + \kappa_b \sum_{\boldsymbol{n}}b_{\boldsymbol{n}}\rho_t b_{\boldsymbol{n}}^\dagger.
\end{align}

\begin{figure}
  \centering
  \includegraphics[width=18cm]{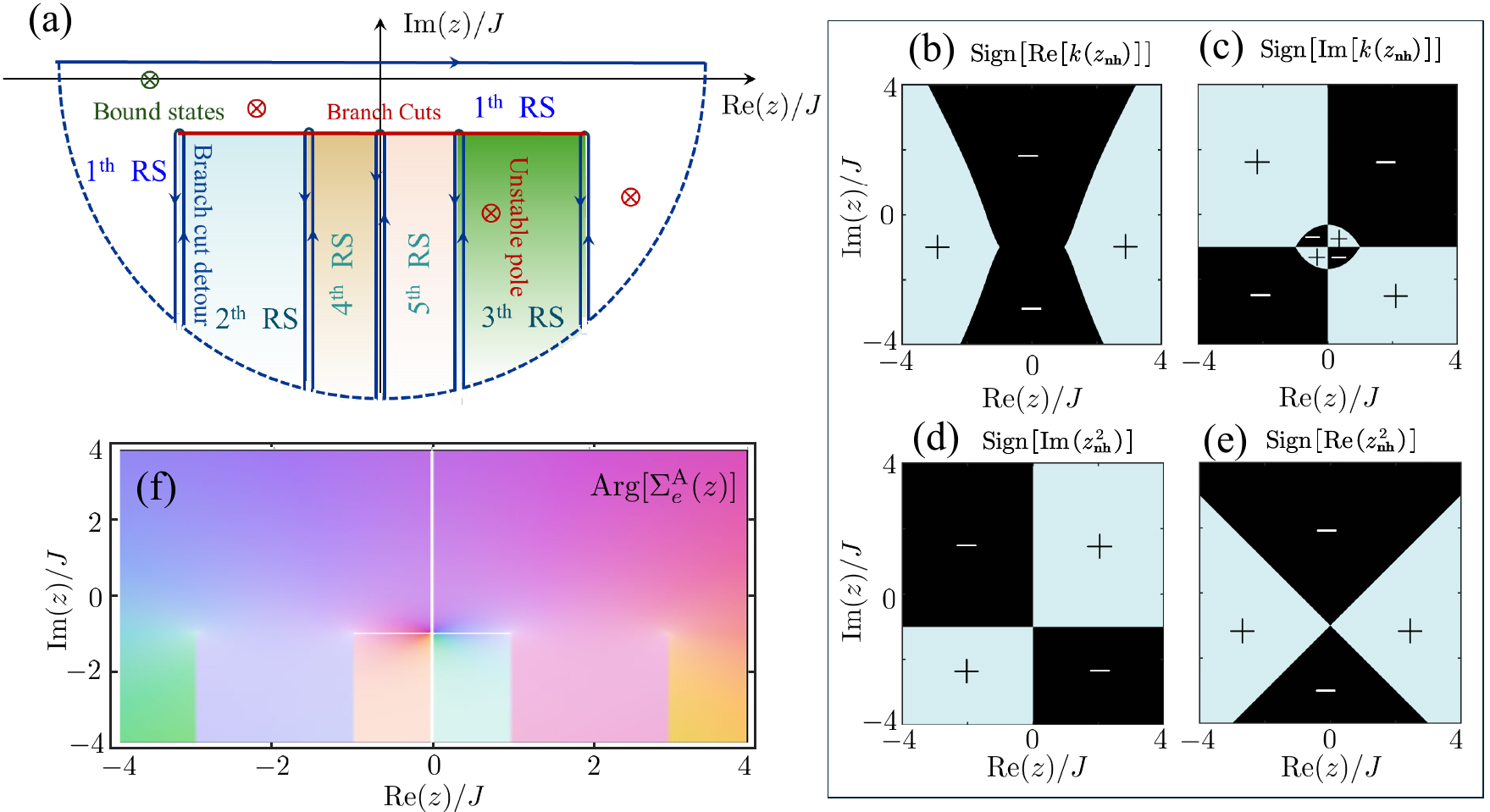}
  \caption{(a) Contour of integration employed to evaluate the amplitude $C_{e}(t)$ for a QE coupled to honeycomb bath with homogeneous dissipation. The associated self-energy $\Sigma_{e}^{\A}(z)$ exhibits discontinuities at $z=\pm 3J-i\kappa_{+},\pm J-i\kappa_{+},-i\kappa_{+},$ necessitating a detour around these non-analytic points and resulting in branch-cut contributions to the quantum dynamics. The poles on the real axis and in the lower half-plane, which correspond to bound states and unstable states on different Riemann sheets, are color-coded in green and red, respectively. The five Riemann sheets are distinguished by coloring. More specifically, at the band edges $E = 3J$ and $E = -3J$, the integration path switches from the first Riemann sheet (white region) to the third (green region), and from the second sheet (blue region) to the first, respectively. At the edges of $E = J$ and $E = -J$, it switches from the third sheet to the fifth (pink region), and from the fourth sheet (brown region) to the second, respectively. At the band center, the path switches from the fifth to the fourth Riemann sheet. (b-e) Sign of the $\mathrm{Re}[k(z_{\mathrm{nh}})]$, $\mathrm{Im}[k(z_{\mathrm{nh}})]$, $\mathrm{Im}(z^{2}_{\mathrm{nh}})$, and $\mathrm{Re}(z^{2}_{\mathrm{nh}})$ serves as the criterion to identify the specific integration domain needed for applying Cauchy's Residue Theorem. (f) Argument $\mathrm{Arg}[\Sigma_{e}^{\text{A}}(z)]$ of the continued atomic self-energy in Eq.\,(\ref{S55}), considering the QE is coupled to the A sublattice. }\label{figS2}
\end{figure}

Here, the dynamics are governed by an effective non-Hermitian Hamiltonian $H_{\mathrm{eff}} = H_{\mathrm{tot}} - i(\kappa_{a}/2)\sum_{\mathbf{n}} a^\dagger_{\mathbf{n}} a^{}_{\mathbf{n}} - i(\kappa_{b}/2) \sum_{\mathbf{n}} b^\dagger_{\mathbf{n}} b^{}_{\mathbf{n}}$ with $H_{\mathrm{tot}}=H_{\text{atom}}+H_{\text{pg}}+H_{\text{int}}$, or more generally, $H_{\mathrm{eff}} = H_{\text{sys}} + H^{\text{eff}}_{\text{pg}} + H_{\text{int}}$, where $H^{\text{eff}}_{\text{pg}} = H_{\text{pg}} - i\sum_{\boldsymbol{n}} (\kappa_a a_{\boldsymbol{n}}^\dagger a^{}_{\boldsymbol{n}} + \kappa_b b_{\boldsymbol{n}}^\dagger b^{}_{\boldsymbol{n}})/2$. The remaining terms, $\sum_{\boldsymbol{n}}a_{\boldsymbol{n}}\rho_t a_{\boldsymbol{n}}^\dagger$ and $\sum_{\boldsymbol{n}}b_{\boldsymbol{n}}\rho_t b_{\boldsymbol{n}}^\dagger$, are the socalled ``recycling" terms that describe quantum jumps. The formal solution of Eq.~(\ref{S8}) can be expressed as
\begin{align}\label{S9}
\rho_t = e^{-i H^{}_{\mathrm{eff}} t} \rho_0 e^{i H_{\mathrm{eff}}^\dagger t} + \int_0^t \dd{s}\,\mathcal{P}\left[e^{-i H^{}_{\mathrm{eff}} s} \rho_0 e^{i H_{\mathrm{eff}}^\dagger s}\right],
\end{align}
where the jump operator $\mathcal{P}[\mathcal{O}] = \sum_{\boldsymbol{n}} a_{\boldsymbol{n}} [\mathcal{O}] a^\dagger_{\boldsymbol{n}} + \sum_{\boldsymbol{n}} b_{\boldsymbol{n}} [\mathcal{O}] b^\dagger_{\boldsymbol{n}}$ has been introduced for conciseness. In the single-excitation subspace, any quantum jump projects the quantum states deterministically into the zero-excitation ground state, $\ket{G} \equiv \ket{g}^{\otimes N_{e}} \otimes \ket{\mathrm{vac}}$ with $N_{e}$ the total number of QEs and $\ket{\mathrm{vac}}$ the vacuum state of the photonic graphene. Therefore, the general solution in Eq.\,(\ref{S9}) simplifies to $\rho_t = \tilde{\rho}_t + p_t \ket{G}\bra{G}$, where $\tilde{\rho}_t$ is the ``no-jump" evolution and $p_t = 1 - \braket{\psi(t)}$ is the total probability that a jump has occurred. This solution has a clear interpretation via the quantum trajectory theory\,\cite{SMPhysRevLett.68.580,SMRevModPhys.70.101}. The norm squared of the state, $\braket{\psi(t)} = \| e^{-i H_{\mathrm{eff}} t} \ket{\psi(0)} \|^{2}$, represents the survival probability of the initial excitation, where $\ket{\Psi(t)} = e^{-i H_{\mathrm{eff}} t} \ket{\Psi(0)}$ is the state evolved under the effective non-Hermitian Hamiltonian $H_{\mathrm{eff}}$. A quantum jump occurs at time $t$ if a random number $\delta_t \in [0,1]$ satisfies $\delta_t > \braket{\psi(t)}$. For an ensemble average over infinitely many trajectories ($N_{t} \to \infty$), the density matrix is given by
\begin{align}\label{S10}
\rho_t = \ket{\psi(t)}\bra{\psi(t)} + p_t \ket{G}\bra{G} = e^{-i H_{\mathrm{eff}} t} \rho_0 e^{i H_{\mathrm{eff}}^\dagger t} + p_t \ket{G}\bra{G}.
\end{align}
For a mixed initial state $\rho_0$, the norm $\braket{\psi(t)}$ is generalized to $\mathrm{Tr}[e^{-i H^{}_{\mathrm{eff}} t} \rho_0 e^{i H_{\mathrm{eff}}^\dagger t}]$. A special case arises when the loss rates for two sublattices are equal ($\kappa_{a} = \kappa_{b} = \kappa$). In this scenario, the effective Hamiltonian $H_{\mathrm{eff}}=H_{\mathrm{tot}}-i\kappa\sum_{\boldsymbol{n}} (a_{\boldsymbol{n}}^\dagger a^{}_{\boldsymbol{n}}+b_{\boldsymbol{n}}^\dagger b^{}_{\boldsymbol{n}})/2$ commutes with $H_{\mathrm{tot}}$, leading to a simple closed-form solution as follows
\begin{align}\label{S11}
\rho(t) = e^{-\kappa t} \ket{\psi_{0}(t)}\bra{\psi_{0}(t)} + (1 - e^{-\kappa t}) \ket{G}\bra{G},
\end{align}
where $\ket{\psi_{0}(t)} = e^{-i H_{\mathrm{tot}} t} \ket{\psi(0)}$ is the state evolved under the unitary, lossless Hamiltonian $H_{\mathrm{tot}}$. These results establish a foundation for quantifying how dissipation shapes the associated quantum dynamics.

We can now elucidate the dissipative graphene's key spectral properties by confining the photonic dynamics to the single-excitation subspace, as governed by the effective bath Hamiltonian $H^{\text{eff}}_{\text{ph}}$. Following the same procedure outlined in Eqs.\,(\ref{S1}-\ref{S4}), the corresponding non-Hermitian kernel Hamiltonian is constructed as
\begin{align}\label{S12}
\boldsymbol{\mathrm{h}}_{\boldsymbol{k}}=\left[\begin{array}{cc}
-i\kappa_{a}/2 & f(\boldsymbol{k})\\
f^{*}(\boldsymbol{k}) & -i\kappa_{b}/2
\end{array}\right]=\Re[f(\boldsymbol{k})]\sigma_x-\Im[f(\boldsymbol{k})]\sigma_y-i\kappa_{-}\sigma_{z}-i\kappa_{+}\sigma_{0},
\end{align}
where the quantities $\kappa_{+}=(\kappa_{a}+\kappa_{b})/4$ and $\kappa_{-}=(\kappa_{a}-\kappa_{b})/4$ have been defined. This expression is exactly the result presented in the main text as Eq.\,(3). Notably, the Bloch Hamiltonian $\boldsymbol{\mathrm{h}}_{\boldsymbol{k}}$ possesses passive PT symmetry, as defined by $\sigma_{x}(\boldsymbol{\mathrm{h}}_{\boldsymbol{k}}+i\kappa_{+}\sigma_{0})^{*}\sigma_{x}=\boldsymbol{\mathrm{h}}_{\boldsymbol{k}}+i\kappa_{+}\sigma_{0}$. The non-Hermiticity of the bath originates from two distinct sources: a uniform background dissipation, $-i\kappa_{+}\sigma_{0}$, and a staggered non-Hermitian term, $-i\kappa_{-}\sigma_{z}$, which introduces balanced gain and loss between the two sublattices. If the background loss term $-i\kappa_{+}\sigma_{0}$ is neglected, the Bloch Hamiltonian is found to possess an exact PT symmetry. Diagonalization of $\boldsymbol{\mathrm{h}}_{\boldsymbol{k}}$ yields the following complex eigenenergies
\begin{align}\label{S13}
\omega_{\pm}(\boldsymbol{k})=-i\kappa_{+}\pm\sqrt{\omega^{2}_{\text{pg}}(\boldsymbol{k})-\kappa_{-}^{2}}.
\end{align}
Therefore, the non-Hermitian effective Hamiltonian $H^{\text{eff}}_{\text{pg}}$ for photonic graphene can be rewritten as
\begin{align}\label{S14}
H^{\text{eff}}_{\text{pg}}=\sum_{\boldsymbol{k}}\left[a_{\boldsymbol{k}}^{\dagger},b_{\boldsymbol{k}}^{\dagger}\right]\left[\begin{array}{cc}
-i\kappa_{a}/2 & f(\boldsymbol{k})\\
f^{*}(\boldsymbol{k}) & -i\kappa_{b}/2
\end{array}\right]\left[\begin{array}{c}
a_{\boldsymbol{k}}\\
b_{\boldsymbol{k}}
\end{array}\right]=\sum_{\boldsymbol{k}}\left(\omega_{+}(\boldsymbol{k})u_{\boldsymbol{k},L}^{\dagger}u^{}_{\boldsymbol{k},R}+
\omega_{-}(\boldsymbol{k})l_{\boldsymbol{k},L}^{\dagger}l^{}_{\boldsymbol{k},R}\right),
\end{align}
where the states $\ket{u_{\boldsymbol{k}, R}} \equiv u_{\boldsymbol{k}, R}^{\dagger} \ket{\mathrm{vac}}$ and $\ket{l_{\boldsymbol{k}, R}} \equiv l_{\boldsymbol{k}, R}^{\dagger} \ket{\mathrm{vac}}$ represent the right eigenvectors corresponding to the upper and lower energy bands, respectively. Their associated left eigenvectors are given by $\bra{u_{\boldsymbol{k}, L}} \equiv \bra{\mathrm{vac}} u^{}_{\boldsymbol{k}, L}$ and $\bra{l_{\boldsymbol{k}, L}} \equiv \bra{\mathrm{vac}} l^{}_{\boldsymbol{k}, L}$. They are determined by the eigenvalue equations for the upper band: $H^{\text{eff}}_{\text{pg}}\ket{u_{\boldsymbol{k},R}}=\omega_{+}(\boldsymbol{k})\ket{u_{\boldsymbol{k},R}}$ and $(H^{\text{eff}}_{\text{pg}})^{\dagger}\ket{u_{\boldsymbol{k},L}}=\omega^{*}_{+}(\boldsymbol{k})\ket{u_{\boldsymbol{k},L}}$. The equations for the lower band are analogous. More concretely, they have the forms of
\begin{align}\label{S15}
\ket{u_{\boldsymbol{k},R}}=&\left[1,\frac{\epsilon_{\boldsymbol{k}}+i\kappa_{-}}{f(\boldsymbol{k})}\right]^{{\rm T}}/\sqrt{2},\,\,\,\,\,\,\bra{u_{\boldsymbol{k},L}}=\left[\frac{\epsilon_{\boldsymbol{k}}-i\kappa_{-}}{\epsilon_{\boldsymbol{k}}},\frac{f(\boldsymbol{k})}{\epsilon_{\boldsymbol{k}}}\right]/\sqrt{2},\nonumber\\
\ket{l_{\boldsymbol{k},R}}=&\left[-1,\frac{\epsilon_{\boldsymbol{k}}-i\kappa_{-}}{f(\boldsymbol{k})}\right]^{{\rm T}}/\sqrt{2},\,\,\,\bra{l_{\boldsymbol{k},L}}=\left[-\frac{\epsilon_{\boldsymbol{k}}+i\kappa_{-}}{\epsilon_{\boldsymbol{k}}},\frac{f(\boldsymbol{k})}{\epsilon_{\boldsymbol{k}}}\right]/\sqrt{2},
\end{align}
where $\epsilon_{\boldsymbol{k}}\equiv\sqrt{\omega^{2}_{\text{pg}}(\boldsymbol{k})-\kappa_{-}^{2}}$ has been introduced. Obviously, the above eigenspectrum can be directly traced back to the dissipation-free case, i.e., $\omega_{\pm}(\boldsymbol{k})=\omega_{\text{pg}}(\boldsymbol{k}), u_{\boldsymbol{k},R}=u_{\boldsymbol{k},L}=u_{\boldsymbol{k}},l_{\boldsymbol{k},R}=l_{\boldsymbol{k},L}=l_{\boldsymbol{k}}$, when $\kappa_{a}=\kappa_{b}=0$.
\begin{figure}
  \centering
  \includegraphics[width=18cm]{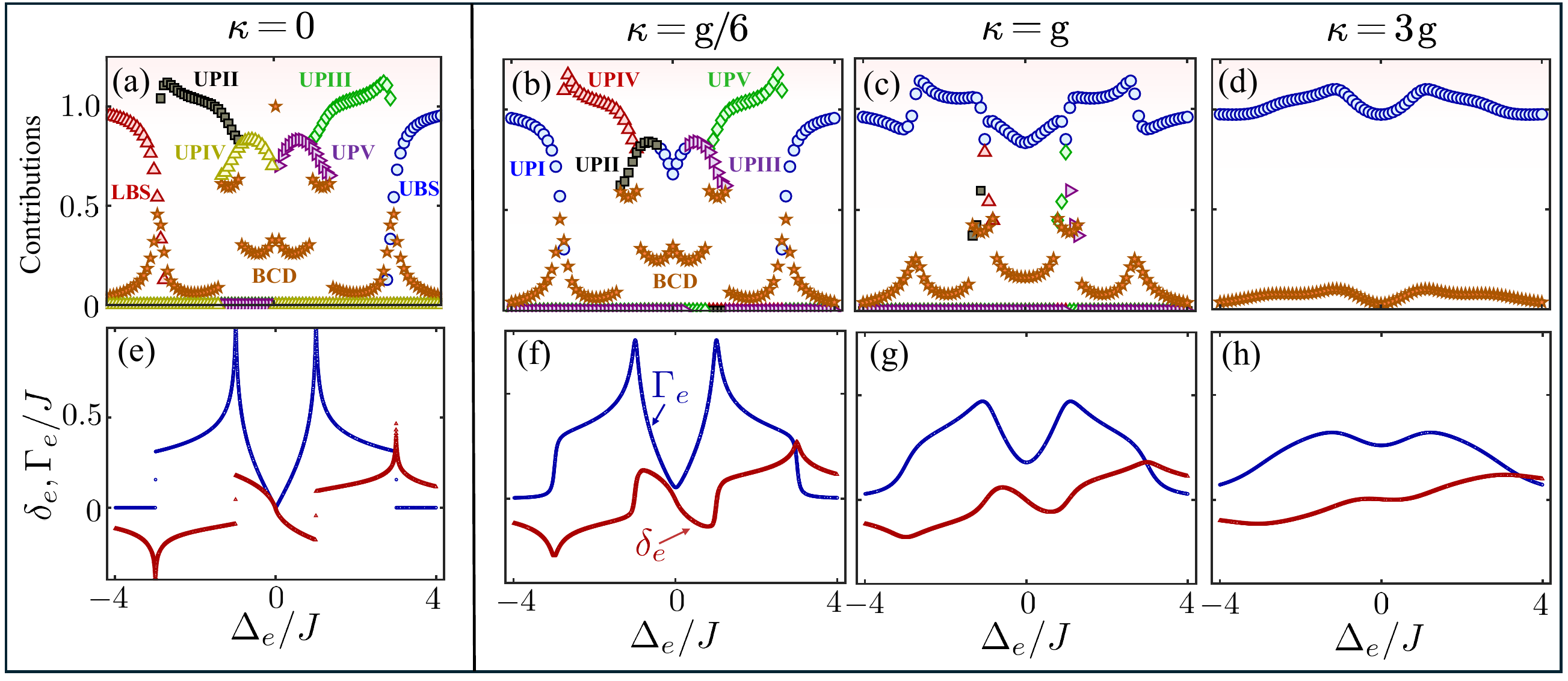}
  \caption{(a) Absolute values of the contributions to $C_{e}^{\text{A}}(0)$ without dissipation: LBS (red triangles), the lower bound state; UBS (blue circles), the upper bound state; UPII-V, unstable poles on Riemann sheets II-V (see legend for markers); BCD (brown stars), the branch-cut detours. Also, dynamics contributions of $C_{e}^{\text{A}}(t)$ at time $t = 0$  are shown as a function of detuning $\Delta_{e}$ for different homogeneous dissipation strengths: (b) $\kappa=\text{g}/6$, (c) $\kappa=\text{g}$, and (d) $\kappa=3\text{g}$. Symbols denote: UPI (blue circles), UPII (black squares), UPIII (purple triangles), UPIV (red triangles), UPIV (green diamonds), and BCD (brown stars). The light-matter coupling strength implemented here reads $\text{g}=0.6J$. Panels (e-h) present the Markovian decay rates (blue lines) and frequency shifts (red lines) as functions of detuning $\Delta_{e}$, corresponding to the dissipation strengths in panels (a-d), respectively.}\label{figS3}
\end{figure}
\subsection{The calculation of self-energy for a single quantum emitter}\label{IC}
We proceed to give an analytical description of the quantum dynamics governing the interaction between a single two-level quantum emitter with a photonic graphene. The effective Hamiltonian of the total system plus photonic environment reads $H_{\text{tot}} = H_{\text{atom}} + H^{\text{eff}}_{\text{pg}} + H_{\text{int}}$. Here, $H_{\text{atom}}=\omega_{e}\sigma^{\dagger}\sigma$ is the free Hamiltonian of the two-level QE with $\omega_{e}$ the atomic transition frequency, and the effective bath Hamiltonian $H^{\text{eff}}_{\text{pg}}$ is given by Eq.\,(\ref{S14}). The interaction Hamiltonian between the system and the bath under the rotating-wave approximation can be expressed as
\begin{align}\label{S16}
H_{{\rm int}}=\text{g}(c_{\boldsymbol{n}_{e},\alpha}^{\dagger}\sigma+\mathrm{H.c.})=\frac{\text{g}}{N}\sum_{\boldsymbol{k}}(c_{\boldsymbol{k},\alpha}^{\dagger}\sigma e^{-i\boldsymbol{k}\cdot\boldsymbol{n}_{e}}+\mathrm{H.c.}),
\end{align}
where $\boldsymbol{n}_{e}$ is the index identifying the cavity mode to which the QE is coupled and $\text{g}$ the coupling strength. Note that the second subscript $\alpha\in\{\text{A},\text{B}\}$ in
$c_{\boldsymbol{n}_{e},\alpha}$ denotes the sublattice, with $\alpha=\text{A}$ and $\text{B}$ corresponding to $c_{\boldsymbol{n}_{e},\text{A}}=a_{\boldsymbol{n}_{e}}$ and $c_{\boldsymbol{n}_{e},\text{B}}=b_{\boldsymbol{n}_{e}}$, respectively, which is same for $c_{\boldsymbol{k},\alpha}$. Notably, The interaction Hamiltonian $H_{{\rm int}}$ conserves the total number of excitations, and therefore, the state within the single-excitation manifold for $t>0$ is given by
\begin{align}\label{S17}
\ket{\Psi_t} = \left(C_{e}(t)\sigma^{\dagger}+\sum\limits_{\boldsymbol{n}}\left(C^{\text{A}}_{\boldsymbol{n}}(t)a_{\boldsymbol{n}}^{\dagger}+C^{\text{B}}_{\boldsymbol{n}}(t)b_{\boldsymbol{n}}^{\dagger}\right)\right)|g;\text{vac}\rangle,
\end{align}
where $C_{e}(t)$ and $C^{\text{A}}_{\boldsymbol{n}}(t) \left(C^{\text{B}}_{\boldsymbol{n}}(t)\right)$ are the amplitudes of probability at time $t$ to find an excitation populated in the QE and in the bosonic mode located at site $\boldsymbol{n}$ of sublattice A (B), respectively.

The analytical solution for the quantum dynamics is available within the framework in the thermodynamic limit $N\to\infty$. It is under this foundational premise that the system's time-evolution operator,
$U(t)=e^{-iH_{\text{tot}}t}$, is constructed via an inverse Fourier transform of its associated Green function. The explicit integral representation is given as\,\cite{SMCohen_book,SMsciadv0297}
\begin{align}\label{S18}
U(t)=-\frac{1}{2\pi i}\int_{\mathcal{C}}G_{\text{tot}}(z)e^{-izt}\dd{z}=-\frac{1}{2\pi i}\int_{\mathcal{C}}\frac{1}{z-H_{\text{tot}}}e^{-izt}\dd{z},
\end{align}
where the integration contour $\mathcal{C}$ is defined to lie immediately above the real axis in the complex plane, extending from negative to positive infinity as illustrated in Fig.~\ref{figS2}(a). To subsequently investigate the dynamical behavior of the QE, our analysis focuses on the single-excitation subspace, requiring the projection of the full evolution operator $U(t)$ onto the subspace described by
\begin{align}\label{S19}
	\mathcal{P}\equiv \ketbra{e}\otimes\ketbra{\text{vac}},\quad \mathcal{Q}\equiv\ketbra{g}\otimes\sum_{\boldsymbol{n}}\left(a^\dagger_{\boldsymbol{n}}\ketbra{\text{vac}}a_{\boldsymbol{n}}^{}+b^\dagger_{\boldsymbol{n}}\ketbra{\text{vac}}b_{\boldsymbol{n}}^{}\right),
\end{align}
which satisfy $\mathcal{P}+\mathcal{Q}=\mathbb{I}_1$ with $\mathbb{I}_1$ being the identity operator in the single-excitation subspace. As a consequence, the resolvent $G_{{\rm{tot}}}(z) = (z-H_{{\rm{tot}}})^{-1}$ can be projected onto
\begin{align}\label{S20}
\mathcal{P}G_{\text{tot}}(z)\mathcal{P}=\frac{\mathcal{P}}{z-E_{\mathcal{P}}-\Sigma_{\mathcal{P}}(z)},\,\,\,
\mathcal{Q}G_{\text{tot}}(z)\mathcal{P}=\frac{\mathcal{Q}}{z-E_{\mathcal{Q}}-\Sigma_{\mathcal{Q}}(z)}H_{{\rm int}}\frac{\mathcal{P}}{z-E_{\mathcal{P}}-\Sigma_{\mathcal{P}}(z)},
\end{align}
where the notations $E_{X}\equiv X(H_{\text{atom}}+H^{\text{eff}}_{\text{pg}})X$ and $\Sigma_{X}(z)=X\boldsymbol{\Sigma}(z)X$ have been introduced for convenience. Here, $\boldsymbol{\Sigma}(z)$ is the level-shift operator and is given by
\begin{align}\label{S21}
\boldsymbol{\Sigma}(z) = H_{{\rm int}}+H_{{\rm int}}\frac{\mathcal{Q}}{z-E_{\mathcal{Q}}-\mathcal{Q}H_{{\rm int}}\mathcal{Q}}H_{{\rm int}}.
\end{align}

It is now possible to obtain the self-energyies $\Sigma_{e}^{\alpha}(z)=\bra{e}\boldsymbol{\Sigma}(z)\ket{e}$ of a single QE coupled to different sublattices $\alpha=\A$ and $\B$, which have the form of
\begin{align}
\Sigma_{e}^{\A}(z)=&\,\text{g}^{2}\left\langle {\rm vac}\right|c_{\boldsymbol{n}_{e},\A}(z-H^{\text{eff}}_{\text{pg}})^{-1}c_{\boldsymbol{n}_{e},\A}^{\dagger}\left|{\rm vac}\right\rangle =\,\text{g}^{2}\left\langle {\rm vac}\right|c_{\boldsymbol{n}_{e},\A}\mathcal{Q}(z-H^{\text{eff}}_{\text{pg}})^{-1}\mathcal{Q}c_{\boldsymbol{n}_{e},\A}^{\dagger}\left|{\rm vac}\right\rangle \nonumber \\=&\,\text{g}^{2}\left\langle {\rm vac}\right|c_{\boldsymbol{n}_{e},\A}\sum_{\boldsymbol{k}}\left[\frac{\left|u_{\boldsymbol{k},R}\right\rangle \left\langle u_{\boldsymbol{k},L}\right|}{z+i\kappa_{+}-\epsilon_{\boldsymbol{k}}}+\frac{\left|l_{\boldsymbol{k},R}\right\rangle \left\langle l_{\boldsymbol{k},L}\right|}{z+i\kappa_{+}+\epsilon_{\boldsymbol{k}}}\right]c_{\boldsymbol{n}_{e},\A}^{\dagger}\left|{\rm vac}\right\rangle\nonumber \\=&\frac{\text{g}^{2}}{N^{2}}\sum_{\boldsymbol{k},\boldsymbol{k}',\boldsymbol{k}^{\prime\prime}}\left\langle {\rm vac}\right|a_{\boldsymbol{k}'}\left[\frac{\left|u_{\boldsymbol{k},R}\right\rangle \left\langle u_{\boldsymbol{k},L}\right|}{z+i\kappa_{+}-\epsilon_{\boldsymbol{k}}}+\frac{\left|l_{\boldsymbol{k},R}\right\rangle \left\langle l_{\boldsymbol{k},L}\right|}{z+i\kappa_{+}+\epsilon_{\boldsymbol{k}}}\right]a_{\boldsymbol{k}^{\prime\prime}}^{\dagger}\left|{\rm vac}\right\rangle \nonumber \\=&\,\frac{\text{g}^{2}}{2N^{2}}\sum_{\boldsymbol{k},\boldsymbol{k}',\boldsymbol{k}^{\prime\prime}}\left[\frac{\epsilon_{\boldsymbol{k}^{\prime\prime}}-i\kappa_{-}}{z+i\kappa_{+}-\epsilon_{\boldsymbol{k}}}
+\frac{\epsilon_{\boldsymbol{k}^{\prime\prime}}+i\kappa_{-}}{z+i\kappa_{+}+\epsilon_{\boldsymbol{k}}}\right]\frac{\delta_{\boldsymbol{k},\boldsymbol{k}'}
\delta_{\boldsymbol{k},\boldsymbol{k}^{\prime\prime}}}{\epsilon_{\boldsymbol{k}^{\prime\prime}}}\nonumber\\
=&\,\frac{\text{g}^{2}}{N^{2}}\sum_{\boldsymbol{k}}\frac{z+i\kappa_{b}/2}{(z+i\kappa_{+})^{2}-\epsilon_{\boldsymbol{k}}^{2}}=
\frac{\text{g}^{2}}{N^{2}}\sum_{\boldsymbol{k}}\frac{z+i\kappa_{b}/2}{z_{{\rm nh}}^{2}-\omega^{2}_{\text{pg}}({\boldsymbol{k}})}=\,\text{g}^{2}\iint\frac{\dd{\boldsymbol{k}}}{(2\pi)^{2}}\frac{z+i\kappa_{b}/2}{z_{{\rm nh}}^{2}-\omega^{2}_{\text{pg}}(\boldsymbol{k})},\label{S22}\\
\Sigma_{e}^{\B}(z)=&\,\text{g}^{2}\left\langle {\rm vac}\right|c_{\boldsymbol{n}_{e},\B}(z-H^{\text{eff}}_{\text{pg}})^{-1}c_{\boldsymbol{n}_{e},\B}^{\dagger}\left|{\rm vac}\right\rangle =\,\text{g}^{2}\iint\frac{\dd{\boldsymbol{k}}}{(2\pi)^{2}}\frac{z+i\kappa_{a}/2}{z_{{\rm nh}}^{2}-\omega^{2}_{\text{pg}}({\boldsymbol{k}})},\label{S23}
\end{align}
where we have defined $z^{2}_{{\rm nh}}\equiv(z+i\kappa_{+})^{2}+\kappa_{-}^{2}$. We proceed to derive the analytical expressions of these self-energies by employing a definite integral, given by
\begin{align}\label{S24}
\int_{d}^{c}\frac{\dd{x}}{\sqrt{(a-x)(b-x)(c-x)(x-d)}}=\frac{2}{\sqrt{(a-c)(b-d)}}K(k^2),
\end{align}
where $K(m)$ is the complete elliptic integral of the first kind defined by $K(m) = \int_{0}^{\pi/2}d\theta\left[1-m\sin^{2}(\theta)\right]^{-1/2}$. Substituting Eq.\,(\ref{S24}) into Eqs.\,(\ref{S22}) and (\ref{S23}) yields
\begin{align}
\Sigma_{e}^{\text{A}}(z)=&\,\frac{(z+i\kappa_{b}/2)\text{g}^{2}}{2\pi}\int_{-\pi}^{\pi}\frac{\dd{k_{1}}}{\sqrt{(z_{{\rm nh}}^{2}-3J^{2}-2J^{2}\cos k_{1})^{2}-(2J^{2}\cos\frac{k_{1}}{2})^{2}}}\nonumber\\
=&\,\frac{(z+i\kappa_{b}/2)\text{g}^{2}}{2\pi J^{2}}\int_{-1}^{1}\frac{\dd{x}}{\sqrt{(1-x^{2})(\frac{(z_{{\rm nh}}^{2}-J^{2})+2J\sqrt{z_{{\rm nh}}^{2}}}{2J^{2}}-x)(\frac{(z_{{\rm nh}}^{2}-J^{2})-2J\sqrt{z_{{\rm nh}}^{2}}}{2J^{2}}-x)}}\nonumber\\
=&\,\frac{\text{g}^{2}(z+i\kappa_{b}/2)}{4\pi}\frac{8}{(\sqrt{z_{{\rm nh}}^{2}}+3J)^{1/2}(\sqrt{z_{{\rm nh}}^{2}}-J)^{3/2}}K\left[k(z_{\text{nh}})^2\right],\label{S25}\\
\Sigma_{e}^{\text{B}}(z)=&\,\frac{\text{g}^{2}(z+i\kappa_{a}/2)}{4\pi}\frac{8}{(\sqrt{z_{{\rm nh}}^{2}}+3J)^{1/2}(\sqrt{z_{{\rm nh}}^{2}}-J)^{3/2}}K\left[k(z_{\text{nh}})^2\right],\label{S26}
\end{align}
where the parameter $k(z_{\text{nh}})$ reads
\begin{align}\label{S27}
k(z)=\sqrt{\frac{16\sqrt{z^{2}}J^{3}}{(\sqrt{z^{2}}+3J)(\sqrt{z^{2}}-J)^{3}}}=\frac{4\sqrt[4]{z^{2}}J^{3/2}}{(\sqrt{z^{2}}+3J)^{1/2}(\sqrt{z^{2}}-J)^{3/2}}=\frac{C(z)\sqrt[4]{z^{2}}J^{3/2}}{2},
\end{align}
which allows the above self-energies to be further simplified as
\begin{align}\label{S28}
\Sigma_{e}^{\text{A}}(z)=\frac{\text{g}^{2}(z+i\kappa_{b}/2)}{4\pi}C(z_{{\rm nh}})K\left[k(z_{{\rm nh}})^{2}\right],\,\,\,\,\Sigma_{e}^{\text{B}}(z)=\frac{\text{g}^{2}(z+i\kappa_{a}/2)}{4\pi}C(z_{{\rm nh}})K\left[k(z_{{\rm nh}})^{2}\right].
\end{align}
The availability of these analytical self-energies enables us to determine the relevant probability amplitudes according to the resolvent operator technique, namely: $C_{e}^{\text{A}}(t)$, the excitation amplitude that an emitter coupled to the sublattice A remains in its excited state at time $t$ after spontaneous emission; $C^{\text{A},\text{A(B)}}_{\boldsymbol{n}}(t)$, the probability amplitude at time $t$ to find an excitation in the bosonic mode located at site $\boldsymbol{n}$ of sublattice A(B), given that the QE couples to sublattice A; $C_{e}^{\text{B}}(t)$, the corresponding excited-state amplitude for an emitter coupled to the sublattice B; and $C^{\text{B},\text{A(B)}}_{\boldsymbol{n}}(t)$, the probability amplitude at time $t$ to find an excitation in the bosonic mode located at site $\boldsymbol{n}$ of sublattice A(B), given that the QE couples to sublattice B. It is found that they have the following compact forms
\begin{align}
C_{e}^{\text{A}}(t)&=-\frac{1}{2\pi i}\int_{-\infty}^{\infty}\dd{E}\,\frac{1}{E-\Delta_{e}-\Sigma^{\text{A}}_{e}(E+i0^{+})}e^{-iEt},\label{S29}\\
C_{e}^{\text{B}}(t)&=-\frac{1}{2\pi i}\int_{-\infty}^{\infty}\dd{E}\,\frac{1}{E-\Delta_{e}-\Sigma^{\text{B}}_{e}(E+i0^{+})}e^{-iEt},\label{S30}\\
C^{\text{A,A}}_{\boldsymbol{n}}(t)&=-\frac{1}{2\pi i}\int_{-\infty}^{\infty}\dd{E} \int_{-\pi}^{\pi}\int_{-\pi}^{\pi} \frac{\dd{\boldsymbol{k}}}{(2\pi)^{2}}\,\frac{\text{g}(z+i\kappa_{b}/2)}{\left[z_{{\rm nh}}^{2}-\omega_{\text{pg}}^{2}(\boldsymbol{k})\right]\left[E-\Delta_{e}-\Sigma^{\text{A}}_{e}(E+i0^{+})\right]}e^{i(\boldsymbol{k}\cdot\boldsymbol{n}-E)t},\label{S31}\\
C^{\text{A,B}}_{\boldsymbol{n}}(t)&=-\frac{1}{2\pi i}\int_{-\infty}^{\infty}\dd{E} \int_{-\pi}^{\pi}\int_{-\pi}^{\pi} \frac{\dd{\boldsymbol{k}}}{(2\pi)^{2}}\,\frac{\text{g}(\epsilon_{\boldsymbol{k}}^{2}+\kappa_{-}^{2})/f(\boldsymbol{k})}{\left[z_{{\rm nh}}^{2}-\omega_{\text{pg}}^{2}(\boldsymbol{k})\right]\left[E-\Delta_{e}-\Sigma^{\text{A}}_{e}(E+i0^{+})\right]}e^{i(\boldsymbol{k}\cdot\boldsymbol{n}-E)t},\label{S32}\\
C^{\text{B,A}}_{\boldsymbol{n}}(t)&=-\frac{1}{2\pi i}\int_{-\infty}^{\infty}\dd{E} \int_{-\pi}^{\pi}\int_{-\pi}^{\pi} \frac{\dd{\boldsymbol{k}}}{(2\pi)^{2}}\,\frac{\text{g} f(\boldsymbol{k})}{\left[z_{{\rm nh}}^{2}-\omega_{\text{pg}}^{2}(\boldsymbol{k})\right]\left[E-\Delta_{e}-\Sigma^{\text{B}}_{e}(E+i0^{+})\right]}e^{i(\boldsymbol{k}\cdot\boldsymbol{n}-E)t},\label{S33}\\
C^{\text{B,B}}_{\boldsymbol{n}}(t)&=-\frac{1}{2\pi i}\int_{-\infty}^{\infty}\dd{E} \int_{-\pi}^{\pi}\int_{-\pi}^{\pi} \frac{\dd{\boldsymbol{k}}}{(2\pi)^{2}}\,\frac{\text{g}(z+i\kappa_{a}/2)}{\left[z_{{\rm nh}}^{2}-\omega_{\text{pg}}^{2}(\boldsymbol{k})\right]\left[E-\Delta_{e}-\Sigma^{\text{B}}_{e}(E+i0^{+})\right]}e^{i(\boldsymbol{k}\cdot\boldsymbol{n}-E)t}.\label{S34}
\end{align}
Note that in the above calculations, we have adopted a rotating frame with respect to the resonant frequency of the cavity modes, $\omega_c$. This frequency serves as the energy reference for the problem, such that the effective atomic energy can be expressed as $\Delta_{e} = \omega_{e} - \omega_{c}$. Based on these useful analytical expressions for the self-energies in Eq.\,(\ref{S25}) and Eq.\,(\ref{S26}), the time evolution of both the atomic and photonic parts are, in principle, fully determined, thereby enabling the extraction of any dynamical properties of interest. Notice, the expression in Eqs.\,(\ref{S28}) and (\ref{S29}) are exactly the Eqs.\,(5) and (4) presented in the main text, respectively.

\subsection{The calculation of self-energy for two quantum emitters}\label{ID}
In this subsection, we present a detailed dynamical description for two QEs coupled to a photonic graphene. To maintain full generality, we explicitly consider that the two QEs are coupled to sites $\boldsymbol{n}_{1}$ and $\boldsymbol{n}_{2}$, belonging to sublattices $\alpha$ and $\beta$, respectively. For simplicity, all emitters are assumed to share the same energy detuning $\Delta_{e}$. The QEs and their photonic surroundings, are now captured by the total Hamiltonian $H_{\text{tot}} = H_{\text{atom}} + H^{\text{eff}}_{\text{pg}} + H_{\text{int}}$, where
\begin{align}\label{S35}
H_{\text{atom}}=\Delta_{e}(\sigma_{1}^{\dagger}\sigma_{1}^{}+\sigma_{2}^{\dagger}\sigma_{2}^{})+\Omega(\sigma_{1}^{\dagger}\sigma_{2}^{}+\mathrm{H.c.}),\,\,\,H_{{\rm int}}=\text{g}(c_{\boldsymbol{n}_{1},\alpha}^{\dagger}\sigma_{1}^{}+c_{\boldsymbol{n}_{2},\beta}^{\dagger}\sigma_{2}^{}+\mathrm{H.c.}).
\end{align}
Here, the direct coupling between QEs has been considered with strength $\Omega$, and $H^{\text{eff}}_{\text{pg}}$ corresponds to the structured photon bath as defined in Eq.\,(\ref{S14}), and $H_{\text{int}}$ describes the light-matter interaction under the rotating-wave approximation. This interaction Hamiltonian also conserves the total number of excitations, and therefore, the state within the single-excitation manifold at $t>0$ is given by
\begin{align}\label{eq36}
\ket{\Psi_t} = \left(e_{1}(t)\sigma_{1}^{\dagger}+e_{2}(t)\sigma_{2}^{\dagger}+\sum\limits_{\boldsymbol{n}}\left(C^{\text{A}}_{\boldsymbol{n}}(t)a_{\boldsymbol{n}}^{\dagger}+C^{\text{B}}_{\boldsymbol{n}}(t)b_{\boldsymbol{n}}^{\dagger}\right)\right)|g,g;\text{vac}\rangle,
\end{align}
where $e_{\ell}(t)$ and $C^{\text{A (B)}}_{\boldsymbol{n}}(t)$ are the probability amplitudes at time $t$ to find an excitation populated in the $\ell$th QE and in the cavity mode located at site $\boldsymbol{n}$ of sublattice A (B), respectively.

The time-evolution operator for the multi-emitter scenario retains a structure analogous to that of Eq.~(\ref{S18}), with the essential modification that the total Hamiltonian is now given by the multi-emitter form in Eq.~(\ref{S35}). This extension allows the formalism to capture the collective behaviors for the emitters coupled to this photonic graphene. We will focus our analysis on the single-excitation subspace, paying particular attention to inter-emitter excitation transfer. This requires the projection of the full evolution operator $U(t)$ onto the following subspace:
\begin{align}\label{S37}
	\mathcal{P}\equiv (\ketbra{e,g}+\ketbra{g,e})\otimes\ketbra{\text{vac}},\quad \mathcal{Q}\equiv\ketbra{g,g}\otimes\sum_{\boldsymbol{n}}\left(a^\dagger_{\boldsymbol{n}}\ketbra{\text{vac}}a_{\boldsymbol{n}}^{}+b^\dagger_{\boldsymbol{n}}\ketbra{\text{vac}}b_{\boldsymbol{n}}^{}\right),
\end{align}
which satisfy $\mathcal{P}+\mathcal{Q}=\mathbb{I}_1$, with $\mathbb{I}_1$ being the identity operator in the single-excitation subspace. In the basis spanned by $\ket{e,g;{\rm vac}}\equiv\ket{s_{1}},\ket{g,e;{\rm vac}}\equiv\ket{s_{2}}$, the associated Green function $G_{{\rm{tot}}}(z) = (z-H_{{\rm{tot}}})^{-1}$ takes the following matrix form
\begin{align}\label{S38}
\left[\begin{array}{cc}
G_{11}(z) & G_{12}(z)\\
G_{21}(z) & G_{22}(z)
\end{array}\right]	=	\left[\begin{array}{cc}
z-\Delta_{e}-\Sigma_{11}(z) & -\Omega-\Sigma_{12}(z)\\
-\Omega-\Sigma_{21}(z) & z-\Delta_{e}-\Sigma_{22}(z)
\end{array}\right]^{-1},
\end{align}
where $G_{mn}(z)=\bra{e_{m}}\mathcal{P}G_{{\rm tot}}(z)\mathcal{P}\ket{e_{n}},\Sigma_{mn}(z)=\bra{e_{m}}\mathcal{P}\boldsymbol{\Sigma}(z)\mathcal{P}\ket{e_{n}}$. According to Eqs.\,(\ref{S21}) and (\ref{S35}), the self-energies in Eq.\,(\ref{S38}) can be expressed as
\begin{align}
\Sigma_{11}^{\alpha\alpha}(z)=\Sigma_{11}(z)=\text{g}^{2}\langle{\rm vac}|c_{\boldsymbol{n}_{1},\alpha}(z-H^{\text{eff}}_{\text{pg}})^{-1}c_{\boldsymbol{n}_{1},\alpha}^{\dagger}|{\rm vac}\rangle=\text{g}^{2}\langle\boldsymbol{n}_{1},\alpha|(z-H^{\text{eff}}_{\text{pg}})^{-1}|\boldsymbol{n}_{1},\alpha\rangle,
\label{S39}\\
\Sigma_{12}^{\alpha\beta}(z)=\Sigma_{12}(z)=\text{g}^{2}\langle{\rm vac}|c_{\boldsymbol{n}_{1},\alpha}(z-H^{\text{eff}}_{\text{pg}})^{-1}c_{\boldsymbol{n}_{2},\beta}^{\dagger}|{\rm vac}\rangle=\text{g}^{2}\langle\boldsymbol{n}_{1},\alpha|(z-H^{\text{eff}}_{\text{pg}})^{-1}|\boldsymbol{n}_{2},\beta\rangle,
\label{S40}\\
\Sigma_{21}^{\beta\alpha}(z)=\Sigma_{21}(z)=\text{g}^{2}\langle{\rm vac}|c_{\boldsymbol{n}_{2},\beta}(z-H^{\text{eff}}_{\text{pg}})^{-1}c_{\boldsymbol{n}_{1},\alpha}^{\dagger}|{\rm vac}\rangle=\text{g}^{2}\langle\boldsymbol{n}_{2},\beta|(z-H^{\text{eff}}_{\text{pg}})^{-1}|\boldsymbol{n}_{1},\alpha\rangle,
\label{S41}\\
\Sigma_{22}^{\beta\beta}(z)=\Sigma_{22}(z)=\text{g}^{2}\langle{\rm vac}|c_{\boldsymbol{n}_{2},\beta}(z-H^{\text{eff}}_{\text{pg}})^{-1}c_{\boldsymbol{n}_{2},\beta}^{\dagger}|{\rm vac}\rangle=\text{g}^{2}\langle\boldsymbol{n}_{2},\beta|(z-H^{\text{eff}}_{\text{pg}})^{-1}|\boldsymbol{n}_{2},\beta\rangle.\label{S42}
\end{align}
Relying on these general expressions for the self-energies of QEs, the projected time-evolution operator can be directly obtained from Eqs.\,(\ref{S18}),(\ref{S37}) and (\ref{S38}), yielding
\begin{align}\label{S43}
\mathcal{P}U(t)\mathcal{P}&=-\frac{1}{2\pi i}\int_{\mathcal{C}}\mathcal{P}G_{\text{tot}}(z)\mathcal{P}e^{-izt}\dd{z}\nonumber\\
&=-\frac{1}{2\pi i}\int_{\mathcal{C}}dz\frac{e^{-izt}}{\mathscr{D}(z)}\left[\begin{array}{c}
|s_{1}\rangle\\
|s_{2}\rangle
\end{array}\right]^{T}\left[\begin{array}{cc}
z-\Delta_{e}-\Sigma_{22}^{\beta\beta}(z) & \Omega+\Sigma_{21}^{\beta\alpha}(z)\\
\Omega+\Sigma_{12}^{\alpha\beta}(z) & z-\Delta_{e}-\Sigma_{11}^{\alpha\alpha}(z)
\end{array}\right]\left[\begin{array}{c}
\langle s_{1}|\\
\langle s_{2}|
\end{array}\right],
\end{align}
where the determinant $\mathscr{D}(z)$ reads
\begin{align}\label{S44}
\mathscr{D}(z)=[z-\Delta_{e}-\Sigma_{11}^{\alpha\alpha}(z)][z-\Delta_{e}-\Sigma_{22}^{\beta\beta}(z)]-[\Omega+\Sigma_{12}^{\alpha\beta}(z)][\Omega+\Sigma_{21}^{\beta\alpha}(z)].
\end{align}
From now on, we can describe the dynamics of graphenic QED containing two QEs. For instance, if the first emitter is initially excited, after evolving over a duration $t$, the probability amplitudes for the population at each emitter are given by
\begin{align}
e_{1}(t)=&\langle s_{1}|PU(t)P|s_{1}\rangle=-\frac{1}{2\pi i}\int_{\mathcal{C}}\frac{z-\Delta_{e}-\Sigma_{22}^{\beta\beta}(z)}{\mathscr{D}(z)}e^{-izt}\dd{z},\label{S45}\\
e_{2}(t)=&\langle s_{2}|PU(t)P|s_{1}\rangle=-\frac{1}{2\pi i}\int_{\mathcal{C}}\frac{\Omega+\Sigma_{12}^{\alpha\beta}(z)}{\mathscr{D}(z)}e^{-izt}\dd{z}.\label{S46}
\end{align}

Upon specifying the given atomic coupling configuration, the expressions for the two-emitter self-energies in Eqs.\,(\ref{S39}-\ref{S42}) can be further evaluated. We consider three distinct coupling scenarios: ($i$) two QEs couple to the A sublattice, ($ii$) both couple to the B sublattice, and ($iii$) one emitter couple to the A sublattice and the other to the B sublattice. For the case ($i$), they have the following simple forms
\begin{align}
\Sigma_{11}^{\A\A}(z)=&\,\text{g}^{2}\left\langle {\rm vac}\right|c_{\boldsymbol{n}_{1},\A}(z-H^{\text{eff}}_{\text{pg}})^{-1}c_{\boldsymbol{n}_{1},\A}^{\dagger}\left|{\rm vac}\right\rangle =\,\text{g}^{2}\left\langle {\rm vac}\right|c_{\boldsymbol{n}_{1},\A}\mathcal{Q}(z-H^{\text{eff}}_{\text{pg}})^{-1}\mathcal{Q}c_{\boldsymbol{n}_{1},\A}^{\dagger}\left|{\rm vac}\right\rangle \nonumber \\=&\,\text{g}^{2}\left\langle {\rm vac}\right|c_{\boldsymbol{n}_{1},\A}\sum_{\boldsymbol{k}}\left[\frac{\left|u_{\boldsymbol{k},R}\right\rangle \left\langle u_{\boldsymbol{k},L}\right|}{z+i\kappa_{+}-\epsilon_{\boldsymbol{k}}}+\frac{\left|l_{\boldsymbol{k},R}\right\rangle \left\langle l_{\boldsymbol{k},L}\right|}{z+i\kappa_{+}+\epsilon_{\boldsymbol{k}}}\right]c_{\boldsymbol{n}_{1},\A}^{\dagger}\left|{\rm vac}\right\rangle\nonumber \\=&\frac{\text{g}^{2}}{N^{2}}\sum_{\boldsymbol{k},\boldsymbol{k}',\boldsymbol{k}^{\prime\prime}}\left\langle {\rm vac}\right|a_{\boldsymbol{k}'}\left[\frac{\left|u_{\boldsymbol{k},R}\right\rangle \left\langle u_{\boldsymbol{k},L}\right|}{z+i\kappa_{+}-\epsilon_{\boldsymbol{k}}}+\frac{\left|l_{\boldsymbol{k},R}\right\rangle \left\langle l_{\boldsymbol{k},L}\right|}{z+i\kappa_{+}+\epsilon_{\boldsymbol{k}}}\right]a_{\boldsymbol{k}^{\prime\prime}}^{\dagger}\left|{\rm vac}\right\rangle \nonumber \\=&\,\frac{\text{g}^{2}}{2N^{2}}\sum_{\boldsymbol{k},\boldsymbol{k}',\boldsymbol{k}^{\prime\prime}}\left[\frac{\epsilon_{\boldsymbol{k}^{\prime\prime}}-i\kappa_{-}}{z+i\kappa_{+}-\epsilon_{\boldsymbol{k}}}
+\frac{\epsilon_{\boldsymbol{k}^{\prime\prime}}+i\kappa_{-}}{z+i\kappa_{+}+\epsilon_{\boldsymbol{k}}}\right]\frac{\delta_{\boldsymbol{k},\boldsymbol{k}'}
\delta_{\boldsymbol{k},\boldsymbol{k}^{\prime\prime}}}{\epsilon_{\boldsymbol{k}^{\prime\prime}}}\nonumber\\
=&\,\frac{\text{g}^{2}}{N^{2}}\sum_{\boldsymbol{k}}\frac{z+i\kappa_{b}/2}{(z+i\kappa_{+})^{2}-\epsilon_{\boldsymbol{k}}^{2}}=
\frac{\text{g}^{2}}{N^{2}}\sum_{\boldsymbol{k}}\frac{z+i\kappa_{b}/2}{z_{{\rm nh}}^{2}-\omega^{2}_{\text{pg}}({\boldsymbol{k}})}=\,\text{g}^{2}\iint\frac{\dd{\boldsymbol{k}}}{(2\pi)^{2}}\frac{z+i\kappa_{b}/2}{z_{{\rm nh}}^{2}-\omega^{2}_{\text{pg}}({\boldsymbol{k}})},\label{S47}\\
\Sigma_{12}^{\A\A}(z)=&\,\text{g}^{2}\left\langle {\rm vac}\right|c_{\boldsymbol{n}_{1},\A}(z-H^{\text{eff}}_{\text{pg}})^{-1}c_{\boldsymbol{n}_{2},\A}^{\dagger}\left|{\rm vac}\right\rangle
=\,\text{g}^{2}\iint\frac{\dd{\boldsymbol{k}}}{(2\pi)^{2}}\frac{z+i\kappa_{b}/2}{z_{{\rm nh}}^{2}-\omega^{2}_{\text{pg}}({\boldsymbol{k}})}e^{i\boldsymbol{k}\cdot(\boldsymbol{n}_{1}-\boldsymbol{n}_{2})},\label{S48}\\
\Sigma_{21}^{\A\A}(z)=&\,\text{g}^{2}\left\langle {\rm vac}\right|c_{\boldsymbol{n}_{2},\A}(z-H^{\text{eff}}_{\text{pg}})^{-1}c_{\boldsymbol{n}_{1},\A}^{\dagger}\left|{\rm vac}\right\rangle
=\,\text{g}^{2}\iint\frac{\dd{\boldsymbol{k}}}{(2\pi)^{2}}\frac{z+i\kappa_{b}/2}{z_{{\rm nh}}^{2}-\omega^{2}_{\text{pg}}({\boldsymbol{k}})}e^{-i\boldsymbol{k}\cdot(\boldsymbol{n}_{1}-\boldsymbol{n}_{2})}.\label{S49}
\end{align}
For the case ($ii$), the results are analogous to those in case ($i$), with $\kappa_{b}$ replaced by $\kappa_{a}$ in the final expressions. Therefore, we proceed to evaluate the self-energies for case ($iii$), obtaining
\begin{align}
\Sigma_{12}^{\A\B}(z)=&\,\text{g}^{2}\left\langle {\rm vac}\right|c_{\boldsymbol{n}_{1},\A}(z-H^{\text{eff}}_{\text{pg}})^{-1}c_{\boldsymbol{n}_{2},\B}^{\dagger}\left|{\rm vac}\right\rangle =\,\text{g}^{2}\left\langle {\rm vac}\right|c_{\boldsymbol{n}_{1},\A}\mathcal{Q}(z-H_{\text{B}})^{-1}\mathcal{Q}c_{\boldsymbol{n}_{2},\B}^{\dagger}\left|{\rm vac}\right\rangle \nonumber \\=&\,\text{g}^{2}\left\langle {\rm vac}\right|c_{\boldsymbol{n}_{1},\A}\sum_{\boldsymbol{k}}\left[\frac{\left|u_{\boldsymbol{k},R}\right\rangle \left\langle u_{\boldsymbol{k},L}\right|}{z+i\kappa_{+}-\epsilon_{\boldsymbol{k}}}+\frac{\left|l_{\boldsymbol{k},R}\right\rangle \left\langle l_{\boldsymbol{k},L}\right|}{z+i\kappa_{+}+\epsilon_{\boldsymbol{k}}}\right]c_{\boldsymbol{n}_{2},\B}^{\dagger}\left|{\rm vac}\right\rangle\nonumber \\=&\frac{\text{g}^{2}}{N^{2}}\sum_{\boldsymbol{k},\boldsymbol{k}',\boldsymbol{k}^{\prime\prime}}e^{i\boldsymbol{k}'\cdot\boldsymbol{n}_{1}}e^{-i\boldsymbol{k}^{\prime\prime}\cdot\boldsymbol{n}_{2}}\left\langle {\rm vac}\right|a_{\boldsymbol{k}'}\left[\frac{\left|u_{\boldsymbol{k},R}\right\rangle \left\langle u_{\boldsymbol{k},L}\right|}{z+i\kappa_{+}-\epsilon_{\boldsymbol{k}}}+\frac{\left|l_{\boldsymbol{k},R}\right\rangle \left\langle l_{\boldsymbol{k},L}\right|}{z+i\kappa_{+}+\epsilon_{\boldsymbol{k}}}\right]b_{\boldsymbol{k}^{\prime\prime}}^{\dagger}\left|{\rm vac}\right\rangle \nonumber \\=&\,\frac{\text{g}^{2}}{2N^{2}}\sum_{\boldsymbol{k},\boldsymbol{k}',\boldsymbol{k}^{\prime\prime}}e^{i\boldsymbol{k}'\cdot\boldsymbol{n}_{1}}e^{-i\boldsymbol{k}^{\prime\prime}\cdot\boldsymbol{n}_{2}}\left[\frac{f(\boldsymbol{k}^{\prime\prime})}{z+i\kappa_{+}-\epsilon_{\boldsymbol{k}}}
-\frac{f(\boldsymbol{k}^{\prime\prime})}{z+i\kappa_{+}+\epsilon_{\boldsymbol{k}}}\right]\frac{\delta_{\boldsymbol{k},\boldsymbol{k}'}
\delta_{\boldsymbol{k},\boldsymbol{k}^{\prime\prime}}}{\epsilon_{\boldsymbol{k}^{\prime\prime}}}\nonumber\\
=&\,\frac{\text{g}^{2}}{N^{2}}\sum_{\boldsymbol{k}}\frac{f(\boldsymbol{k})}{(z+i\kappa_{+})^{2}-\epsilon_{\boldsymbol{k}}^{2}}e^{i\boldsymbol{k}\cdot(\boldsymbol{n}_{1}-\boldsymbol{n}_{2})}=\,\text{g}^{2}\iint\frac{\dd{\boldsymbol{k}}}{(2\pi)^{2}}\frac{\omega_{\text{pg}}(\boldsymbol{k})e^{i\phi(\boldsymbol{k})}}{z_{{\rm nh}}^{2}-\omega^{2}_{\text{pg}}({\boldsymbol{k}})}e^{i\boldsymbol{k}\cdot(\boldsymbol{n}_{1}-\boldsymbol{n}_{2})},\label{S50}\\
\Sigma_{12}^{\B\A}(z)=&\,\text{g}^{2}\left\langle {\rm vac}\right|c_{\boldsymbol{n}_{1},\B}(z-H^{\text{eff}}_{\text{pg}})^{-1}c_{\boldsymbol{n}_{2},\A}^{\dagger}\left|{\rm vac}\right\rangle
=\,\text{g}^{2}\iint\frac{\dd{\boldsymbol{k}}}{(2\pi)^{2}}\frac{\omega_{\text{pg}}(\boldsymbol{k})e^{-i\phi(\boldsymbol{k})}}{z_{{\rm nh}}^{2}-\omega^{2}_{\text{pg}}({\boldsymbol{k}})}e^{i\boldsymbol{k}\cdot(\boldsymbol{n}_{1}-\boldsymbol{n}_{2})},\label{S51}\\
\Sigma_{21}^{\B\A}(z)=&\,\text{g}^{2}\left\langle {\rm vac}\right|c_{\boldsymbol{n}_{2},\B}(z-H^{\text{eff}}_{\text{pg}})^{-1}c_{\boldsymbol{n}_{1},\A}^{\dagger}\left|{\rm vac}\right\rangle
=\,\text{g}^{2}\iint\frac{\dd{\boldsymbol{k}}}{(2\pi)^{2}}\frac{\omega_{\text{pg}}(\boldsymbol{k})e^{-i\phi(\boldsymbol{k})}}{z_{{\rm nh}}^{2}-\omega^{2}_{\text{pg}}({\boldsymbol{k}})}e^{-i\boldsymbol{k}\cdot(\boldsymbol{n}_{1}-\boldsymbol{n}_{2})},\label{S52}\\
\Sigma_{21}^{\A\B}(z)=&\,\text{g}^{2}\left\langle {\rm vac}\right|c_{\boldsymbol{n}_{2},\A}(z-H^{\text{eff}}_{\text{pg}})^{-1}c_{\boldsymbol{n}_{1},\B}^{\dagger}\left|{\rm vac}\right\rangle
=\,\text{g}^{2}\iint\frac{\dd{\boldsymbol{k}}}{(2\pi)^{2}}\frac{\omega_{\text{pg}}(\boldsymbol{k})e^{i\phi(\boldsymbol{k})}}{z_{{\rm nh}}^{2}-\omega^{2}_{\text{pg}}({\boldsymbol{k}})}e^{-i\boldsymbol{k}\cdot(\boldsymbol{n}_{1}-\boldsymbol{n}_{2})}.\label{S53}
\end{align}
Substituting the integrated results of these self-energy functions into Eq.\,(\ref{S43}) yields, in principle, a complete description of the two-emitter dynamics. The approach can be generalized to handle the dynamics of multiple QEs coupled to this photonic graphene by simply redefining the corresponding projection operators.
\section{Light-Matter Interactions for a Single Quantum Emitter with Homogeneous Dissipation}\label{II}
\renewcommand\theequation{S\arabic{equation}}
\makeatletter
\renewcommand{\thefigure}{S\@arabic\c@figure}
\makeatother
In this section, we consider a single QE coupled to a photonic graphene, where specifically, the external dissipation strength is identical for each sublattice, i.e., $\kappa_{a}=\kappa_{b}=\kappa$. This trivial dissipative configuration simply shifts the entire energy spectrum of the original non-dissipative setup uniformly downward in the complex plane by a constant. We begin with an analytical description of the self-energies for a single QE. And then, we detail the various dynamical contributions enabling several key dynamical predictions.
\begin{figure}
  \centering
  \includegraphics[width=18cm]{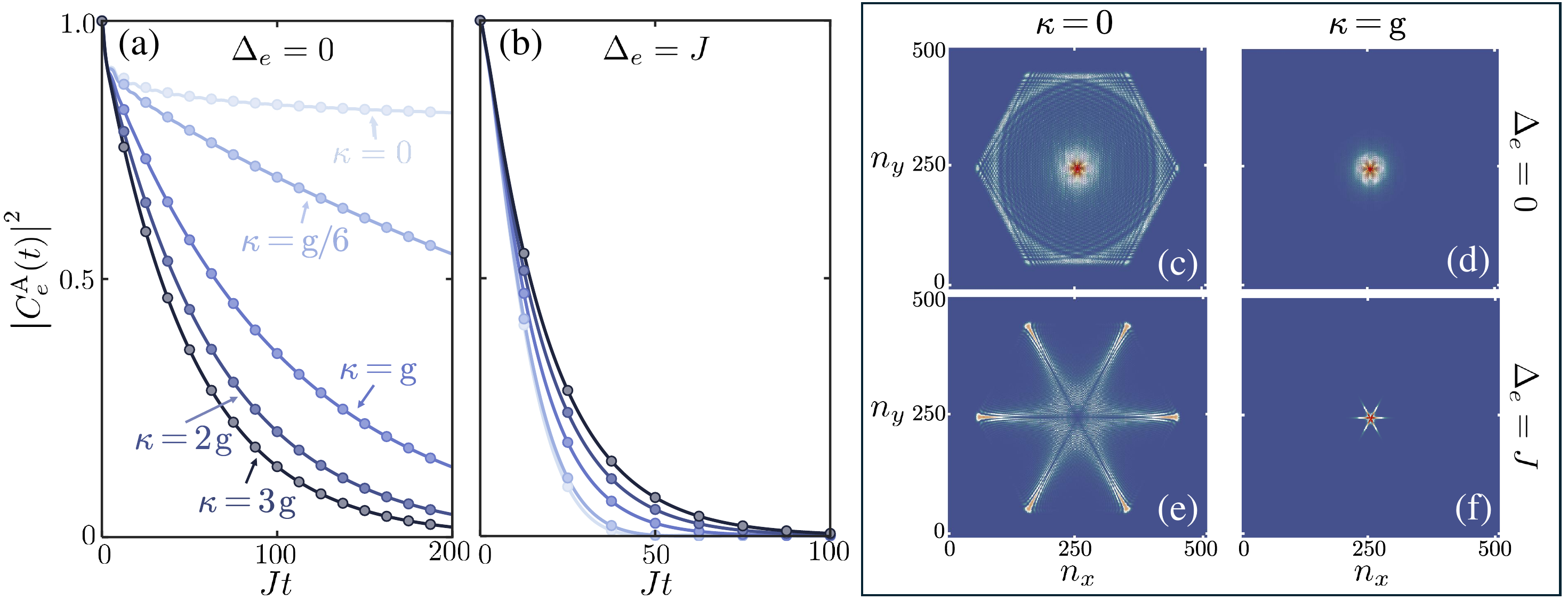}
  \caption{(a-b) Excited-state population $|C_{e}^{\text{A}}(t)|^{2}$ of a single QE coupled to the central site of sublattice A in a photonic graphene with size $N=512$. The results, plotted as a function of scaled time $Jt$, are shown for detunings $\Delta_{e}=0$ (a) and $\Delta_{e}=J$ (b), under uniform dissipation $\kappa_{a}=\kappa_{b}=\kappa$. Notably, the results for different dissipation rates $\kappa = 0$, $\text{g}/6$, $\text{g}$, $2\text{g}$, and $3\text{g}$ are distinguished by line color. (c-f) Bath population $|C^{\text{A,B}}_{\boldsymbol{n}}(t)|^{2}$ in real space at $tJ = 200$, arranged in a $2\times2$ layout. The left (right) column corresponds to the case of $\kappa=0$ ($\kappa=\text{g}$), and the top (bottom) row to the parameter $\Delta_e=0$ ($\Delta_e=J$). The light-matter coupling strength is chosen as $\text{g}=0.2J$ for all panels.}\label{figS4}
\end{figure}

To obtain the exact quantum dynamics of the considered model according to Eqs.\,(\ref{S29}-\ref{S34}), a crucial step is to derive an accurate expression of the relevant self-energies. Fortunately, in the continuum limit, the integral expression for $\Sigma^{\text{A}}_{e}(z)$ can be rigorously computed over the spectral range $(-\infty-i\kappa/2, -3J-i\kappa/2) \cup (3J-i\kappa/2, \infty-i\kappa/2)$. Note that in this case, the dynamical behavior obtained when the QE couples to sublattice A is identical to that when it couples to sublattice B. This is simply because the dissipation is applied symmetrically to both sublattices. The definitions of the self-energy function in other regions of the complex plane can be obtained via analytic continuation. In the scenario considered in this section, the time-evolution of atomic population amplitude is given by
\begin{align}\label{S54}
C^{\text{A}}_{e}(t)=\,-\frac{1}{2\pi i}\int_{-\infty}^{\infty}\dd{E}\frac{1}{E-\Delta_{e}-\Sigma_{e}^{\text{A}}(E+i0^{+})}e^{-iEt},
\end{align}
where the analytical form of $\Sigma_{e}^{\text{A}}(z)$ is defined by Eq.\,(\ref{S28}). To ensure the analyticity of the self-energy $\Sigma^{\text{A}}_{e}(z)$ in it's own domain, we define it as follows:
\begin{align}\label{S55}
\Sigma_{e}^{\text{A}}(z)=\,\frac{\text{g}^{2}(z+i\kappa_{b}/2)}{4\pi}\frac{8}{(\sqrt{z_{{\rm nh}}^{2}}+3J)^{1/2}(\sqrt{z_{{\rm nh}}^{2}}-J)^{3/2}}K^{\text{I}}\left[k(z_{\text{nh}})^2\right],
\end{align}
where $z_{{\rm nh}}=z+i\frac{\kappa}{2}$ and the elliptic integral function $K^{{\rm I}}(m)$ is expressed as
\begin{align}
K^{{\rm I}}(m)=&\,K(m),\text{if}\,\,\text{Im}[k(z)]\mathrm{Im}[z^{2}]<0,\label{S56}\\
K^{{\rm I}}(m)=&\,K(m)+2iK(1-m),\text{if}\,\,\text{Im}[k(z)],\mathrm{Im}[z^{2}]>0,\label{S57}\\
K^{{\rm I}}(m)=&\,K(m)-2iK(1-m),\text{if}\,\,\text{Im}[k(z)],\mathrm{Im}[z^{2}]<0.\label{S58}
\end{align}

The next step is the analytic continuation of $\Sigma^{\text{A}}_{e}(z)$ in Eq.\,(\ref{S55}) throughout the lower complex half-plane. We accomplish this through a detailed analysis of the branch-cut structure in the lower complex half-plane. We first recall that for the non-dissipative honeycomb bath, the branch points defining the cut endpoints lie on the real axis at: the band edges $z = \pm 3J$; the points $z = \pm J$ where the density of states diverges; and the band center. In contrast, for the uniformly dissipative honeycomb bath, the corresponding branch points are shifted into the lower half of the complex plane. To be precise, these branch points are given by the roots of the equation $(\sqrt{z^{2}}+3J)^{1/2}(\sqrt{z^{2}}-J)^{3/2}=0$, yielding
\begin{align}
z_{1}=&-\sqrt{(3J)^{2}-\kappa_{-}^{2}}-i\kappa_{+}=-3J-i\kappa/2,\,\,\,\,\,\,\,\,\,\,\,\,\,z_{2}=-\sqrt{J^{2}-\kappa_{-}^{2}}-i\kappa_{+}=-J-i\kappa/2,\label{S59}\\
z_{3}=&\sqrt{J^{2}-\kappa_{-}^{2}}-i\kappa_{+}=J-i\kappa/2,\,\,\,\,\,\,\,\,\,\,\,\,\,\,\,\,\,\,\,\,\,\,\,\,\,\,\,\,\,\,\,\,\,\,\,\,z_{4}=\sqrt{(3J)^{2}-\kappa_{-}^{2}}-i\kappa_{+}=3J-i\kappa/2,\label{S60}
\end{align}
from which one can clearly see that the branch points for the homogeneous dissipation case are shifted downward in the complex plane by a constant $\kappa_{+}$ relative to the non-dissipative case.

\begin{figure}
  \centering
  \includegraphics[width=15cm]{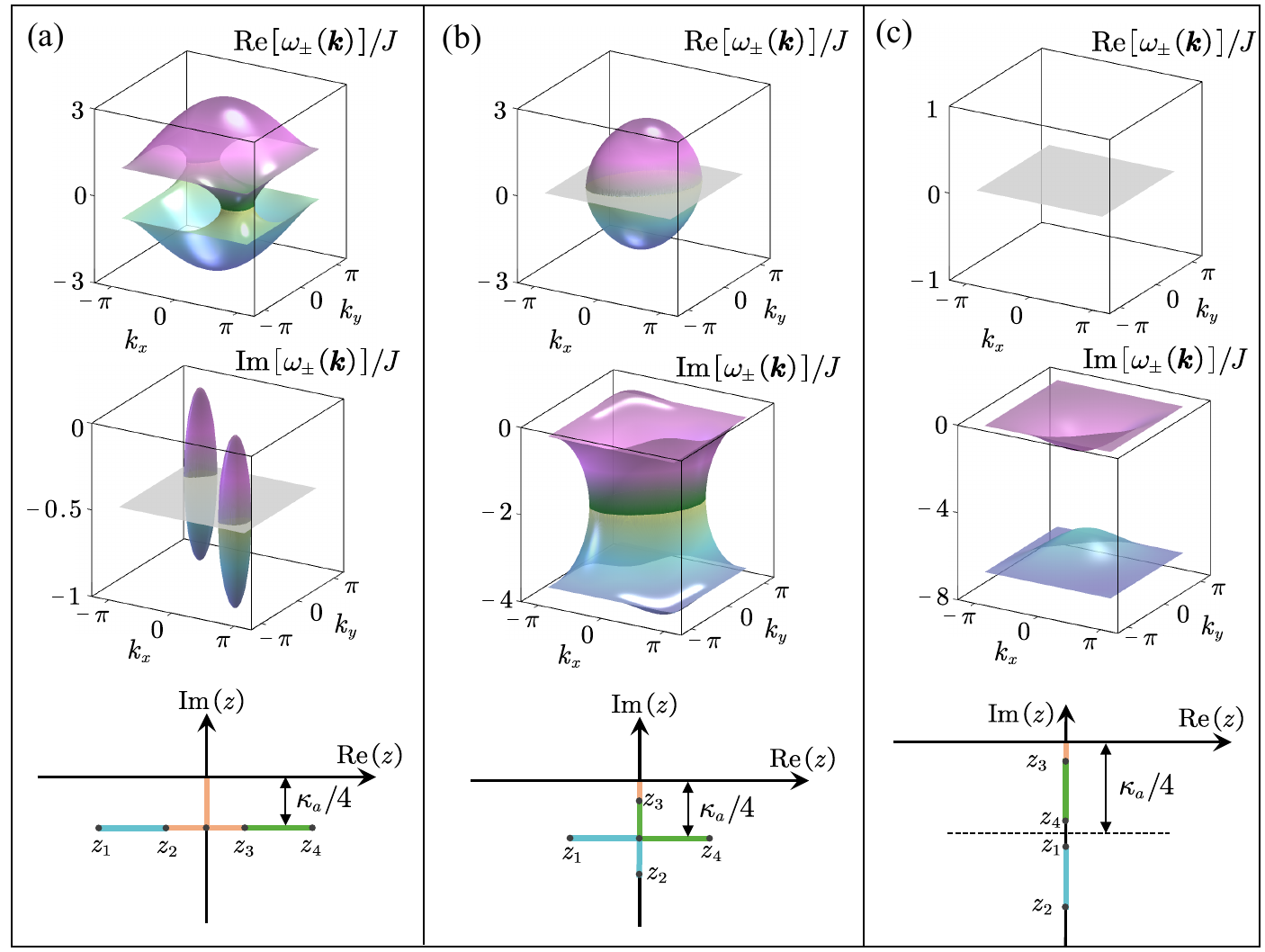}
  \caption{Plots of the complex band structure $\omega_{\pm}(\boldsymbol{k})/J$, showing its real part $\mathrm{Re}[\omega_{\pm}]/J$ and imaginary part $\mathrm{Im}[\omega_{\pm}]/J$, for a photonic graphene with single-sublattice dissipation ($\kappa_a$ variable, $\kappa_b=0$). Results are presented for three values of $\kappa_a/J$: (a) 2, (b) 8, (c) 14. The lower row in each panel illustrates the corresponding distribution of branch cuts in the complex plane under the given dissipation parameters. These branch cuts are depicted by blue, green, and orange lines, which respectively represent the connections between the branch point pairs $z_{1}$ and $z_{2}$, $z_{3}$ and $z_{4}$, as well as the remaining branch cut segments.}\label{figS5}
\end{figure}
To evaluate the dynamical integral in Eq.\,(\ref{S54}), we perform contour integration in the lower-half complex plane, as illustrated in Fig.\,\ref{figS2}(a). Notice, according to the residue theorem, the integral can be decomposed into contributions from poles and branch cuts. While pole contributions are readily obtained by calculating residues, the branch cut treatment requires more sophisticated approaches. We present two principal methods for handling the branch cuts. The first involves direct integration along a tightly enclosing contour around the branch cuts. The second, and our preferred approach, employs analytic continuation by detouring around band edges into other Riemann sheets of the integrand, as illustrated by the branch cut detours in Fig.~\ref{figS2}(a). This method reveals that the integrand may possess unstable poles in other Riemann sheets, thereby decomposing the branch cut contribution into parts from these poles and branch cut detours.  The analytical expression for the self-energy as presented in Eq.~(\ref{S55}), corresponds to the integrand in the first Riemann sheet. The chosen contour in Fig.~\ref{figS2}(a) includes detours around the singularities at complex energies $z_{1},z_{2},z_{3},z_{4}$, and $-i\kappa/2$. Hence, we must analytically continue $\Sigma_{e}^{\text{A}}(z)$ into the entire lower-half plane. We accomplish this by plotting in Fig.~\ref{figS2}(b-e) the sign functions of $\mathrm{Re}[k(z_{\mathrm{nh}})]$, $\mathrm{Im}[k(z_{\mathrm{nh}})]$, $\mathrm{Im}(z^{2}_{\mathrm{nh}})$, and $\mathrm{Re}(z^{2}_{\mathrm{nh}})$, thereby fully specifying the phase of the elliptic integral functions' argument. Based on these phase information derived from these arguments, we have identified not only the first Riemann sheet of the self-energy function as defined in Eq.\,(\ref{S55}), but also the second, third, fourth, and fifth Riemann sheets in regions $\text{II-V}$, and have presented the analytic continuation results of $\Sigma^{\text{A}}_{e}(z)$ on the corresponding Riemann sheets, i.e., replacing the elliptic integral function $K^{{\rm I}}(m)$ in Eq.\,(\ref{S55}) with
\begin{align}
K^{{\rm II}}(m)	=\,&(-3)K(m)+2iK(1-m),\text{if}\,\,\text{Re}[k(z)]>(<)0,\label{S61}\\
K^{{\rm III}}(m)	=\,&(-3)K(m)-2iK(1-m),\text{if}\,\,\text{Re}[k(z)]>(<)0,\label{S62}\\
K^{{\rm IV}}(m)	=\,	&3K(m)+2(-4)iK(1-m),\text{if}\,\,\text{Im}[k(z)]>(<)0,\label{S63}\\
K^{{\rm V}}(m)	=\,	&3K(m)-2(+4)iK(1-m),\text{if}\,\,\text{Im}[k(z)]<(>)0.\label{S64}
\end{align}
These specific linear combinations of elliptic integrals are carefully selected to ensure the self-energy remains continuous across the entire integration contour\,\cite{SMPhysRevA.97.043831}. This is captured by the fact that the argument of the analytically continued $\Sigma_{e}^{\text{A}}(z)$ varies continuously along the integration contour, as shown in Fig.\,\ref{figS2}(f).

With the analytical continuation $\Sigma_{e}^{\text{A}}(z)$ now well-defined throughout the entire Riemann sheets, the atomic dynamics is fully characterized by bound states (BSs), unstable poles (UPs), and branch-cut-induced detours (BCDs) contributions:
\begin{align}\label{S65}
C_{e}^{\text{A}}(t)=\sum_{z_{k}\in{\rm BS}}{\rm Res}[G_{e}^{\text{A}}(z),z_{k}]e^{-iz_{k}t}+\sum_{z_{k}\in{\rm UP}}{\rm Res}[G_{e}^{\text{A}}(z),z_{k}]e^{-iz_{k}t}+\sum_{k=1}^{5}C_{{\rm BCD}}^{k}(t),
\end{align}
where $G_{e}^{\text{A}}(z)=\left[z-\Delta_{e}-\Sigma_{e}^{\text{A}}(z)\right]^{-1}$ represents the single-emitter Green function. Here,  $R_{\mathrm{BS}}={\rm Res}[G_{e}^{\text{A}}(z),z_{\mathrm{BS}}]$ and $R_{\mathrm{UP}}={\rm Res}[G_{e}^{\text{A}}(z),z_{\mathrm{UP}}]$ are the residues at the bound state and a given unstable pole, respectively. They can be calculated via the contour-integration techniques, yielding $R_{\mathrm{BS(UP)}} = \left[1-\partial_{z}\Sigma^{\text{A}}_{e}(z)\right]^{-1}_{z=z_{\mathrm{BS(UP)}}}$.
\begin{figure}
  \centering
  \includegraphics[width=18cm]{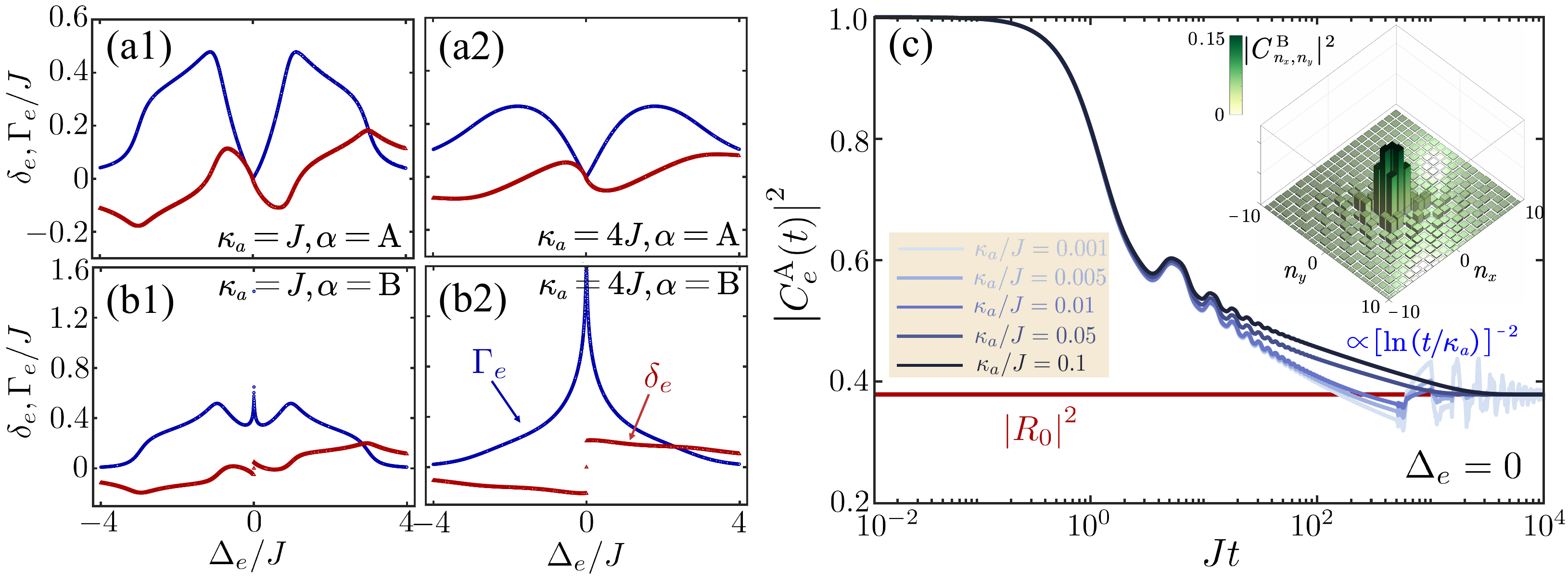}
  \caption{The Markovian decay rates (blue lines) and frequency shifts (red lines) are shown for a single QE coupled to sublattice A (a1–a2) and sublattice B (b1–b2) of a photonic graphene, respectively, plotted against the detuning $\Delta_{e}/J$ under single-sublattice dissipation. (c) The excited-state population $\left|C_{e}^{\text{A}}(t)\right|^{2}$ of a single QE coupled to the center of sublattice A is plotted as a function of the scaled time $Jt$, for a bath of size $N=512$ with detuning $\Delta_{e}=0$. Simulation results are color-coded by the dissipation strength $\kappa_a/J$ (values: $0.001, 0.005, 0.01, 0.05, 0.1$), while $\kappa_b$ is held at zero. It is noted that the excited-state population predicted under different dissipation strengths eventually stabilizes at an asymptotic value $\left|R_{0}\right|^{2}$ that is independent of dissipation, as indicated by the red line.}\label{figS6}
\end{figure}
Physically, the magnitudes $R_{\mathrm{BS}}$ and $R_{\mathrm{UP}}$ quantify the strength of the initial state's projection onto the discrete bound state and the radiative state, respectively. Mathematically, the corresponding poles $z_{{\rm BS}}$ and $z_{{\rm UP}}$ are determined by solving the equations $z_{{\rm BS}}-\Delta_{e}-\Sigma_{e}(z_{{\rm BS}})=0$ and $z_{{\rm UP}}-\Delta_{e}-\Sigma_{e}(z_{{\rm UP}})=0$, respectively. Note that for brevity, the Riemann sheet indices are omitted in these eigenvalue equations; in practice, care must be taken to use the analytically continued self-energies from the corresponding Riemann sheets. The same principle holds for all subsequent dynamical descriptions. Finally, the explicit forms of branch-cut-induced detours $C^{k}_{\mathrm{BCD}}(t)$ are given by
\begin{align}\label{S66}
C^{k}_{\mathrm{BCD}}(t) =\frac{1}{2\pi}\int_{-\infty}^{-\kappa/2}\dd{y} \left[G_{e}^{\mathrm{A},r_{k}}(x_{k}+iy)-G_{e}^{\mathrm{A},l_{k}}(x_{k}+iy)\right]e^{-i(x_{k}+iy)t},
\end{align}
where the values $x_{1}=-3J$, $x_{2}=-J$, $x_{3}=0$, $x_{4}=J$, and $x_{5}=3J$ mark the five singularities. The second superscript index pair $(r_i, l_i)$ on the Green's function identifies the Riemann sheet, where $(r_1, l_1) = (\mathrm{II}, \mathrm{I})$, $(r_2, l_2) = (\mathrm{IV}, \mathrm{II})$, $(r_3, l_3) = (\mathrm{V}, \mathrm{IV})$, $(r_4, l_4) = (\mathrm{III}, \mathrm{V})$, and $(r_5, l_5) = (\mathrm{I}, \mathrm{III})$.

We are now readily to characterize the single-atom dynamics through a series of summations over the amplitudes given in Eq.\,(\ref{S65}). In Fig.\,\ref{figS3}(a), we first plot these contributions in the case without dissipation. We use different marker styles (see legend) distinguish the various components: red triangles for the lower bound state (LBS), blue circles for the upper bound state (UBS), black squares for the poles on the second Riemann sheet (UPII), green diamonds for the third (UPIII), yellow triangles for the fourth (UPIV), purple triangles for the fifth (UPV), and brown stars for the branch-cut-induced detours (BCD). We can simply summary several observations from Fig.\,\ref{figS3}(a) as follows: ($i$) For large detunings $|\Delta_e|$ far from the band edges, the dynamics is governed by either the lower or upper bound state, suppressing any decay in the QE because it is detuned from the continuum. Furthermore, because of the divergence of $\delta_e(E)$ [the real part of self-energy $\Sigma_{e}^{\text{A}}(E+i0^{+})$] at $E = \pm 3J$ [see Fig.\,\ref{figS3}(e)], these bound states are persist for all $\Delta_e$, even within the band. ($ii$) When the QE transition frequency moves close to the band edges, the bound states contribution progressively declines as the BCDs gains prominence. ($iii$) Owing to the divergence in the imaginary part of $\Sigma_{e}^{\text{A}}(E+i0^{+})$ at $E =- J$ ($E = J$), the contributions of two types of UPs from Riemann sheets II and IV (III and V) coexist for a range of $\delta_e(-J+0^{+})-\delta_e(-J-0^{+})$ [$\delta_e(J+0^{+})-\delta_e(J-0^{+})$]. (iv) A key feature near the Dirac point is the discontinuous jump in the middle BCD contribution [with $k=3$ in Eq.\,(\ref{S66})] at $\Delta_{e} = 0$. This jump coincides with the vanishing of all UP contributions, while the middle BCD component itself scales as $\left|C^{3}_{\mathrm{BCD}}(t)\right|\propto 1/\mathrm{ln}(t)$.

\begin{figure}
  \centering
  \includegraphics[width=17.5cm]{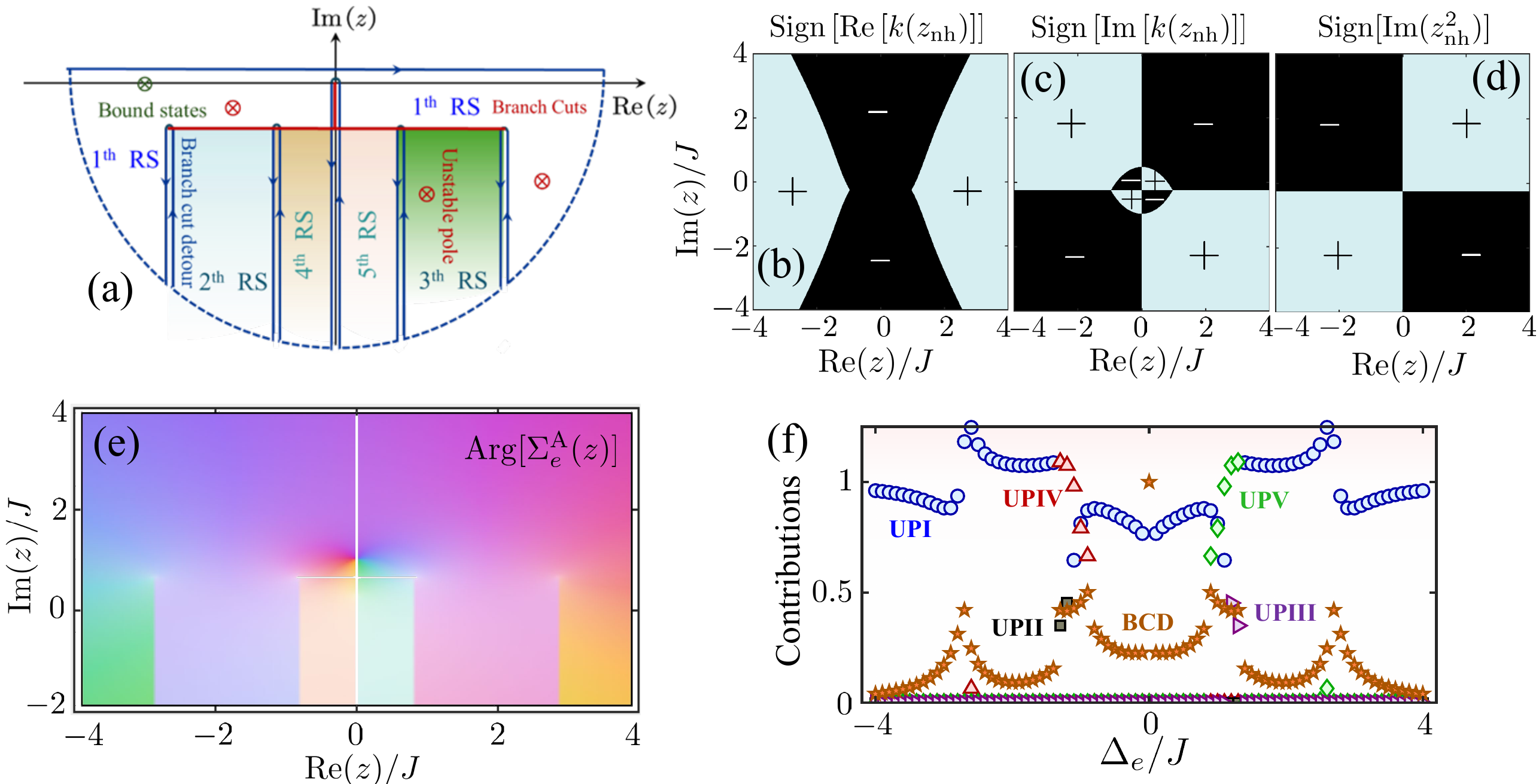}
  \caption{(a) Contour of integration employed to evaluate the probability amplitude $C^{\A}_{e}(t)$ in Eq.\,(\ref{S73}) for a single QE coupled to a photonic graphene with single-sublattice dissipation. The self-energy $\Sigma_{e}^{\text{A}}(z)$ exhibits discontinuities at $z=\pm\sqrt{(3J)^{2}-(\kappa_a/4)^{2}}-i\kappa_a/4,\pm\sqrt{J^{2}-(\kappa_a/4)^{2}}-i\kappa_a/4$, and $0$, necessitating detours around these non-analytic energies and resulting in branch-cut contributions to the quantum dynamics. Poles located on the real axis (corresponding to bound states) are marked by green dots, while those in the lower half of the complex plane (associated with unstable or resonance poles) are denoted by red dots. The five Riemann sheets are distinguished by coloring. More specifically, at the band edges $E=\sqrt{(3J)^{2}-(\kappa_a/4)^{2}}$ and $E=-\sqrt{(3J)^{2}-(\kappa_a/4)^{2}}$, the integration path switches from the first Riemann sheet (white region) to the third (green region), and from the second sheet (blue region) to the first, respectively. At the energy edges of $E=\sqrt{J^{2}-(\kappa_a/4)^{2}}$ and $E=-\sqrt{J^{2}-(\kappa_a/4)^{2}}$, it switches from the third sheet to the fifth (pink region), and from the fourth sheet (brown region) to the second, respectively. At the band center, the path switches from the fifth to the fourth Riemann sheet. (b-d) Sign of the $\mathrm{Re}\left[k(z_{\mathrm{nh}})\right]$, $\mathrm{Im}\left[k(z_{\mathrm{nh}})\right]$, and $\mathrm{Im}(z^{2}_{\mathrm{nh}})$ serves as the criterion to identify the specific Riemann sheets. (e) The argument $\mathrm{Arg}[\Sigma_{e}^{\text{A}}(z)]$ of the analytically continued self-energy $\Sigma_{e}^{\text{A}}(z)$ according to Eq.\,(\ref{S75}). (f) Dynamics contributions of $C_{e}^{\text{A}}(t)$ at time $t = 0$ as a function of detuning $\Delta_{e}$. Symbols denote: UPI (blue circles), UPII (black squares), UPIII (purple triangles), UPIV (red triangles), UPV (green diamonds), and BCD (brown stars). The light-matter coupling strength implemented here is $\text{g}=0.6J$ and the dissipation strength is $\kappa_a=J,\kappa_b=0$.}\label{figS7}
\end{figure}

We then proceed by selecting three different dissipation strengths, namely $\kappa = \text{g}/6,\,\text{g}$, and $3\text{g}$, and plotting the various contributions to $C_{e}^{\text{A}}(t)$ at the initial time $t=0$, as illustrated in Figs.\,\ref{figS3}(b-d). Different marker styles are used to distinguish these contributions. Specifically, the unstable poles contribution from the first Riemann sheet is marked with blue circles (labeled by UPI), those from the second Riemann sheet with black squares (labeled by UPII), the third with purple triangles (labeled by UPIII), the fourth with red triangles (labeled by UPIV), the fifth with green diamonds (labeled by UPV), and the contribution from branch-cut-induced detours is marked by brown stars (labeled by BCD), from which we obtain the following key observations: ($i$) The contributions from the upper and lower bound states, which exist outside the band in the dissipationless case, vanish completely. This occurs as their corresponding poles, originally located on the real axis, move downward into the lower half of the complex plane and acquire a finite imaginary part. ($ii$) The distribution of different dynamical contributions shift with increasing dissipation. More concretely, under weak dissipation ($\kappa=\text{g}/6$), unstable poles on the first Riemann sheet dominate outside and at the center of the energy band. As dissipation increases to a moderate strength ($\kappa=\text{g}$), UP contributions from sheets II–V are gradually taken over by UPI, while poles located farther from $|E|=J$ recede more rapidly. When the dissipation is enhanced into the strong regime ($\kappa=3\text{g}$), the ensuing dynamics is dictated exclusively by the contributions from BCDs and UPI. ($iii$) The discontinuity in the middle BCD contribution at the band center, present in the non-dissipative case [Fig.~\ref{figS3}(a)], is absent when dissipation is introduced. ($iv$) Finally, the Markovian decay rate at the Dirac point becomes non-zero and is gradually elevated as the dissipation strength increases. Meanwhile, the originally divergent DOS around $|E| = J$ becomes finite. This implies that the corresponding radiative modes, which initially exhibited infinitesimal lifetimes, now acquire progressively longer lifetimes. Furthermore, the discontinuous Lamb shift $\delta_{e}$ present in the dissipationless case now turns into a continuous one.

\begin{figure}
  \centering
  \includegraphics[width=18cm]{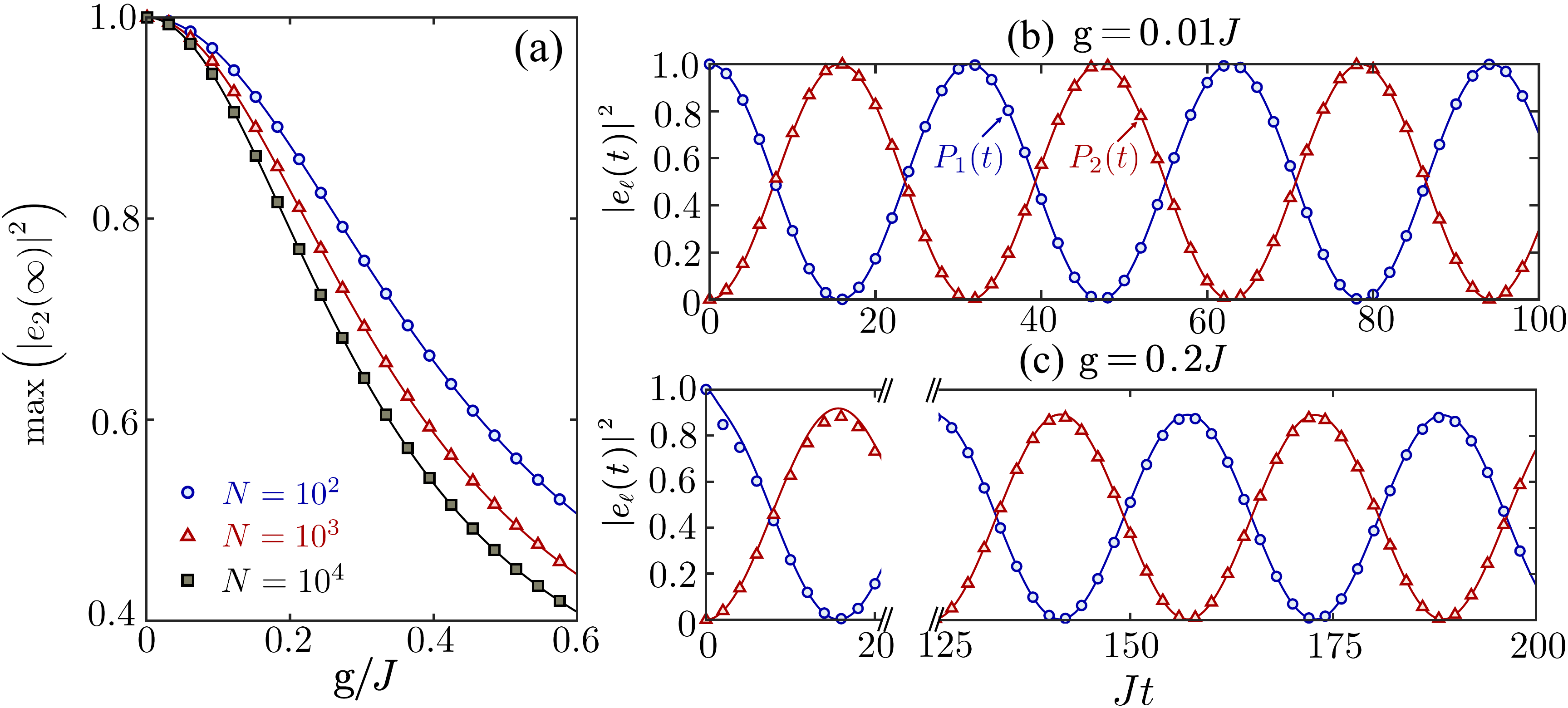}
  \caption{The maximal excitation probability of the acceptor QE, $\max\left(\left|e_{2}(\infty)\right|^{2}\right)$, as a function of the light-matter coupling strength $\text{g}$. The excitation is transferred from a resonant donor QE ($\Delta_{e}=0$) to this acceptor QE that is co-located within the same cavity in sublattice A. We simulate the dependence of $\max\left(\left|e_{2}(\infty)\right|^{2}\right)$ on $\text{g}$ for different lattice sizes, employing the following markers: $N=10^{2}$ (blue circles), $N=10^{3}$ (red triangles), and $N=10^{4}$ (black squares). Panels (b) and (c) show the time evolution of the atomic excitation probabilities for coupling strengths $\text{g}=0.01J$ and $\text{g}=0.2J$, respectively, at a fixed lattice size $N=2^{6}$. The populations $P_1(t) = |e_1(t)|^2$ and $P_2(t) = |e_2(t)|^2$ are shown as blue and red lines, where solid curves denote exact numerical results and markers correspond to asymptotic analytical solutions in the long-time limit. They show excellent agreement in the long-time regime.}\label{figS8}
\end{figure}

After thoroughly analyzing all dynamical contributions to the atomic temporal evolution, we can now directly present the effect of uniform dissipation on the dynamics. As shown in Figs.\,\ref{figS4} (a) and (b), we simulate the time evolution of $C_{e}^{\text{A}}(t)$ for detunings of $\Delta_{e}=0$ and $\Delta_{e}=J$, respectively, using a range of dissipation strengths: $\kappa=0, \text{g}/6, \text{g}, 2\text{g}, 3\text{g}$. These dynamical results clearly reveal that a larger dissipation strength enhances the spontaneous emission for $\Delta_{e}=0$ but suppresses it for $\Delta_{e}=J$, directly corroborating the above contribution analysis. The solid lines in Figs.\,\ref{figS4} (a) and (b) represent the analytical solutions according to Eq.\,(\ref{S65}), while the marked lines correspond to the numerical results obtained from the full real-space Hamiltonian, whose detailed procedures will be provided later. The numerical and analytical results are in excellent agreement. We also plot the bath population $C^{\text{A,B}}_{\boldsymbol{n}}(t)$ in real space at time $tJ = 200$ for the detunings $\Delta_{e}=0$ [Figs.\,\ref{figS4} (c) and (d)] and $\Delta_{e}=J$ [Figs.\,\ref{figS4} (e) and (f)]. In the dissipationless case, for a detuning of $\Delta_{e}=0$, the emission propagates through the optical lattice in a regular hexagonal pattern, with the bath population primarily concentrated at the bath center, exhibiting a plum-blossom-like distribution. For a detuning of $\Delta_{e}=J$, the radiated optical field shows strong anisotropy and propagates mainly along three specific directions. In the presence of dissipation, the energy radiated into the lattice dissipates into the external environment, resulting in a rapidly narrowing spatial extent of the field within the bath [see Figs.\,\ref{figS4} (d) and (f)].
\section{Light-Matter Interactions in Dirac Conelike Bath with Single-Sublattice Dissipation}\label{III}
\renewcommand\theequation{S\arabic{equation}}
\makeatletter
\renewcommand{\thefigure}{S\@arabic\c@figure}
\makeatother
In the previous section, we have briefly discussed the case of energy leakage occurring on both two sublattices. Now, we turn to the dynamics of one and two QEs coupled to a photonic graphene with single-sublattice dissipation, e.g., $\kappa_{a}=\kappa,\kappa_{b}=0$.  This dissipation configuration will be adopted as the default in the following analysis in this regime. We begin in subsection \ref{IIIA} with a detailed analytical treatment of the single-emitter dynamics. This is followed in subsection \ref{IIIB} by the derivation of an analytical solution for the vacancy-like dressed bound state (VDS) of the considered model. Subsequently, in subsection \ref{IIIC}, we present an exact theoretical description of the two-emitter dynamics and elucidate the mechanism by which decoherence-free interactions between two QEs emerge under single-sublattice dissipation.

\subsection{Single-emitter dynamics in graphenic QED}\label{IIIA}
In this subsection, we will focus on a more interesting and non-trivial dissipative configuration where energy leakage occurs only in one of the two sublattices, termed single-sublattice dissipation. We will develop an exact analytical description of the dynamics for a single QE that is coupled to such a photonic environment. Without loss of generality, we consider the case where sublattice A has dissipation ($\kappa_{a}\neq 0$) and sublattice B is dissipation-free ($\kappa_{b}=0$), such that $\kappa_{+}=\kappa_{-}=\kappa_a/4$. Unless otherwise specified, the subsequent discussions on single-sublattice dissipation will default to this specific configuration. In order to determine the quantum dynamics via the resolvent method, following the similar procedure used for the uniform dissipation case (where the branch cuts are simply shifted downward), we will analyze how the branch cuts distribute as the dissipation strength $\kappa_a$ gradually increases.

Next, we present the branch cut distributions across three dissipation regions and characterize the corresponding band features. ($i$) For a weak dissipation ($\kappa_a < 4J$), the branch cuts shorten compared to the dissipationless case and shift downward by $\kappa_a/4$ [see Fig.\,\ref{figS5}(a)], resulting in the corresponding endpoints that are given by
\begin{align}
z_{1}=&-\sqrt{(3J)^{2}-(\kappa_a/4)^{2}}-i\kappa_a/4,\,\,\,\,\,\,\,\,\,\,\,\,\,z_{2}=-\sqrt{J^{2}-(\kappa_a/4)^{2}}-i\kappa_a/4,\label{S67}\\
z_{3}=&\sqrt{J^{2}-(\kappa_a/4)^{2}}-i\kappa_a/4,\,\,\,\,\,\,\,\,\,\,\,\,\,\,\,\,\,\,\,\,\,\,\,\,\,\,\,\,z_{4}=\sqrt{(3J)^{2}-(\kappa_a/4)^{2}}-i\kappa_a/4.\label{S68}
\end{align}
In contrast to the dissipationless case where the band touching occurs at two Dirac points,  the eigenvalues $\omega_{\pm}(\boldsymbol{k})$ become complex, and the Dirac points $\boldsymbol{{\rm{K}}}_{\pm} = 2\pi(\pm 1, \mp 1)/3$ (present for $\kappa_{a,b}=0$) morph into a pair of exceptional rings defined by $\left|f(\boldsymbol{k})\right|=\kappa_{-}$. On these rings, the eigenvalues take a fixed value of $-i\kappa_{a}/4$, and the two eigenstates coalesce. Additionally, the upper band $\omega_{+}(\boldsymbol{k})$ vanishes at $\boldsymbol{{\rm{K}}}_{\pm}$ (within the rings) [see Fig.\,\ref{figS5}(a)], which signals the emergence of dissipation-free photon modes. ($ii$) For a moderate dissipation ($4J < \kappa < 12J$), the branch cuts connected by $[z_{1}, z_{2}]$ and $[z_{3}, z_{4}]$ undergo a transformation compared to the dissipationless case: they shorten, shift downward by $\kappa_a/4$, and partially appear on the imaginary axis [see Fig.\,\ref{figS5}(b)], yielding the endpoints
\begin{align}
z_{1}=&-\sqrt{(3J)^{2}-(\kappa_a/4)^{2}}-i\kappa_a/4,\,\,\,\,\,\,\,\,\,\,\,\,\,z_{2}=-i\frac{\kappa_a}{4}\left[1+\sqrt{1-\left(\frac{4J}{\kappa_a}\right)^{2}}\right],\label{S69}\\
z_{3}=&-i\frac{\kappa_a}{4}\left[1-\sqrt{1-\left(\frac{4J}{\kappa_a}\right)^{2}}\right],\,\,\,\,\,\,\,\,\,\,\,\,\,z_{4}=\sqrt{(3J)^{2}-(\kappa_a/4)^{2}}-i\kappa_a/4.\label{S70}
\end{align}
Under this parameter regime, the upper and lower bands touch each other along a single exceptional ring. On this ring, the complex eigenvalues remain fixed at $-i\kappa_{a}/4$ and the two eigenstates coalesce. The real part of the complex energy spectrum remains zero outside the exceptional ring but becomes nonzero inside; meanwhile, the imaginary part is fixed at $-\kappa_a/4$ within the ring, which contrasts with its behavior in the weak dissipation regime. ($iii$) For a strong dissipation ($\kappa_a > 12J$), the branch cuts connected by $[z_{1}, z_{2}]$ and $[z_{3}, z_{4}]$ are both located on the imaginary axis and are separated from each other [see Fig.\,\ref{figS5}(c)]. Their endpoints can be written as
\begin{align}
z_{1}=&-i\frac{\kappa_a}{4}\left[1+\sqrt{1-\left(\frac{12J}{\kappa_a}\right)^{2}}\right],\,\,\,\,\,\,\,\,\,\,\,\,\,z_{2}=-i\frac{\kappa_a}{4}\left[1+\sqrt{1-\left(\frac{4J}{\kappa_a}\right)^{2}}\right],\label{S71}\\
z_{3}=&-i\frac{\kappa_a}{4}\left[1-\sqrt{1-\left(\frac{4J}{\kappa_a}\right)^{2}}\right],\,\,\,\,\,\,\,\,\,\,\,\,\,\,\,\,z_{4}=-i\frac{\kappa_a}{4}\left[1-\sqrt{1-\left(\frac{12J}{\kappa_a}\right)^{2}}\right].\label{S72}
\end{align}
The real part of the complex energy in this regime is zero in the entire first Brillouin zone, while the imaginary parts of the upper and lower bands are separated. Note that the branch cuts formed by these branch points terminate at the origin of the complex plane for all three types of parameter regimes. Moreover, it is noteworthy that the bath with single-sublattice dissipation hosts dissipation-free photon modes at any dissipation strength.

Having these detailed analyses of the distribution of the branch cuts in hand, we can now readily obtain the exact quantum dynamics of the considered model according to Eqs.\,(\ref{S29}-\ref{S34}). For the subsequent analysis, we focus on the weak dissipation regime to enable an analytical dynamical description. Under the thermodynamic limit, the integral expressions for the relevant self-energies, which are key to an analytical description of the dynamics, can be rigorously derived following the procedures presented in Sec.\ref{II}. In the single-sublattice dissipative configuration, the dynamical behavior varies depending on whether the QE couples to sublattice A or B. This difference is reflected in the time-evolution of the atomic excited-state amplitudes for the two cases:
\begin{align}
C^{\text{A}}_{e}(t)=\,-\frac{1}{2\pi i}\int_{-\infty}^{\infty}\dd{E}\frac{1}{E-\Delta_{e}-\Sigma_{e}^{\text{A}}(E+i0^{+})}e^{-iEt},\label{S73}\\
C^{\text{B}}_{e}(t)=\,-\frac{1}{2\pi i}\int_{-\infty}^{\infty}\dd{E}\frac{1}{E-\Delta_{e}-\Sigma_{e}^{\text{B}}(E+i0^{+})}e^{-iEt},\label{S74}
\end{align}
where the analytical forms of $\Sigma_{e}^{\text{A}}(z)$ and $\Sigma_{e}^{\text{B}}(z)$ are defined by Eq.\,(\ref{S28}). Their explicit analytical forms are given by
\begin{align}
\Sigma_{e}^{\text{A}}(z)=\frac{\text{g}^{2}z}{4\pi}C(z_{{\rm nh}})K^{{\rm I}}[k(z_{{\rm nh}})^{2}],\,\,\,\,\Sigma_{e}^{\text{B}}(z)=\frac{\text{g}^{2}(z+i\kappa/2)}{4\pi}C(z_{{\rm nh}})K^{{\rm I}}[k(z_{{\rm nh}})^{2}],\label{S75}
\end{align}
where the function $K^{{\rm I}}[k(z_{{\rm nh}})^{2}]$ are defined according to Eqs.\,(\ref{S56}-\ref{S58}) to maintain the self-energy's analyticity within its domain.

To elucidate the underlying dynamics, it is instructive to begin by examining the Markovian picture. The atomic dynamics predicted in the Markovian framework can be described simply by the self-energy above the real axis:
\begin{align}\label{S76}
\Sigma_e(E+i0^+)= \delta_e(E) - i\frac{\Gamma_e(E)}{2},\,\,\,\,\,\,\,\,\,\,
C_{e}(t) \approx \exp\left[-i\left[\Delta_{e} + \delta_e(\Delta_{e})\right]t - \frac{\Gamma_e(\Delta_{e})}{2}t\right],
\end{align}
where the $C_{e}(t)$ is calculated by approximately replacing the self-energy $\Sigma_{e}(E+i0^{+})$ in Eqs.\,(\ref{S73}-\ref{S74}) with $\Sigma_{e}(\Delta_{e}+i0^{+})$. In Figs.\,\ref{figS6} (a) and (b), we present the numerical results for $\delta_e$ and $\Gamma_e$ as functions of the detuning $\Delta_e$. These are shown for a single QE coupled to either the A [panels (a1, a2)] or B [panels (b1, b2)] sublattice, with different dissipation strengths $\kappa_a = J, 4J$ as indicated in the legend. When the QE is coupled to the dissipation-free sublattice B, a radiative mode can always be identified across the entire band region. This implies that the excitation energy stored in the QE will be completely released into the environment in the long-time limit. Nevertheless, a dissipation-free mode emerges when the QE resonantly couples to the dissipative sublattice A at the band center, specifically, where $\left.\mathrm{Im}\left[{\Sigma}^{\text{A}}_{e}(E+i0^+)\right]\right|_{E=0}=0$. Notably, this mode remains entirely unaffected by dissipation. To elucidate this exotic behavior, we perform an expansion of the self-energy $\Sigma_{e}^{\text{A}}(E+i0^{+})$ in the vicinity of the band center ($E \approx 0$), which yields
\begin{align}\label{S77}
\Sigma_{e}^{\text{A}}(E+i0^{+})\approx\frac{g^{2}}{\sqrt{3}J^{2}\pi}\left[-\frac{\pi|E|}{2}i+E\log\left(\frac{|E|\kappa_a}{18J^{2}}\right)\right],
\end{align}
from which we find that both $\delta_e(E)$ and $\Gamma_e(E)$ are continuous at $E = 0$, but exhibit a discontinuity in their derivatives. Compared to the dissipation-free case, the radiative modes with infinitesimal lifetime are absent in this scenario. Furthermore, the Lamb shift $\delta_{e}$ near the band center exhibits continuity or discontinuity depending on whether the QE is coupled to sublattice A or B, respectively. Note that the Eq.\,(\ref{S76}) is exactly the Eq.\,(6) in the main text.

As shown in Fig.\,\ref{figS6}(c), while perturbation theory predicts a dissipation-free mode, the actual dynamics of a single QE in the single-sublattice dissipative regime exhibits logarithmic relaxation. Besides, the atomic population in a finite photonic graphene exhibits a persistent oscillation around a constant $\left|R_0\right|^{2}$ in the long-time limit [cf. the red line in Fig.\,\ref{figS6}(c)]. This behavior leads to a breakdown of the Fermi golden rule, thereby requiring a more sophisticated, non-perturbative approach that relies on an exact solution of Eqs.\,(\ref{S73}) and (\ref{S74}).

We evaluate the excitation amplitude of a QE coupled to the dissipative sublattice A by performing the contour integral in Eq.\,\eqref{S73}, with the integration path shown in Fig.\,\ref{figS7}(a). While in the weak dissipation regime ($\kappa_a < 4J$) all four branch points $z_1, z_2, z_3, z_4$ shift downward in the complex plane, as described by Eqs.\,(\ref{S67}–\ref{S68}) and similar to the homogeneous dissipation case, the branch point near the band center remains close to the origin. According to the residue theorem, the contour integral can be decomposed into contributions from poles and branch cuts. The residues at the poles are straightforward to compute. In contrast, evaluating the branch cut contribution requires analytic continuation: the integration path must be detoured around the branch points and onto other Riemann sheets, as illustrated in Fig.~\ref{figS7}(a). This method reveals that the integrand in Eq.\,\eqref{S73} may possess unstable poles in other Riemann sheets, thereby decomposing the branch cut contribution into parts from these poles and branch cut detours. The analytical expression of the self-energy in the first Riemann sheet is presented in Eq.~(\ref{S75}). The chosen contour in Fig.~\ref{figS7}(a) includes detours around the singularities at energies $z_{1},z_{2},z_{3},z_{4}$, and $0$, and passes through other undefined regions. To evaluate the contour integral in Eq.\,\eqref{S73}, the self-energy $\Sigma_{e}^{\text{A}}(z)$ must be analytically continued throughout the entire lower-half plane. Following the similar strategy, we accomplish this by plotting in Fig.~\ref{figS7}(b-d) the sign functions of $\mathrm{Re}[k(z_{\mathrm{nh}})]$, $\mathrm{Im}[k(z_{\mathrm{nh}})]$, and $\mathrm{Im}(z^{2}_{\mathrm{nh}})$, thereby fully specifying the phase of the elliptic integral functions' argument. Based on these phase information, we identify five Riemann sheets defined in regions $\text{I-V}$. Notice, the analytic continuation results of $\Sigma^{\text{A}}_{e}(z)$ on the corresponding Riemann sheets are same as that in Eqs.\,(\ref{S61}-\ref{S64}). Owing to these specific, carefully chosen linear combinations of elliptic integrals, the self-energy remains continuous across the entire integration contour. This continuity is reflected in the smooth variation of the phase of the analytically continued self-energy $\Sigma_{e}^{\text{A}}(z)$ along the contour, as shown in Fig.~\ref{figS7}(e).

Subsequently, the atomic dynamics is fully determined by the contributions from BSs, UPs, and BCDs as formally expressed in Eq.\,(\ref{S65}), with the BCDs now defined as
\begin{align}
C^{k\neq 3}_{\mathrm{BCD}}(t) =&\,\,\frac{1}{2\pi}\int_{-\infty}^{-\kappa_a/4}\dd{y} \left[G_{e}^{\mathrm{A},r_{k}}(x_{k}+iy)-G_{e}^{\mathrm{A},l_{k}}(x_{k}+iy)\right]e^{-i(x_{k}+iy)t},\label{S78}\\
C^{k= 3}_{\mathrm{BCD}}(t) =&\,\,\frac{1}{2\pi}\int_{-\infty}^{0}\dd{y} \left[G_{e}^{\mathrm{A},r_{k}}(x_{k}+iy)-G_{e}^{\mathrm{A},l_{k}}(x_{k}+iy)\right]e^{-i(x_{k}+iy)t}.\label{S79}
\end{align}

We then proceed by plotting the contributions of the BSs, UPs, and BCDs to $C_{e}^{\text{A}}(t)$ at the instant $t=0$ with $\text{g}=0.6J$, as illustrated in Fig.\,\ref{figS7} (f). Different colors and marker styles are used to distinguish the various contributions: blue circles for unstable poles in sheet I (UPI), black squares for unstable poles in sheet II (UPII), purple triangles for unstable poles in sheet III (UPIII), red triangles for unstable poles in sheet IV (UPIV), green diamonds for unstable poles in sheet V (UPV), and brown stars for branch-cut induced detours (BCD). We obtain the following key observations from Fig.\,\ref{figS7}(f): (i) The introduction of single-sublattice dissipation leads to the vanishing of bound states outside the band. Their corresponding poles migrate from the real axis into the lower half-plane, gaining a finite negative imaginary part. (ii) When the detuning is far from the band edges, the primary contribution originates from UPI. When $\Delta_{e}$ approaches the band edge, satisfying $\left|\Delta_{e} \pm \text{Re}(z_{1})\right| \ll J$, the contribution from UPI gradually transforms into that from BCDs. When $\Delta_{e}$ approaches $\text{Re}(z_{2})$  [$\text{Re}(z_{3})$], the contributions from UPI, UPII, and UPIV [UPI, UPIII, and UPV] coexist. In the central region of the band, UPI and the BCDs dominate the contribution. (iii) The BCD contribution shows a discontinuity at the middle of the band, inheriting this characteristic from and thus mirroring the behavior in the dissipation-free case.

The single-emitter logarithmic relaxation shown in Fig.\,\ref{figS6} (c) stems predominantly from the contribution of the middle BCD $C^{k= 3}_{\mathrm{BCD}}(t)$, as clearly demonstrated in Fig.~\ref{figS7}(f). It can be computed by analytically expanding the self-energy $\Sigma_{e}^{\text{A}}(z)$ around the middle of band ($E\approx0$), obtaining
\begin{align}
\Sigma_{e}^{\text{A}}(0^{-}-iy)\approx-\frac{5\text{g}^2y}{\sqrt{3}J^{2}}+\frac{\text{g}^{2}yi}{\sqrt{3}J^{2}\pi}\ln(\frac{\kappa_a y}{18J^{2}}),\,\,\,\,\,\,\,\,\,\,\,\,
\Sigma_{e}^{\text{A}}(0^{+}-iy)\approx\frac{5\text{g}^2y}{\sqrt{3}J^{2}}+\frac{\text{g}^{2}yi}{\sqrt{3}J^{2}\pi}\ln(\frac{\kappa_a y}{18J^{2}}).\label{S80}
\end{align}
Inserting Eqs\,(\ref{S80}) and (\ref{S81}) into Eq\,(\ref{S79}), we have
\begin{align}
\lim_{t\rightarrow\infty}C^{k= 3}_{\mathrm{BCD}}(t)\approx\underset{t\rightarrow\infty}{\mathrm{lim}}\frac{5\sqrt{3}J^{2}\pi}{ \text{g}^{2}}\int_{0}^{\infty}\dd{y}\frac{e^{-yt}}{y\left(\ln(\frac{\kappa_a y}{18J^{2}})\right)^{2}}\approx \frac{5\sqrt{3}J^{2}\pi }{\text{g}^{2}}\frac{1}{\ln(\frac{18J^{2}}{\kappa_a}t)},\label{S81}
\end{align}
which is exactly the Eq.\,(7) in the main text. Notably, the asymptotic expansion of Laplace transform of functions with logarithmic singularities has been utilized\,\cite{SMWONG1978173} in deriving Eq.\,(\ref{S81}), i.e.,
\begin{align}\label{S82}
 \lim_{t\rightarrow\infty}\int_0^\infty dy e^{-yt} y^{\alpha-1} (-\log(y))^\beta =\frac{1}{t^\alpha}\sum_{k=0}^\infty (-1)^k\binom{\beta}{k}\Gamma^{(k)}(\alpha)\big(\log(t))^{\beta-k}.
\end{align}

\subsection{Quasilocalized states in graphenic QED}\label{IIIB}
We recall from subsection \ref{IIIA} that a quasilocalized state (QLS), robust against single-sublattice dissipation, is identified. This robustness is manifested in the atomic dynamics where the population relaxes to a dissipation-independent constant and oscillates around it in the long-time limit. The goal of this subsection is to derive analytical expressions for this constant and the wavefunction of the QLS $\ket{\Psi_{\mathrm{QLS}}}$. Again, we consider here the single-sublattice dissipation configuration with $\kappa_{a}\neq 0,\kappa_{b}=0$.

We are now in a position to quantitatively study the QLS that emerges when the QE is coupled to the sublattice A, with its transition frequency tuned exactly to the middle of the band. The projection of the initial wavefunction $\ket{\Psi_{t=0}}=\sigma^{\dagger}\ket{g;\mathrm{vac}}$ onto the QLS is given by $R_{0}=\braket{\Psi_{t=0}}{\Psi_{\mathrm{QLS}}}$, determining the contribution of this quasibound state to the resulting steady state. The residue $R_{0}$ can be determined as follows
\begin{align}\label{S83}
R_{0}=\left.\frac{1}{1-\partial_{z}\Sigma_{e}^{\text{A}}(z)}\right|_{z=i0^{+}}=\frac{1}{1-\left.\left(\frac{\text{g}^{2}}{N^{2}}\sum\limits_{\boldsymbol{k}}\frac{1}{z_{{\rm nh}}^{2}-|f(\boldsymbol{k})|^{2}}\right)\right|_{z=i0^{+}}}=\frac{1}{1+\left(\frac{\text{g}^{2}}{N^{2}}\sum\limits_{\boldsymbol{k}}\frac{1}{|f(\boldsymbol{k})|^{2}}\right)}=\frac{1}{1+\frac{\text{g}^{2}}{J^{2}}\mathcal{G}(N)},
\end{align}
where the size-dependent quantity $\mathcal{G}(N)=\frac{J^{2}}{N^{2}}\sum\limits_{\boldsymbol{k}}\frac{1}{|f(\boldsymbol{k})|^{2}}$ is introduced for convenience. Given that $1/|f(\boldsymbol{k})|^{2}$ diverges near the Dirac points $\boldsymbol{\mathrm{K}}_{\pm}$, the primary contributions to $\mathcal{G}(N)$ can be approximated by the momentum modes around $\boldsymbol{\mathrm{K}}_{\pm}$, under which
\begin{align}\label{S84}
\mathcal{G}(N)\approx &\,\frac{J^{2}}{N^{2}}\sum_{\boldsymbol{q}}\frac{1}{\left|f(\boldsymbol{\mathrm{K}}_{+}+\boldsymbol{q})\right|^{2}}+\frac{J^{2}}{N^{2}}\sum_{\boldsymbol{q}}
\frac{1}{\left|f(\boldsymbol{\mathrm{K}}_{-}+\boldsymbol{q})\right|^{2}},\nonumber\\
=&\,\frac{2J^{2}}{N^{2}}\sum_{\boldsymbol{q}}\frac{1}{\left|J\boldsymbol{w}_{\pm}\cdot\boldsymbol{q}\right|^{2}}=\frac{2J^{2}}{N^{2}}\sum_{\boldsymbol{q}}\frac{1}{\left|Ji\left(\exp(\frac{2\pi i}{3})q_{1}+\exp(-\frac{2\pi i}{3})q_{2}\right)\right|^{2}},\nonumber\\
=&\,\frac{2}{N^{2}}\sum_{\boldsymbol{q}}\frac{1}{q_{1}^{2}+q_{2}^{2}-q_{1}q_{2}}=\frac{2}{(2\pi)^{2}}\int_{-\pi}^{\pi}\int_{-\pi}^{\pi}\dd{q}_{1}\dd{q}_{2}\frac{1}{q_{1}^{2}+q_{2}^{2}-q_{1}q_{2}}.
\end{align}
Through the coordinate transformation $q_{1,2}=\frac{3}{2}(p_{1}\pm\frac{1}{\sqrt{3}}p_{2})$, it follows that $q_{1}^{2}+q_{2}^{2}-q_{1}q_{2}=\frac{9}{4}(p_{1}^{2}+p_{2}^{2})$ and $\dd{q}_{1}\dd{q}_{2}=\frac{3\sqrt{3}}{2}\dd{q}_{1}\dd{q}_{2}$. This transformation yields an isotropic dispersion relation $\omega_{\text{pg}}(\boldsymbol{k})$ and, furthermore, leads to a more compact expression for $\mathcal{G}(N)$, i.e.,
\begin{align}\label{S85}
\mathcal{G}(N)= &\,\frac{1}{\sqrt{3}\pi{}^{2}}\int_{-\pi}^{\pi}\int_{-\pi}^{\pi}\dd{p}_{1}\dd{p}_{2}\frac{1}{(p_{1}^{2}+p_{2}^{2})}
=\frac{2}{\sqrt{3}\pi}\int_{p_{{\rm min}}}^{p_{c}}\frac{\dd{p}}{p}=\frac{2}{\sqrt{3}\pi}\ln p_{c}+\frac{2}{\sqrt{3}\pi}\ln N,
\end{align}
where the divergences of the integral are regularized by introducing both a minimum $p_{{\rm min}}$ and a maximum cutoff $p_{c}$. The lower cutoff is naturally imposed by the minimum momentum discreteness of the lattice $p_{{\rm min}}\approx 1/N$, and the constant $C=\frac{2}{\sqrt{3}\pi}\ln p_{c}$ can be determined numerically by evaluating the exact double summation $\mathcal{G}(N)$, yielding $C\approx0.2057$. The estimated atomic excitation probability $\left|R_{0}\right|^{2}$ in the long time limit, obtained by substituting Eq.\,\eqref{S85} into Eq.\,\eqref{S83}, agrees well with numerical simulations across a range of dissipation strengths, as clearly demonstrated in Fig.\,\ref{figS6} (c).

We proceed by examining the explicit form of QLS $\ket{\Psi_{\mathrm{QLS}}}$. To this end, we employ the eigenequation $H_{\text{eff}}\ket{\Psi_{\mathrm{BS}}} = E_{\mathrm{BS}}\ket{\Psi_{\mathrm{BS}}}$ to seek its functional expression, where the effective Hamiltonian in the real space with single-sublattice dissipation reads
\begin{align}\label{S86}
H_{{\text {eff}}}=&\,\Delta_{e}\sigma^{\dagger}\sigma+\text{g}(\sigma^{\dagger}a_{n_{e},n_{e}}+\sigma a^{\dagger}_{n_{e},n_{e}})-\frac{\kappa_a}{2}i\sum_{n_{x},n_{y}}a_{n_{x},n_{y}}^{\dagger}a_{n_{x},n_{y}}+J\sum_{n_{x},n_{y}}(a_{n_{x},n_{y}}^{\dagger}b_{n_{x},n_{y}}+a_{n_{x},n_{y}}^{\dagger}b_{n_{x}+1,n_{y}}\nonumber\\
&\,+a_{n_{x},n_{y}}^{\dagger}b_{n_{x},n_{y}+1}
+b_{n_{x},n_{y}}^{\dagger}a_{n_{x},n_{y}}+b_{n_{x}+1,n_{y}}^{\dagger}a_{n_{x},n_{y}}+b_{n_{x},n_{y}+1}^{\dagger}a_{n_{x},n_{y}}),
\end{align}
and the general ansatz for the bound state in the single-excitation manifold takes the form of
\begin{align}\label{S87}
\ket{\Psi_{\mathrm{BS}}}=\left[C_{e}\sigma^{\dagger}+\sum_{n_{x},n_{y}}\left(C^{\A}_{n_{x},n_{y}}a_{n_{x},n_{y}}^{\dagger}+C^{\B}_{n_{x},n_{y}}b_{n_{x},n_{y}}^{\dagger}\right)\right]\ket{g;\mathrm{vac}}.
\end{align}
Substituting Eqs.\,(\ref{S86}) and (\ref{S87}) into the Schr\"{o}dinger equation, $H_{\text{eff}}\ket{\Psi_{\mathrm{BS}}} = E_{\mathrm{BS}}\ket{\Psi_{\mathrm{BS}}}$, yields the following coupled equations
\begin{align}
\Delta_{e}C_{e}+\text{g}C^{\A}_{n_{e},n_{e}}=& \,E_{\mathrm{BS}} C_{e},\label{S88}\\
J(C^{\B}_{n_{x},n_{y}}+C^{\B}_{n_{x}+1,n_{y}}+C^{\B}_{n_{x},n_{y}+1})-\frac{\kappa_a}{2}iC^{\A}_{n_{x},n_{y}}+\text{g}C_{e}\delta_{n_{x},0}\delta_{n_{y},0}=& \,E_{\mathrm{BS}} C^{\B}_{n_{x},n_{y}},\label{S89}\\
J(C^{\A}_{n_{x},n_{y}}+C^{\A}_{n_{x}-1,n_{y}}+C^{\A}_{n_{x},n_{y}-1})=& \,E_{\mathrm{BS}} C^{\A}_{n_{x},n_{y}}.\label{S90}
\end{align}
The condition $\Delta_e =E_{\mathrm{BS}}= 0$ immediately leads to $C^{\A}_{n_e, n_e} = 0$ according to Eq.\,(\ref{S88}). This result, in turn, forces the entire photon field to vanish across the sublattice A, i.e., $C^{\A}_{n_x, n_y} = 0$ for all sites $(n_x, n_y)$, as dictated by the homogeneous difference equation for excitation amplitudes of sites in sublattice A [see Eq.\,(\ref{S90})]. Accordingly, the amplitudes $C^{\B}_{n_{x}, n_{y}}$ are determined by solving the difference equation with a source term, $J(C^{\B}_{n_{x},n_{y}} + C^{\B}_{n_{x}+1,n_{y}} + C^{\B}_{n_{x},n_{y}+1}) = -\text{g}C_{e}\delta_{n_{x},0}\delta_{n_{y},0}$, which in the thermodynamic limit read
\begin{align}\label{S91}
C^{\B}_{n_{x}, n_{y}}=-\frac{\text{g}C_{e}}{J}\int_{-\pi}^{\pi}\frac{\dd{k_{x}}}{2\pi}\int_{-\pi}^{\pi}\frac{\dd{k_{y}}}{2\pi}\frac{e^{i(k_{x}n_{x}+k_{y}n_{y})}}{1+e^{ik_{x}}+e^{ik_{y}}}.
\end{align}
Via the technique of contour integration, this integral admits an analytical evaluation as follows
\begin{align}
C^{\B}_{n_{x}, n_{y}}=&-\frac{\text{g}C_{e}}{J}\int_{|k_{x}|\ge\frac{2\pi}{3}}\frac{\dd{k}_{x}}{2\pi}e^{in_{x}k_{x}}(-1-e^{ik_{x}})^{n_{y}-1}\mathcal{H}(n_{y}-1)+\frac{\text{g}C_{e}}{J}
\int_{|k_{x}|<\frac{2\pi}{3}}\frac{\dd{k}_{x}}{2\pi}e^{in_{x}k_{x}}(-1-e^{ik_{x}})^{n_{y}-1}\mathcal{H}(-n_{y})\nonumber\\
=&-\frac{\text{g}C_{e}}{J}\int_{\frac{2\pi}{3}}^{\pi}\frac{\dd{k}_{x}}{2\pi}\left(e^{in_{x}k_{x}}+e^{-i(n_{x}+n_{y}-1)k_{x}}\right)D_{k_{x}}^{n_{y}-1}\mathcal{H}(n_{y}-1)
+\frac{\text{g}C_{e}}{J}\int_{-\frac{2\pi}{3}}^{\frac{2\pi}{3}}\frac{\dd{k}_{x}}{2\pi}e^{in_{x}k_{x}}D_{k_{x}}^{n_{y}-1}\mathcal{H}(-n_{y})\nonumber\\
=&\sum_{j=0}^{n_{y}-1}C_{e}C_{n_{y}-1}^{j}[G(n_{x}+j)+G(j-n_{x}-n_{y}+1)]\mathcal{H}(n_{y}-1)
+\frac{\text{g}C_{e}}{J}\int_{-\frac{2\pi}{3}}^{\frac{2\pi}{3}}\frac{\dd{k}_{x}}{2\pi}e^{in_{x}k_{x}}D_{k_{x}}^{n_{y}-1}\mathcal{H}(-n_{y}),\label{S92}
\end{align}
where $\mathcal{H}(x) \equiv \Theta(x+0^{+})$ with $\Theta(x)$ the unit Heaviside function, $D_{k_{x}} \equiv -1 - e^{ik_{x}}$, and $G(q)$ is defined by $-(-1)^{n_{y}-1}\text{g}/6J$ for $q=0$ and $-(-1)^{n_{y}-1}\text{g}\left(e^{iq\pi} - e^{iq\frac{2\pi}{3}}\right)/2\pi qJi$ otherwise. Notably, $C_{m}^{n}$ denotes the binomial coefficient. Up to now, we have determined the atomic and photonic amplitudes of the QLS. Then, we demonstrate that this quasibound state is essentially a vacancy-like dressed state (VDS), $\ket{\Psi_{\mathrm{VDS}}}=\left(R_{0}\sigma^{\dagger}+\sum_{n_{x},n_{y}}C^{\B}_{n_{x},n_{y}}b_{n_{x},n_{y}}^{\dagger}\right)\ket{g;\mathrm{vac}}$, which satisfies $H_{{\text{ eff}}}\ket{\Psi_{\mathrm{VDS}}}=\Delta_{e}\ket{\Psi_{\mathrm{VDS}}}$. This state forms a coherent superposition characterized by two key features: (i) its energy is precisely equal to the atomic transition energy $\Delta_{e}$; and (ii) its occupancy on the atom-coupled sublattice is strictly zero.
\begin{figure}
  \centering
  \includegraphics[width=10cm]{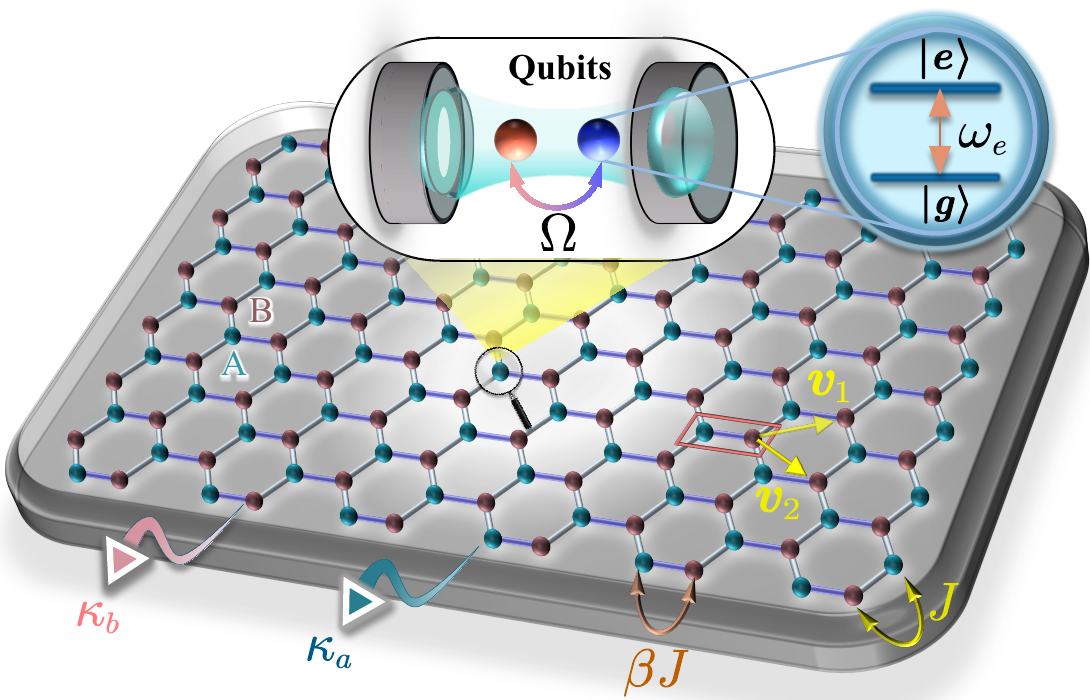}
  \caption{Schematic of quantum emitters coupled to dissipative and anisotropic photonic graphene. The bath consists of two interspersed triangular sublattices, corresponding to sites A and B and represented by dark blue and dark red spheres, respectively. Each sublattice supports a bosonic cavity mode. The primitive vectors $\boldsymbol{v}_{1}$ and $\boldsymbol{v}_{2}$ of the Bravais lattice are shown in cyan. Sites A and B are coupled via inhomogeneous hoppings, denoted by double-headed arrows. The intracell hopping (purple) has strength $\beta J$, whereas the intercell hopping (white) is of strength $J$. A representative unit cell is highlighted by a gray box in the diagram. Each sublattice has its own dissipation rate, with $\kappa_a$ for sublattice A and $\kappa_b$ for sublattice B. A pair of directly coupled qubits, represented by blue and red spheres with coupling strength $\Omega$, are connected to a single site. Each qubit is modeled as a two-level system with ground ($|g\rangle$) and excited ($|e\rangle$) states.}\label{figS9}
\end{figure}
Here, we outline several salient properties of the vacancy-like dressed state. The local coupling of an emitter to the bath creates a local defect in the bath's otherwise periodic structure, thereby giving rise to bound states localized around the emitter. Notably, these emitter-seeded bound states share a one-to-one correspondence with those induced by an on-site potential. Among them, the VDS is characterized by a photonic component identical to that of a bound state arising from an infinite on-site potential (a vacancy). This specific state emerges only when the emitter's transition frequency takes certain values. For a precise definition, applicable also to multiple emitters, we decompose the bath Hamiltonian into $H_{\text{B}}=H^{\circ}_{\text{B}}+H^{\bullet}_{\text{B}}$. The term $H^{\bullet}_{\text{B}}$ consists exclusively of terms involving the sites where the quantum emitters are coupled, whereas $H^{\circ}_{\text{B}}$ encompasses the rest of the terms. Then, a VDS is identified if there exists a single-particle eigenstate
$\ket{\Psi_{\mathrm{VB}}}$ of $H^{\circ}_{\text{B}}$ with energy $E_{\mathrm{VB}}$ that corresponds to a composite eigenstate $\ket{\Psi_{\mathrm{VDS}}}=\sum^{N_{e}}_{\ell=1}e_{\ell}\sigma^{\dagger}_{\ell}\ket{g}^{\otimes N_{e}}\ket{\mathrm{vac}}+\lambda\ket{\Psi_{\mathrm{VB}}}$ of the total Hamiltonian $H_{\mathrm{tot}}$ at the same energy $E_{\mathrm{VDS}}=E_{\mathrm{VB}}$, under the condition $\Delta_{e}=E_{\mathrm{VB}}$.

\subsection{Decoherence-free interactions between two quantum emitters coupled to anisotropic photonic graphene}\label{IIIC}
In this subsection, we will provide a detailed description of the dynamics for two QEs coupled to a photonic graphene with single-sublattice dissipation. Our focus lies in identifying a dissipation-immune physical mechanism that enables the perfect transfer of an initial atomic excitation from one emitter to the other within such a dissipative environment. On the one hand, in a dissipation-free photonic graphene, it has been demonstrated that the excitation stored in one quantum emitter can be transferred almost completely to another at appropriate parameter settings\,\cite{SMPhysRevA.97.043831}. Remarkably, when atomic frequency $\Delta_{e}$ is pinned to the Dirac point, this efficient energy transfer is mediated by the quasi-bound state, enabling it to occur regardless of the distance between QEs. On the other hand, for a photonic graphene with homogeneous dissipation, it can be readily shown that no decoherence-free interaction exists between the emitters. This is because the bound-state poles predicted in the dissipation-free case move into the lower half of the complex plane, causing the QEs’ stored energy to gradually radiate away.

To facilitate perfect energy transfer between QEs within this special dissipative environment, we leverage atomic dark states by co-localizing two QEs inside a single optical cavity belonging to sublattice A, as illustrated in Fig.\,\ref{figS1}. Our analysis begins with the excitation transfer dynamics when the initial excitation resides in the first QE (the donor). The corresponding dynamical process is governed by Eq.\,\eqref{S46}, which is explicitly given as
\begin{align}\label{S93}
e_{2}(t)=&\,\langle s_{2}|\mathcal{P}U(t)\mathcal{P}|s_{1}\rangle=\frac{1}{2\pi i}\int_{\mathcal{C}}\frac{\Omega+\Sigma_{12}^{\A\A}(z)}{\mathscr{D}_{\text{A}}(z)}e^{-izt}\dd{z}\nonumber\\
=&\,\frac{1}{2\pi i}\int_{\mathcal{C}}\frac{\Omega+\Sigma_{12}^{\A\A}(z)}{[z-\Delta_{e}-\Sigma_{11}^{\A\A}(z)][z-\Delta_{e}-\Sigma_{22}^{\A\A}(z)]-[\Omega+\Sigma_{12}^{\A\A}(z)][\Omega+\Sigma_{21}^{\A\A}(z)]}e^{-izt}\dd{z}\nonumber\\
=&\,\frac{1}{2\pi i}\int_{\mathcal{C}}\frac{\Omega+\Sigma_{11}^{\A\A}(z)}{[z-\Delta_{e}-\Sigma_{11}^{\A\A}(z)]^{2}-[\Omega+\Sigma_{11}^{\A\A}(z)]^{2}}e^{-izt}\dd{z}\nonumber\\
=&\,\frac{1}{2\pi i}\int_{\mathcal{C}}\frac{\Omega+\Sigma_{11}^{\A\A}(z)}{[z-\Delta_{e}+\Omega]\times[z-\Delta_{e}-\Omega-2\Sigma_{11}^{\A\A}(z)]}e^{-izt}\dd{z}\nonumber\\
=&\,\frac{1}{2\pi i}\int_{\mathcal{C}}\frac{\Omega+\Sigma_{11}^{\A\A}(z)}{\mathscr{D}_{\text{A}}(z)}e^{-izt}\dd{z}=\frac{1}{2\pi i}\int_{\mathcal{C}}\mathscr{L}_{2}(z)e^{-izt}\dd{z},
\end{align}
exploiting the identity $\Sigma_{12}^{\A\A}(z)=\Sigma_{21}^{\A\A}(z)=\Sigma_{11}^{\A\A}(z)$ for two QEs coupled to a common cavity. The pole equation $\mathscr{D}_{\text{A}}(z)=0$ gives the solutions $z_{1}=\Delta_{e}-\Omega$ and $z_{2}=\Delta_{e}+\Omega+2\Sigma_{11}^{\mathrm{AA}}(z_{2})$. These correspond physically to two coherent bound states: an environment-decoupling dark state with energy $E_{\text{dark}}=\Delta_{e}-\Omega$, and a QLS that emerges only when $\Delta_{e}+\Omega=0$, whose energy is $E_{\text{QLS}}=0$. In the long-time limit, the contribution from radiative states becomes negligible, and the dynamics of the second (acceptor) atom is dominated by the bound states, i.e.,
\begin{align}\label{S94}
e_{2}(t)={\rm Res}[\mathscr{L}_{2}(z),z_{1}]e^{-iz_{1}t}+{\rm Res}[\mathscr{L}_{2}(z),z_{2}]e^{-iz_{2}t},
\end{align}
where the Residues for the dark state pole $z_{1}=\Delta_{e}-\Omega$ and the QLS pole $z_{2}=0$ are respectively given by
\begin{align}
{\rm Res}[\mathscr{L}_{2}(z),z_{1}]=&\,\left.\frac{\Omega+\Sigma_{11}^{\A\A}(z)}{\mathrm{d}\mathscr{D}_{\text{A}}(z)/\mathrm{d}z}\right|_{z=z_{1}}=
\left.\frac{\Omega+\Sigma_{11}^{\A\A}(z)}{z-\Delta_{e}-\Omega-2\Sigma_{11}^{\A\A}(z)}\right|_{z=z_{1}}=-\frac{1}{2},\label{S95}\\
{\rm Res}[\mathscr{L}_{2}(z),z_{2}]=&\,\left.\frac{\Omega+\Sigma_{11}^{\A\A}(z)}{\mathrm{d}\mathscr{D}_{\text{A}}(z)/\mathrm{d}z}\right|_{z=z_{2}=0}=
\left.\frac{\Omega+\Sigma_{11}^{\A\A}(z)}{[z-2\Sigma_{11}^{\A\A}(z)]+(z-\Delta_{e}+\Omega)(1-2\frac{\mathrm{d}\Sigma_{11}^{\A\A}(z)}{\mathrm{d}z})}\right|_{z=0}\nonumber\\
=&\,\left.\frac{\Omega+\Sigma_{11}^{\A\A}(z)}{[(z+2\Omega)(1-2\frac{\mathrm{d}\Sigma_{11}^{\A\A}(z)}{\mathrm{d}z})}\right|_{z=0}=\left.\frac{1}{2(1-2\frac{\mathrm{d}\Sigma_{11}^{\A\A}(z)}{\mathrm{d}z})}\right|_{z=0}
=\frac{1}{2\left(1+\frac{2\text{g}^{2}}{J^{2}}\mathcal{G}(N)\right)}.\label{S96}
\end{align}
Similarly, the temporal evolution of the excitation probability amplitude for the donor QE is governed by
\begin{align}\label{S97}
e_{1}(t)=&\,\langle s_{1}|\mathcal{P}U(t)\mathcal{P}|s_{1}\rangle=\frac{1}{2\pi i}\int_{\mathcal{C}}\frac{z-\Delta_{e}-\Sigma_{22}^{\A\A}(z)}{\mathscr{D}_{\text{A}}(z)}e^{-izt}\dd{z}\nonumber\\
=&\,\frac{1}{2\pi i}\int_{\mathcal{C}}\frac{z-\Delta_{e}-\Sigma_{22}^{\A\A}(z)}{[z-\Delta_{e}-\Sigma_{11}^{\A\A}(z)][z-\Delta_{e}-\Sigma_{22}^{\A\A}(z)]-[\Omega+\Sigma_{12}^{\A\A}(z)][\Omega+\Sigma_{21}^{\A\A}(z)]}e^{-izt}\dd{z}\nonumber\\
=&\,\frac{1}{2\pi i}\int_{\mathcal{C}}\frac{z-\Delta_{e}-\Sigma_{11}^{\A\A}(z)}{[z-\Delta_{e}-\Sigma_{11}^{\A\A}(z)]^{2}-[\Omega+\Sigma_{11}^{\A\A}(z)]^{2}}e^{-izt}\dd{z}\nonumber\\
=&\,\frac{1}{2\pi i}\int_{\mathcal{C}}\frac{z-\Delta_{e}-\Sigma_{11}^{\A\A}(z)}{[z-\Delta_{e}+\Omega]\times[z-\Delta_{e}-\Omega-2\Sigma_{11}^{\A\A}(z)]}e^{-izt}\dd{z}\nonumber\\
=&\,\frac{1}{2\pi i}\int_{\mathcal{C}}\frac{z-\Delta_{e}-\Sigma_{11}^{\A\A}(z)}{\mathscr{D}_{\text{A}}(z)}e^{-izt}\dd{z}=\frac{1}{2\pi i}\int_{\mathcal{C}}\mathscr{L}_{1}(z)e^{-izt}\dd{z},
\end{align}
which shares the same pole equation and its associated bound states as $e_{2}(t)$. In the long-time limit, the dynamics of the donor QE is dominated by bound-state contributions and can be expressed as
\begin{align}\label{S98}
e_{1}(t)={\rm Res}[\mathscr{L}_{1}(z),z_{1}]e^{-iz_{1}t}+{\rm Res}[\mathscr{L}_{1}(z),z_{2}]e^{-iz_{2}t},
\end{align}
where the Residues for the dark state pole $z_{1}=\Delta_{e}-\Omega$ and the QLS pole $z_{2}=0$ are respectively given by
\begin{align}
{\rm Res}[\mathscr{L}_{1}(z),z_{1}]=&\,\left.\frac{z-\Delta_{e}-\Sigma_{11}^{\A\A}(z)}{\mathrm{d}\mathscr{D}_{\text{A}}(z)/\mathrm{d}z}\right|_{z=z_{1}}=
\left.\frac{z-\Delta_{e}-\Sigma_{11}^{\A\A}(z)}{z-\Delta_{e}-\Omega-2\Sigma_{11}^{\A\A}(z)}\right|_{z=z_{1}}=\frac{1}{2},\label{S99}\\
{\rm Res}[\mathscr{L}_{1}(z),z_{2}]=&\,\left.\frac{z-\Delta_{e}-\Sigma_{11}^{\A\A}(z)}{\mathrm{d}\mathscr{D}_{\text{A}}(z)/\mathrm{d}z}\right|_{z=z_{2}=0}=
\left.\frac{z-\Delta_{e}-\Sigma_{11}^{\A\A}(z)}{[z-2\Sigma_{11}^{\A\A}(z)]+(z-\Delta_{e}+\Omega)(1-2\frac{\mathrm{d}\Sigma_{11}^{\A\A}(z)}{\mathrm{d}z})}\right|_{z=0}\nonumber\\
=&\,\left.\frac{z-\Delta_{e}-\Sigma_{11}^{\A\A}(z)}{[(z+2\Omega)(1-2\frac{\mathrm{d}\Sigma_{11}^{AA}(z)}{\mathrm{d}z})}\right|_{z=0}=\left.\frac{1}{2(1-2\frac{\mathrm{d}\Sigma_{11}^{\A\A}(z)}{\mathrm{d}z})}\right|_{z=0}
=\frac{1}{2\left(1+\frac{2\text{g}^{2}}{J^{2}}\mathcal{G}(N)\right)}.\label{S100}
\end{align}
Inserting Eqs.\,(\ref{S95}-\ref{S96}) and Eqs.\,(\ref{S99}-\ref{S100}) into Eq.\,(\ref{S93}) and Eq.\,(\ref{S98}), respectively, we have
\begin{align}\label{S101}
e_{2}(t)=-\frac{1}{2}e^{2i\Omega t}+\frac{1}{2}[\frac{1}{1+2\text{g}^{2}\mathcal{G}(N)/J^{2}}],\,\,\,\,\,e_{1}(t)=\frac{1}{2}e^{2i\Omega t}+\frac{1}{2}[\frac{1}{1+2\text{g}^{2}\mathcal{G}(N)/J^{2}}].
\end{align}

Analysis of Eq.\,(\ref{S101}) implies that the efficiency of energy transfer between QEs is critically governed by both the light-matter interaction strength $\text{g}$ and the lattice size, as shown in Fig.\,\ref{figS8}(a). Specifically, we find that under a fixed light-matter coupling strength, a larger photonic lattice size leads to a smaller steady-state population in  the acceptor QE. Conversely, the weaker the coupling strength, the larger the steady-state population of the acceptor atom, with the value approaching unity in the limit $\text{g}/J \ll 1$. This behavior is clearly demonstrated by examining the time evolutions of $P_{1}(t) = \left|e_{1}(t)\right|^{2}$ and $P_{2}(t) = \left|e_{2}(t)\right|^{2}$ for a weak coupling ($\text{g}=0.01J$) and a moderate coupling ($\text{g}=0.2J$) in a bath of size $N=2^{6}$, as shown in Figs.\,\ref{figS8}(b) and \ref{figS8}(c), respectively. Notably, the remarkable agreement between the numerical (solid lines) and analytical (markers) solutions for $P_{1}(t)$ and $P_{2}(t)$ confirms the reliability of our approach in capturing the long-time behavior. The analytical solution is obtained straightforwardly by inserting Eq.\,(\ref{S85}) into Eq.\,(\ref{S101}).

We now outline the numerical method employed to simulate the quantum dynamics across the full Hilbert space. Our approach entails solving the time-dependent Schr\"{o}dinger equation for the non-Hermitian effective Hamiltonian $H_{\text{eff}} = H_{\text{atom}} + H^{\text{eff}}_{\text{pg}} + H_{\text{int}}$ directly in the real-space. To do so, we construct a matrix representation $\mathbb{H}$ of the effective full Hamiltonian in a basis comprising $2N^2$ localized bosonic modes and $N_e$ atomic excited states, restricted to the single-excitation subspace. In this representation, $H^{\text{eff}}_{\text{pg}}$ becomes a $2N^2 \times 2N^2$ sparse matrix. After that, the total dynamics is described by $2N^2 + N_\text{e}$ coupled differential equations $\dot{\bm{\psi}}(t) = -i \mathbb{H}\bm{\psi}(t)$, where $N_\text{e}$ denotes the total number of QEs and $\bm{\psi}(t) \in \mathbb{C}^{2N^2 + N_\text{e}}$ is a time-dependent vector defined as $\bm{\psi}(t)=\left(C^{\text{A}}_{\boldsymbol{n}_1}(t),...,C^{\text{A}}_{\boldsymbol{n}_{N^2}}(t),C^{\text{B}}_{\boldsymbol{n}_1}(t),...,C^{\text{B}}_{\boldsymbol{n}_{N^2}}(t),e_1(t),...,e_{N_\text{e}}(t)\right)^T$. The matrix representation $\mathbb{H}$ of the effective full Hamiltonian including both the QEs and their photonic environment can be arranged in a block structure:
\begin{equation}\label{S102}
\mathbb{H} =
\begin{tikzpicture}[baseline=(current bounding box.center)]
\matrix (m)[
    matrix of math nodes,
    left delimiter={(},right delimiter={)},
    inner sep=1pt,
    column sep=0.5pt,row sep=0.5pt
    ]
    {|[inner sep=5mm]|H^{\text{eff}}_{\text{pg}} & |[inner sep=1mm]|H_{\mathrm{int}}\\
     |[inner sep=1mm]|H_{\mathrm{int}} & |[inner sep=1mm]|H_{\text{atom}}\\};

\draw (m-1-1.south west) |- (m-1-1.north east) |- (m-1-1.south west);
\draw (m-2-2.south west) |- (m-2-2.north east) |- (m-2-2.south west);
\draw (m-2-2.north east) |- (m-1-1.north east);
\draw (m-1-1.south west) |- (m-2-2.south west);
\end{tikzpicture},
\end{equation}
where $H^{\text{eff}}_{\text{pg}}$ is sparse, $H_{\text{atom}}$ includes diagonal entries representing the free Hamiltonian of the QEs and may also contain off-diagonal terms if inter‑QE couplings are present, while $H_{\mathrm{int}}$ only has nonzero entries at the lattice sites that are coupled to the QEs. Note that the solution to the differential equation $\dot{\bm{\psi}}(t) = -i \mathbb{H}\bm{\psi}(t)$ is given by the matrix exponential $\bm{\psi}(t) = \exp\left(-i \mathbb{H} t \right) \bm{\psi}(0)$, which can be evaluated efficiently using several state-of-the-art algorithms\,\cite{SMMohy,SMHigham}.

We now proceed to determine the specific forms of the dark state $\ket{\Psi_{\mathrm{Dark}}}$ and QLS $\ket{\Psi_{\mathrm{QLS}}}$. These wavefunctions are obtained by solving the secular equation $H_{\text{eff}}\ket{\Psi} = E\ket{\Psi}$ with the energy $E$ taken as $E_{\mathrm{Dark}}$ and $E_{\mathrm{QLS}}$, respectively. After some algorithms, we have
\begin{align}\label{S103}
\ket{\Psi_{\mathrm{Dark}}}=\frac{1}{\sqrt{2}}(\sigma_{1}^{\dagger}-\sigma_{2}^{\dagger})\ket{g,g;\mathrm{vac}},\,\,\,\,
\ket{\Psi_{\mathrm{QLS}}}=\mathcal{N}\left[(\sigma_{1}^{\dagger}+\sigma_{2}^{\dagger})+\sum_{\boldsymbol{n}}\left(C^{\A}_{\boldsymbol{n}}a_{\boldsymbol{n}}^{\dagger}+C^{\B}_{\boldsymbol{n}}b_{\boldsymbol{n}}^{\dagger}\right)\right]\ket{g,g;\mathrm{vac}},
\end{align}
where $C^{\A}_{n_{x},n_{y}}=0$ for all sites in sublattice A, $\mathcal{N}$ denotes the normalization constant, and $C^{\B}_{n_{x},n_{y}}$ takes the form of
\begin{align}\label{S104}
C^{\B}_{n_{x},n_{y}}=\sum_{j=0}^{n_{y}-1}2C_{n_{y}-1}^{j}[G(n_{x}+j)+G(j-n_{x}-n_{y}+1)]\mathcal{H}(n_{y}-1)
+\frac{2\text{g}}{J}\int_{-\frac{2\pi}{3}}^{\frac{2\pi}{3}}\frac{dk_{x}}{2\pi}e^{in_{x}k_{x}}D_{k_{x}}^{n_{y}-1}\mathcal{H}(-n_{y}),
\end{align}
which is derived in a manner analogous to the single QE case.

We thus end up with a simple demonstration that the coherent energy transfer cannot be achieved between two QEs coupled to the same cavity in dissipativeless sublattice B. The excitation transport dynamics in this scenario is governed by
\begin{align}\label{S105}
e_{2}(t)=&\,\langle e_{2}|\mathcal{P}U(t)\mathcal{P}|e_{1}\rangle=\frac{1}{2\pi i}\int_{\mathcal{C}}\frac{\Omega+\Sigma_{12}^{\B\B}(z)}{\mathscr{D}_{\text{B}}(z)}e^{-izt}\dd{z}\nonumber\\
=&\,\frac{1}{2\pi i}\int_{\mathcal{C}}\frac{\Omega+\Sigma_{12}^{\B\B}(z)}{[z-\Delta_{e}-\Sigma_{11}^{\B\B}(z)][z-\Delta_{e}-\Sigma_{22}^{\B\B}(z)]-[\Omega+\Sigma_{12}^{\B\B}(z)][\Omega+\Sigma_{21}^{\B\B}(z)]}e^{-izt}\dd{z}\nonumber\\
=&\,\frac{1}{2\pi i}\int_{\mathcal{C}}\frac{\Omega+\Sigma_{11}^{\B\B}(z)}{[z-\Delta_{e}-\Sigma_{11}^{\B\B}(z)]^{2}-[\Omega+\Sigma_{11}^{\B\B}(z)]^{2}}e^{-izt}\dd{z}\nonumber\\
=&\,\frac{1}{2\pi i}\int_{\mathcal{C}}\frac{\Omega+\Sigma_{11}^{\B\B}(z)}{[z-\Delta_{e}+\Omega]\times[z-\Delta_{e}-\Omega-2\Sigma_{11}^{\B\B}(z)]}e^{-izt}\dd{z}\nonumber\\
=&\,\frac{1}{2\pi i}\int_{\mathcal{C}}\frac{\Omega+\Sigma_{11}^{\B\B}(z)}{\mathscr{D}_{\text{B}}(z)}e^{-izt}\dd{z}=\frac{1}{2\pi i}\int_{\mathcal{C}}\mathscr{L}_{2}(z)e^{-izt}\dd{z},
\end{align}
exploiting the identity $\Sigma_{12}^{\B\B}(z)=\Sigma_{21}^{\B\B}(z)=\Sigma_{11}^{\B\B}(z)$ for two QEs coupled to a common cavity. The pole equation $\mathscr{D}_{\B}(z)=0$ yields only the atomic dark state of energy $E_{\mathrm{Dark}} = \Delta_{e} - \Omega$ with the corresponding residue ${\rm Res}[\mathscr{L}_{2}(z), z=E_{\mathrm{Dark}}] = -1/2$, leading to $e_{2}(t)=-e^{-i(\Delta_{e}-\Omega)t}/2$. Consequently, at most one-quarter of the donor QE's energy can be transferred to the acceptor QE.

In summary, we find that when two QEs are simultaneously coupled to the same lattice site, an additional dark state—consisting purely of atomic components—appears. Moreover, a QLS with zero energy emerges only when the emitters are coupled to sites of the dissipative sublattice. The coherent superposition of these two bound states enables perfect energy transfer between the QEs, which remains unaffected by environmental dissipation in the weak‑coupling regime. Notably, neither bound state alone is sufficient to realize this effect. In the next section, we will examine how other types of bound states can be harnessed to facilitate energy transfer in this dissipative photonic‑graphene environment.
\begin{figure}
  \centering
  \includegraphics[width=16cm]{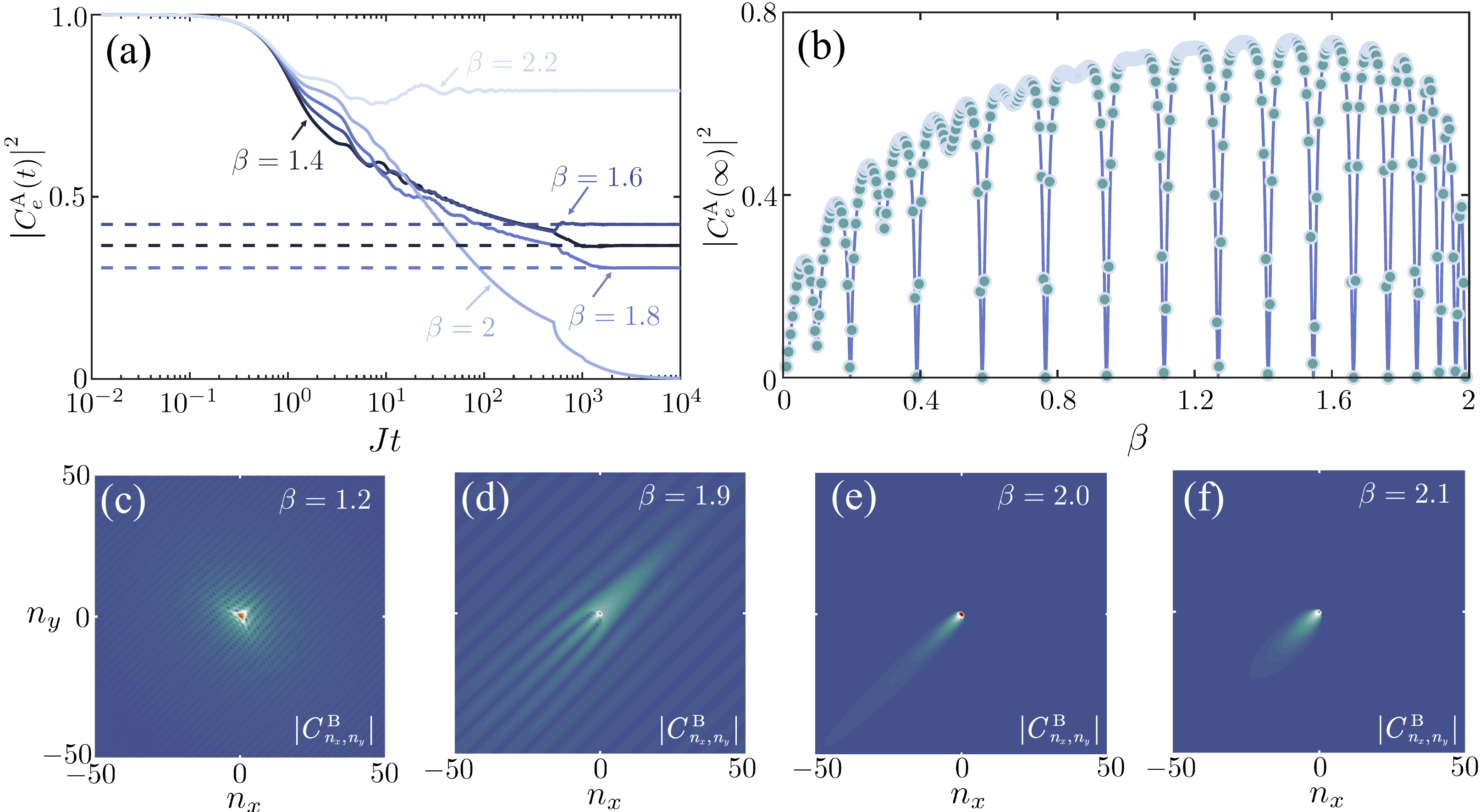}
  \caption{ (a) Time evolution of the excitation probability $\left|C_{e}^{\text{A}}(t)\right|^{2}$ for a single QE coupled to sublattice A, with parameters set to $\text{g}=0.5J$, $\kappa_a=0.01J$, and lattice size $N=512$. The differently colored curves indicate the numerical results for different values of the anisotropy parameter ranging from $\beta=1.4$ to $\beta=2.2$. (b) Variation of the steady-state atomic excitation probability, $\left|C_{e}^{\text{A}}(\infty)\right|^{2}$, as a function of the anisotropy parameter with the implemented parameters $\text{g}=0.5J$ and $N=64$. We present the real‑space distribution of the photonic wave function $\left|C^{\B}_{n_{x},n_{y}}\right|$ in sublattice B and its dependence on the anisotropy parameter $\beta$: (c) $\beta=1.2$, (d) $\beta=1.9$, (e) $\beta=2$, (f) $\beta=2.1$. The wave function is calculated through Eqs.\,(\ref{S132}-\ref{S133}) and evaluated over the spatial region defined by $n_{x}\in[-50,50]$ and $n_{y}\in[-50,50]$. It should be noted that the results in all panels are obtained by setting a resonant atomic frequency $\Delta_{e}=0$.
}\label{figS10}
\end{figure}
%
\section{Quantum Dynamics in an Anisotropic Dirac Photonic Bath with Single-Sublattice Dissipation}\label{IV}
\renewcommand\theequation{S\arabic{equation}}
\makeatletter
\renewcommand{\thefigure}{S\@arabic\c@figure}
\makeatother
In the previous sections, we have investigated isotropic graphenic QED under both homogeneous and single‑sublattice dissipation configurations. In this section, we address how an anisotropic photonic graphene modifies light-matter interactions, focusing on the case of single-sublattice dissipation. We begin in subsection \ref{IVA} with a brief description of the spectral properties for this anisotropic bath. This is followed by the derivation of the self-energy functions for both single- and two-emitter cases. Building upon these useful calculations, we then analyze in Subsection\,\ref{IVB} the single‑emitter dynamics in the anisotropic lattice and examine the associated single‑emitter QLS. Finally, in subsection \ref{IVC}, we demonstrate the feasibility of engineering decoherence‑free interactions between QEs and analyze how these interactions depend on the anisotropy parameter.

\subsection{The non-Hermitian anisotropic photonic graphene}\label{IVA}
We now introduce anisotropy into the photonic graphene by modifying the intracell coupling strength to $\beta J$, while keeping the intercell coupling fixed at $J$, as depicted in Fig.\,\ref{figS9}. All other aspects of the model remain unchanged from the isotropic case, including the lattice structure, the primitive vectors $\boldsymbol{v}{1}=(3,\sqrt{3})/2$ and $\boldsymbol{v}{2}=(3,-\sqrt{3})/2$, and the dissipation rates $\kappa_a$ and $\kappa_b$ for sublattices A and B, respectively. The computational methodology employed in this section bears a strong resemblance to the one rigorously established in Sec.\,\ref{I}. To maintain the narrative flow and avoid undue repetition, we have chosen to omit the step-by-step derivations here. We first present the anisotropic bath Hamiltonian in real space in the absence of photon loss
\begin{align}
H_{\text{pg}}=\omega_{c}\sum\limits_{\boldsymbol{n}}(a^{\dagger}_{\boldsymbol{n}}a^{}_{\boldsymbol{n}}+b^{\dagger}_{\boldsymbol{n}}b^{}_{\boldsymbol{n}})+\sum_{\boldsymbol{n}}\left(J a_{\boldsymbol{n}+\boldsymbol{v}_{1}}^{\dagger}b^{}_{\boldsymbol{n}}+J a_{\boldsymbol{n}+\boldsymbol{v}_{2}}^{\dagger}b^{}_{\boldsymbol{n}}+\beta J a_{\boldsymbol{n}}^{\dagger}b^{}_{\boldsymbol{n}}+{\rm H.c.}\right),\label{S106}
\end{align}
where the operators $a_{\boldsymbol{n}}$ ($a_{\boldsymbol{n}}^{\dagger}$) and $b_{\boldsymbol{n}}$ ($b_{\boldsymbol{n}}^{\dagger}$) denote the annihilation (creation) operators for cavity modes on sublattice A and B at position $\boldsymbol{n}$, respectively, and $\beta$ is the anisotropy parameter. Notably, the configuration with $\beta= 1$ maps exactly onto the original isotropic lattice. This anisotropic photonic graphene, which comprises also two sublattices (A and B), yields a two-band energy structure that can be obtained directly via a discrete Fourier transform under periodic boundary conditions. We similarly begin by rewriting the bath Hamiltonian Eq.\,(\ref{S106}) in a frame rotating at the cavity frequency, $\omega_{c}$, thereby setting $\omega_{c}$ as the energy reference. By performing the transformations in Eq.\,(\ref{S2}), we obtain
\begin{align}
H_{\text{pg}}= &\,J\sum\limits_{\boldsymbol{k}}(\beta+e^{-i\boldsymbol{k}\cdot\boldsymbol{v}_{1}}+e^{-i\boldsymbol{k}\cdot\boldsymbol{v}_{2}})a^{\dagger}_{\boldsymbol{k}}b^{}_{\boldsymbol{k}}+\mathrm{H.c.}=\sum_{\boldsymbol{k}}\left[a_{\boldsymbol{k}}^{\dagger},b_{\boldsymbol{k}}^{\dagger}\right]\left[\begin{array}{cc}
0 & f_{\mathrm{aniso}}(\boldsymbol{k})\\
f_{\mathrm{aniso}}^{*}(\boldsymbol{k}) & 0
\end{array}\right]\left[\begin{array}{c}
a_{\boldsymbol{k}}\\
b_{\boldsymbol{k}}
\end{array}\right],\label{S107}
\end{align}
where the natation $f_{\mathrm{aniso}}(\boldsymbol{k})=1+e^{-i\boldsymbol{k}\cdot\boldsymbol{v}_{1}}+e^{-i\boldsymbol{k}\cdot\boldsymbol{v}_{2}}$ has been introduced for conciseness. The bath Hamiltonian can be expressed as $H_{\text{pg}}=\sum_{\boldsymbol{k}}\boldsymbol{\mathrm{o}}_{\boldsymbol{k}}^\dagger \boldsymbol{\mathrm{h}}^{(0)}_{\boldsymbol{k}}\boldsymbol{\mathrm{o}}^{}_{\boldsymbol{k}}$, where $\boldsymbol{\mathrm{o}}_{\boldsymbol{k}}=[a_{\boldsymbol{k}},b_{\boldsymbol{k}}]^T$. Here, $\boldsymbol{\mathrm{h}}^{(0)}_{\boldsymbol{k}}$ represents the Bloch Hamiltonian (or kernel Hamiltonian), and has the form of
\begin{align}
\boldsymbol{\mathrm{h}}^{(0)}_{\boldsymbol{k}}=\left[\begin{array}{cc}
0 & f_{\mathrm{aniso}}(\boldsymbol{k})\\
f_{\mathrm{aniso}}^{*}(\boldsymbol{k}) & 0
\end{array}\right]=\Re[f_{\mathrm{aniso}}(\boldsymbol{k})]\sigma_x-\Im[f_{\mathrm{aniso}}(\boldsymbol{k})]\sigma_y.\label{S108}
\end{align}
 The Hamiltonian $\boldsymbol{\mathrm{h}}^{(0)}_{\boldsymbol{k}}$ exhibits chiral (sublattice) symmetry, ensuring that all eigenmodes appear in chiral pairs with opposite energies. Upon diagonalization, we obtain
\begin{align}
H_{\text{pg}}=\sum_{\boldsymbol{k}}\left[u_{\boldsymbol{k}}^{\dagger},l_{\boldsymbol{k}}^{\dagger}\right]\left[\begin{array}{cc}
\omega_{\mathrm{aniso}}(\boldsymbol{k}) & 0\\
0 & -\omega_{\mathrm{aniso}}(\boldsymbol{k})
\end{array}\right]\left[\begin{array}{c}
u_{\boldsymbol{k}}\\
l_{\boldsymbol{k}}
\end{array}\right]=\sum_{\boldsymbol{k}}\omega_{\mathrm{aniso}}(\boldsymbol{k})(u_{\boldsymbol{k}}^{\dagger}u_{\boldsymbol{k}}-l_{\boldsymbol{k}}^{\dagger}l_{\boldsymbol{k}}),\label{S109}
\end{align}
where the eigenoperators $u_{\boldsymbol{k}}=[a_{\boldsymbol{k}}+ b_{\boldsymbol{k}}e^{i\phi(\boldsymbol{k})}]/\sqrt{2}$ and $l_{\boldsymbol{k}}=[a_{\boldsymbol{k}}- b_{\boldsymbol{k}}e^{i\phi(\boldsymbol{k})}]/\sqrt{2}$ represent the annihilation operators for the upper and lower band modes, respectively, and the phase factor is defined by $\phi(\boldsymbol{k})\equiv\arctan\left(\Im[f_{\mathrm{aniso}}(\boldsymbol{k})]/\Re[f_{\mathrm{aniso}}(\boldsymbol{k})]\right)$. The dispersion relation $\omega_{\mathrm{aniso}}(\boldsymbol{k})$ for this anisotropic photonic graphene is given by\,\cite{SMRedondo-Yuste_2021}
\begin{align}
\omega_{\mathrm{aniso}}(\boldsymbol{k})=\sqrt{2+\beta^{2}+2\cos(k_{1}-k_{2})+2\beta(\cos k_{1}+\cos k_{2})}.\label{S110}
\end{align}
In the presence of dissipation, the equation of motion for the system plus environment is described by the following Lindblad master equation
\begin{align}\label{S111}
	\dot{\rho}_t=-i[H_{\text{atom}}+H_{\text{pg}}+H_{\text{int}},\rho_t]+\kappa_{a}\sum_{\boldsymbol{n}}\mathcal{D}[a_{\boldsymbol{n}}]\rho_t
+\kappa_{b}\sum_{\boldsymbol{n}}\mathcal{D}[b_{\boldsymbol{n}}]\rho_t,
\end{align}
where $H_{\text{atom}}$ and $H_{\text{int}}$ are the system Hamiltonian containing only the QEs and the light-matter interaction Hamiltonian, respectively, and $H_{\text{pg}}$ is now defined by Eq.~(\ref{S109}). In order to solve Eq.\,\eqref{S111} in the single-excitation subspace, it is instructive to transform the master equation into a more tractable form
\begin{align}\label{S112}
\dot{\rho}_t = -i(H_{\text{eff}} \rho_t - \rho_t H_{\text{eff}}^\dagger) + \kappa_a \sum_{\boldsymbol{n}}a_{\boldsymbol{n}}\rho_t a_{\boldsymbol{n}}^\dagger + \kappa_b \sum_{\boldsymbol{n}}b_{\boldsymbol{n}}\rho_t b_{\boldsymbol{n}}^\dagger.
\end{align}
Observably, the introduction of lattice dissipation leads to a marked departure from the closed system's dynamics, as captured by the effective Hamiltonian $H_{\text{eff}} = H_{\text{tot}} - i(\kappa_{a}/2)\sum_{\boldsymbol{n}} a^\dagger_{\boldsymbol{n}} a^{}_{\boldsymbol{n}} - i(\kappa_{b}/2) \sum_{\boldsymbol{n}} b^\dagger_{\boldsymbol{n}} b^{}_{\boldsymbol{n}}$, where $H_{\text{tot}}=H_{\text{atom}}+H_{\text{pg}}+H_{\text{int}}$. The solution to the master equation (\ref{S112}) can be obtained by proceedings as before, and the details are omitted here for brevity. By following the same procedure presented in Eqs.\,(\ref{S1}-\ref{S4}), we construct the corresponding non-Hermitian Bloch Hamiltonian as
\begin{align}\label{S113}
\boldsymbol{\mathrm{h}}_{\boldsymbol{k}}=\left[\begin{array}{cc}
-i\kappa_{a}/2 & f_{\mathrm{aniso}}(\boldsymbol{k})\\
f_{\mathrm{aniso}}^{*}(\boldsymbol{k}) & -i\kappa_{b}/2
\end{array}\right]=\Re[f_{\mathrm{aniso}}(\boldsymbol{k})]\sigma_x-\Im[f_{\mathrm{aniso}}(\boldsymbol{k})]\sigma_y-i\kappa_{-}\sigma_{z}-i\kappa_{+}\sigma_{0},
\end{align}
where the notations $\kappa_{+}$ and $\kappa_{-}$ have been defined before. We diagonalize the kernel Hamiltonian in Eq.\,(\ref{S110}) to obtain the complex eigenenergies
\begin{align}\label{S114}
\omega_{\pm}(\boldsymbol{k})=-i\kappa_{+}\pm\sqrt{\omega^{2}_{\mathrm{aniso}}(\boldsymbol{k})-\kappa_{-}^{2}}.
\end{align}
After that, the non-Hermitian effective bath Hamiltonian $H^{\text{eff}}_{\text{pg}}$ can be rewritten as
\begin{align}\label{S115}
H^{\text{eff}}_{\text{pg}}=\sum_{\boldsymbol{k}}\left[a_{\boldsymbol{k}}^{\dagger},b_{\boldsymbol{k}}^{\dagger}\right]\left[\begin{array}{cc}
-i\kappa_{a}/2 & f_{\mathrm{aniso}}(\boldsymbol{k})\\
f_{\mathrm{aniso}}^{*}(\boldsymbol{k}) & -i\kappa_{b}/2
\end{array}\right]\left[\begin{array}{c}
a_{\boldsymbol{k}}\\
b_{\boldsymbol{k}}
\end{array}\right]=\sum_{\boldsymbol{k}}\left(\omega_{+}(\boldsymbol{k})u_{\boldsymbol{k},L}^{\dagger}u^{}_{\boldsymbol{k},R}+
\omega_{-}(\boldsymbol{k})l_{\boldsymbol{k},L}^{\dagger}l^{}_{\boldsymbol{k},R}\right),
\end{align}
where the right and left eigenvectors, $\ket{u/l_{\boldsymbol{k},R}}$ and $\bra{u/l_{\boldsymbol{k},L}}$, are defined analogously to Eq.\,(\ref{S15}).

In the following, we investigate the dynamics of QEs coupled to this anisotropic photonic graphene within the single-excitation subspace. This is implemented by projecting the full evolution operator $U(t)$ onto the subspace spanned exclusively by the QEs. The related projection operators are defined by Eqs.\,(\ref{S19}) and (\ref{S37}), which correspond to the single-QE and two-QE scenarios, respectively. Applying the resolvent operator method introduced in Sec.\,\ref{I}, we derive the following self-energies for single and two QEs:
\begin{align}
\Sigma_{e}^{\text{A}}(z)=&\,\text{g}^{2}\left\langle {\rm vac}\right|c_{\boldsymbol{n}_{e},\A}(z-H^{\text{eff}}_{\text{pg}})^{-1}c_{\boldsymbol{n}_{e},\A}^{\dagger}\left|{\rm vac}\right\rangle =\,\text{g}^{2}\iint\frac{\dd{\boldsymbol{k}}}{(2\pi)^{2}}\frac{z+i\kappa_{b}/2}{z_{{\rm nh}}^{2}-\omega^{2}_{\mathrm{aniso}}(\boldsymbol{k})}=\Sigma_{11}^{\text{AA}}(z)=\Sigma_{22}^{\text{AA}}(z),\label{S116}\\
\Sigma_{e}^{\text{B}}(z)=&\,\text{g}^{2}\left\langle {\rm vac}\right|c_{\boldsymbol{n}_{e},\B}(z-H^{\text{eff}}_{\text{pg}})^{-1}c_{\boldsymbol{n}_{e},\B}^{\dagger}\left|{\rm vac}\right\rangle =\,\text{g}^{2}\iint\frac{\dd{\boldsymbol{k}}}{(2\pi)^{2}}\frac{z+i\kappa_{a}/2}{z_{{\rm nh}}^{2}-\omega^{2}_{\mathrm{aniso}}(\boldsymbol{k})}=\Sigma_{11}^{\text{BB}}(z)=\Sigma_{22}^{\text{BB}}(z),\label{S117}\\
\Sigma_{12}^{\text{AA}}(z)=&\,\text{g}^{2}\left\langle {\rm vac}\right|c_{\boldsymbol{n}_{1},\A}(z-H^{\text{eff}}_{\text{pg}})^{-1}c_{\boldsymbol{n}_{2},\A}^{\dagger}\left|{\rm vac}\right\rangle
=\,\text{g}^{2}\iint\frac{\dd{\boldsymbol{k}}}{(2\pi)^{2}}\frac{z+i\kappa_{b}/2}{z_{{\rm nh}}^{2}-\omega^{2}_{\mathrm{aniso}}(\boldsymbol{k})}e^{i\boldsymbol{k}\cdot(\boldsymbol{n}_{1}-\boldsymbol{n}_{2})},\label{S118}\\
\Sigma_{21}^{\text{AA}}(z)=&\,\text{g}^{2}\left\langle {\rm vac}\right|c_{\boldsymbol{n}_{2},\A}(z-H^{\text{eff}}_{\text{pg}})^{-1}c_{\boldsymbol{n}_{1},\A}^{\dagger}\left|{\rm vac}\right\rangle
=\,\text{g}^{2}\iint\frac{\dd{\boldsymbol{k}}}{(2\pi)^{2}}\frac{z+i\kappa_{b}/2}{z_{{\rm nh}}^{2}-\omega^{2}_{\mathrm{aniso}}(\boldsymbol{k})}e^{-i\boldsymbol{k}\cdot(\boldsymbol{n}_{1}-\boldsymbol{n}_{2})},\label{S119}\\
\Sigma_{12}^{\text{BB}}(z)=&\,\text{g}^{2}\left\langle {\rm vac}\right|c_{\boldsymbol{n}_{1},\B}(z-H^{\text{eff}}_{\text{pg}})^{-1}c_{\boldsymbol{n}_{2},\B}^{\dagger}\left|{\rm vac}\right\rangle
=\,\text{g}^{2}\iint\frac{\dd{\boldsymbol{k}}}{(2\pi)^{2}}\frac{z+i\kappa_{a}/2}{z_{{\rm nh}}^{2}-\omega^{2}_{\mathrm{aniso}}(\boldsymbol{k})}e^{i\boldsymbol{k}\cdot(\boldsymbol{n}_{1}-\boldsymbol{n}_{2})},\label{S120}\\
\Sigma_{21}^{\text{BB}}(z)=&\,\text{g}^{2}\left\langle {\rm vac}\right|c_{\boldsymbol{n}_{2},\B}(z-H^{\text{eff}}_{\text{pg}})^{-1}c_{\boldsymbol{n}_{1},\B}^{\dagger}\left|{\rm vac}\right\rangle
=\,\text{g}^{2}\iint\frac{\dd{\boldsymbol{k}}}{(2\pi)^{2}}\frac{z+i\kappa_{a}/2}{z_{{\rm nh}}^{2}-\omega^{2}_{\mathrm{aniso}}(\boldsymbol{k})}e^{-i\boldsymbol{k}\cdot(\boldsymbol{n}_{1}-\boldsymbol{n}_{2})},\label{S121}\\
\Sigma_{12}^{\text{AB}}(z)=&\,\text{g}^{2}\left\langle {\rm vac}\right|c_{\boldsymbol{n}_{1},\A}(z-H^{\text{eff}}_{\text{pg}})^{-1}c_{\boldsymbol{n}_{2},\B}^{\dagger}\left|{\rm vac}\right\rangle
=\,\text{g}^{2}\iint\frac{\dd{\boldsymbol{k}}}{(2\pi)^{2}}\frac{\omega_{\mathrm{aniso}}(\boldsymbol{k})e^{i\phi(\boldsymbol{k})}}{z_{{\rm nh}}^{2}-\omega^{2}_{\mathrm{aniso}}(\boldsymbol{k})}e^{i\boldsymbol{k}\cdot(\boldsymbol{n}_{1}-\boldsymbol{n}_{2})},\label{S122}\\
\Sigma_{12}^{\text{BA}}(z)=&\,\text{g}^{2}\left\langle {\rm vac}\right|c_{\boldsymbol{n}_{1},\B}(z-H^{\text{eff}}_{\text{pg}})^{-1}c_{\boldsymbol{n}_{2},\A}^{\dagger}\left|{\rm vac}\right\rangle
=\,\text{g}^{2}\iint\frac{\dd{\boldsymbol{k}}}{(2\pi)^{2}}\frac{\omega_{\mathrm{aniso}}(\boldsymbol{k})e^{-i\phi(\boldsymbol{k})}}{z_{{\rm nh}}^{2}-\omega^{2}_{\mathrm{aniso}}(\boldsymbol{k})}e^{i\boldsymbol{k}\cdot(\boldsymbol{n}_{1}-\boldsymbol{n}_{2})},\label{S123}\\
\Sigma_{21}^{\text{BA}}(z)=&\,\text{g}^{2}\left\langle {\rm vac}\right|c_{\boldsymbol{n}_{2},\B}(z-H^{\text{eff}}_{\text{pg}})^{-1}c_{\boldsymbol{n}_{1},\A}^{\dagger}\left|{\rm vac}\right\rangle
=\,\text{g}^{2}\iint\frac{\dd{\boldsymbol{k}}}{(2\pi)^{2}}\frac{\omega_{\mathrm{aniso}}(\boldsymbol{k})e^{-i\phi(\boldsymbol{k})}}{z_{{\rm nh}}^{2}-\omega^{2}_{\mathrm{aniso}}(\boldsymbol{k})}e^{-i\boldsymbol{k}\cdot(\boldsymbol{n}_{1}-\boldsymbol{n}_{2})},\label{S124}\\
\Sigma_{21}^{\text{AB}}(z)=&\,\text{g}^{2}\left\langle {\rm vac}\right|c_{\boldsymbol{n}_{2},\A}(z-H^{\text{eff}}_{\text{pg}})^{-1}c_{\boldsymbol{n}_{1},\B}^{\dagger}\left|{\rm vac}\right\rangle
=\,\text{g}^{2}\iint\frac{\dd{\boldsymbol{k}}}{(2\pi)^{2}}\frac{\omega_{\mathrm{aniso}}(\boldsymbol{k})e^{i\phi(\boldsymbol{k})}}{z_{{\rm nh}}^{2}-\omega^{2}_{\mathrm{aniso}}(\boldsymbol{k})}e^{-i\boldsymbol{k}\cdot(\boldsymbol{n}_{1}-\boldsymbol{n}_{2})},\label{S125}
\end{align}
where the definitions of $\Sigma_{e}^{\alpha}(z)$ and $\Sigma_{mn}^{\alpha\beta}(z)$, along with the physical processes they describe, are consistent with those discussed previously. With these useful results, we can, in principle, determine the single- and two-QE dynamics from Eqs.\,(\ref{S29}-\ref{S30}) and (\ref{S45}-\ref{S46}), respectively, thereby revealing a range of interesting anisotropy-dependent phenomena in graphenic QED.
\begin{figure}
  \centering
  \includegraphics[width=18.0cm]{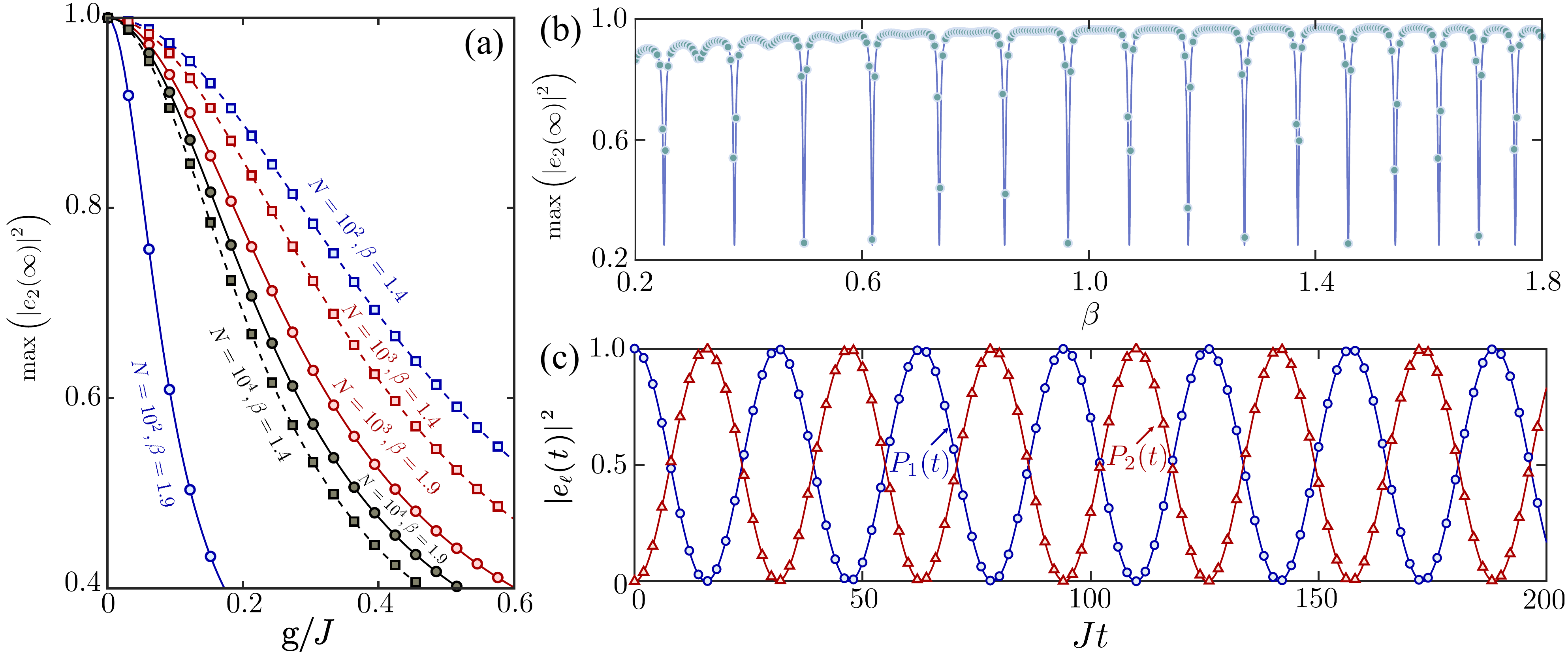}
  \caption{(a) The maximal excitation probability of the acceptor QE, $\max\left(\left|e_{2}(\infty)\right|^{2}\right)$, as a function of the scaled light-matter coupling strength $\text{g}/J$. The excitation is transferred from a resonant donor QE ($\Delta_{e}=0$) to an acceptor QE located in the same cavity of sublattice A. We also show how $\max\left(\left|e_{2}(\infty)\right|^{2}\right)$ depends on $\text{g}$ for different bath sizes, using the following markers: $N=10^{2}$ (blue), $N=10^{3}$ (red), and $N=10^{4}$ (black). Results with different anisotropic parameters are distinguished by solid ($\beta=1.9$) and dashed ($\beta=1.4$) lines. Panel (b) presents the dependence of $\max\left(\left|e_{2}(\infty)\right|^{2}\right)$ on $\beta$ with $N=2^{6}$. (c) Time evolution of the atomic excitation probabilities $P_1(t)=\left|e_{1}(\infty)\right|^{2}$ and $P_2(t)=\left|e_{2}(\infty)\right|^{2}$ for two QEs in anisotropic photonic graphene with coupling strength $\text{g}=0.01J$, lattice size $N=2^6$, and anisotropic parameter $\beta=1.4$. The blue and red lines represent $P_1(t)$ and $P_2(t)$, respectively, where solid lines and markers indicate the numerical and analytical solutions, which show excellent agreement in the long-time limit.
}\label{figS11}
\end{figure}
\subsection{Quasilocalized state in anisotropic graphenic QED}\label{IVB}
As established in subsection \ref{IIIA}, coupling a single QE to an isotropic photonic graphene with single-sublattice dissipation leads to the formation of QLS that robust against dissipation. More concretely, the dynamical population of the QE relaxes to a dissipation-independent constant and oscillates around it in the long-time limit. In this subsection, we aims to address how anisotropy affects the steady-state atomic population and the bound-state wavefunction in anisotropic graphenic QED with single-sublattice dissipation, i.e., $\kappa_{a}\neq 0,\kappa_{b}=0$. To this end, we derive analytical expressions for these long-time asymptotic dynamical properties, specifically for the atomic excitation probability and modified QLS $\ket{\Psi_{\mathrm{QLS}}}$.

In Fig.\,\ref{figS10}(a), we plot the excited-state population $\left|C_{e}^{\A}(t)\right|^{2}$ for an initially excited QE coupled to the center of the dissipative sublattice A with strength $\text{g} = 0.5J$. The graphene lattice has a size of $N=512$, and sublattice A has a dissipation rate of $\kappa_a = 0.01J$. The various curves are obtained by varying the anisotropy parameter $\beta$ from $1.4$ to $2.2$ (see legend). As shown in Fig.\,\ref{figS10}(a), these single-emitter relaxation dynamics remain robust against sublattice dissipation, with the dissipation primarily controlling the rate at which the excitation probability reaches a steady state. For $\beta \in (1, 2)$, the population $|C_{e}^{\text{A}}(t)|^{2}$ undergoes saturation at sufficiently long times and enters a persistent oscillatory regime around a steady-state value, analogous to the isotropic case. This asymptotic value is determined by the overlap $|R_0|^2$ between the QE and the QLS modified by the anisotropy. We formalize this by expressing $R_0$ as a function of both $N$ and $\beta$, i.e., $R_0 \equiv R_0(\beta,N)$, with its explicit form given subsequently.

A key difference from the isotropic case emerges: for a fixed $\beta$, $|R_0|^{2}$ shows no monotonic dependence on the system size $N$ as presented in Fig.\,\ref{figS8}(a), but rather displays pronounced oscillations that strongly depend on the value of $\beta$. A similar oscillatory dependence on $\beta$ is observed at fixed $N$ in Fig.\,\ref{figS10}(b). The steady-state excitation probability $\left|C_{e}^{\text{A}}(\infty)\right|^{2}$ overall decreases as $\beta$ approaches the critical boundaries (0 and 2). Superimposed on this trend, it attains a maximum near $\beta \approx 1.4$ and shows periodic near-zero minima. Moreover, the overlap function vanishes at criticality, with $R_0(\beta=2, N) = 0$ for all lattice sizes. This is attributed to a resonant interaction with a zero-energy extended mode that appears for the specific boundary conditions used. This resonance prevents the formation of a stationary bound state by enabling a complete transfer of the excitation from the emitter into the extended mode [see Fig.\,\ref{figS10}(a)]. When the parameter $\beta$ exceeds 2, the QED system is found to support robust atom-photon bound states. Within this regime, while the excitation population shows an initial decay, it eventually saturates at a steady value that does not depend on system size, demonstrating characteristics analogous to those found in conventional band gaps\,\cite{SMPhysRevA.50.1764}.

We now analyze quantitatively the QLS formed by a single QE coupled to this isotropic lattice. Our setup consists of an initially excited QE in dissipative sublattice A, with its transition frequency resonant at the middle of the photonic band. The projection of the initial state $\ket{\Psi_{t=0}} = \sigma^\dagger \ket{g;\mathrm{vac}}$ onto the bound state $\ket{\Psi_{\mathrm{QLS}}}$ is given by the residue $R_{0}=\braket{\Psi_{t=0}}{\Psi_{\mathrm{QLS}}}$. This quantity dictates the bound state's weight in the steady state and can be determined as follows
\begin{align}\label{S126}
R_{0}=\left.\frac{1}{1-\partial_{z}\Sigma_{e}^{\text{A}}(z)}\right|_{z=i0^{+}}=\frac{1}{1-\left.\left(\frac{\text{g}^{2}}{N^{2}}\sum\limits_{\boldsymbol{k}}\frac{1}{z_{{\rm nh}}^{2}-\omega^{2}_{\mathrm{aniso}}(\boldsymbol{k})}\right)\right|_{z=i0^{+}}}=\frac{1}{1+\left(\frac{\text{g}^{2}}{N^{2}}\sum\limits_{\boldsymbol{k}}\frac{1}{\omega^{2}_{\mathrm{aniso}}(\boldsymbol{k})}\right)}=\frac{1}{1+\frac{\text{g}^{2}}{J^{2}}\mathcal{G}(\beta,N)},
\end{align}
where the factor $\mathcal{G}(\beta,N) = \frac{J^{2}}{N^{2}} \sum_{\boldsymbol{k}} \omega_{{\rm aniso}}^{-2}(\boldsymbol{k})$ implies its dependence on both the lattice size $N$ and the anisotropic parameter $\beta$. Given that $\omega_{{\rm aniso}}^{-2}(\boldsymbol{k})$ diverges near the modified Dirac points $\boldsymbol{\mathrm{K}}_{\pm}=\left(\pm\arccos(-\frac{\beta}{2}),\mp\arccos(-\frac{\beta}{2})\right)$, the asymptotic expansion of the $f_{{\rm aniso}}(\boldsymbol{k})$ in the vicinity of the momentum modes $\boldsymbol{\mathrm{K}}_{\pm}$ is given by $f_{{\rm aniso}}(\boldsymbol{\mathrm{K}}_{\pm}+\boldsymbol{q})=J\boldsymbol{w}_{\pm}\cdot\boldsymbol{q}$ with $\boldsymbol{w}_{\pm}=i\left[\exp\left(\pm\arccos(-\frac{\beta}{2})i\right),\exp\left(\mp\arccos(-\frac{\beta}{2})i\right)\right]$. Therefore, the primary contributions to $\mathcal{G}(\beta,N)$ can be approximated by the momentum modes around $\boldsymbol{\mathrm{K}}_{\pm}$, under which
\begin{align}\label{S127}
\mathcal{G}(\beta,N)\approx &\,\frac{J^{2}}{N^{2}}\sum_{\boldsymbol{q}}\frac{1}{\left|f(\boldsymbol{\mathrm{K}}_{+}+\boldsymbol{q})\right|^{2}}+\frac{J^{2}}{N^{2}}\sum_{\boldsymbol{q}}
\frac{1}{\left|f(\boldsymbol{\mathrm{K}}_{-}+\boldsymbol{q})\right|^{2}},\nonumber\\
=&\,\frac{2J^{2}}{N^{2}}\sum_{\boldsymbol{q}}\frac{1}{\left|J\boldsymbol{w}_{\pm}\cdot\boldsymbol{q}\right|^{2}}=\frac{2J^{2}}{N^{2}}\sum_{\boldsymbol{q}}\frac{1}{\left|Ji\left
[\exp\left(\arccos(-\frac{\beta}{2})i\right)q_{1}+\exp\left(-\arccos(-\frac{\beta}{2})i\right)q_{2}\right]\right|^{2}},\nonumber\\
=&\,\frac{2}{N^{2}}\sum_{\boldsymbol{q}}\frac{1}{q_{1}^{2}+q_{2}^{2}+Z_{\beta}q_{1}q_{2}}=\frac{2}{(2\pi)^{2}}\int_{-\pi}^{\pi}\int_{-\pi}^{\pi}\dd{q}_{1}\dd{q}_{2}\frac{1}{q_{1}^{2}+q_{2}^{2}+Z_{\beta}q_{1}q_{2}},
\end{align}
where $Z_{\beta}=2\cos\left[2\arccos(-\frac{\beta}{2})\right]$ is defined. Through the coordinate transformation $q_{1,2}=\frac{3}{2}(\frac{1}{\sqrt{2+Z_{\beta}}}p_{1}\pm\frac{1}{\sqrt{2-Z_{\beta}}}p_{2})$, it follows that $q_{1}^{2}+q_{2}^{2}+Z_{\beta}q_{1}q_{2}=\frac{9}{4}(p_{1}^{2}+p_{2}^{2})$ and $\dd{q}_{1}\dd{q}_{2}=\frac{9}{2\sqrt{4-Z_{\beta}^{2}}}\dd{p}_{1}\dd{p}_{2}$. This transformation yields an anisotropic dispersion relation $\omega_{\mathrm{aniso}}(\boldsymbol{k})$, thereby leading to a more compact expression for $\mathcal{G}(\beta,N)$, namely
\begin{align}\label{S128}
\mathcal{G}(\beta,N)= &\,\frac{1}{\sqrt{4-Z_{\beta}^{2}}\pi^{2}}\iint_{-\pi}^{\pi}\dd{p}_{1}\dd{p}_{2}\frac{1}{(p_{1}^{2}+p_{2}^{2})}
=\frac{2}{\sqrt{4-Z_{\beta}^{2}}\pi}\int_{p_{{\rm min}}}^{p_{c}}\frac{\dd{p}}{p}=\frac{2}{\sqrt{4-Z_{\beta}^{2}}\pi}\ln p_{c}+\frac{2}{\sqrt{4-Z_{\beta}^{2}}\pi}\ln N,
\end{align}
where the integral is regularized against divergences by introducing two cutoffs: a high-momentum cutoff $p_c$ and a lower cutoff $p_{\mathrm{min}} \approx 1/N$ set by the lattice discreteness. Clearly, the constant $2\ln p_{c}/\sqrt{4-Z_{\beta}^{2}}\pi$ depends on both the maximum cutoff $p_{c}$ and anisotropy parameter $\beta$, and can be determined numerically by exactly evaluating the double summation $\mathcal{G}(\beta,N)$. As shown in Fig.\,\ref{figS10}(a), the steady-state excitation probabilities $\left|R_{0}\right|^{2}$ estimated from Eq.\,(\ref{S126}) using the approximation $\mathcal{G}(\beta,N)$ (dashed lines) agree well with numerical simulations across a wide range of $\beta$, thereby validating our theoretical approach. Note that our analytical and numerical verification of the atomic excitation population $\left|C_{e}^{\text{A}}(\infty)\right|^{2}$ is limited to the regime where the anisotropy parameter satisfies $\beta<2$. Beyond this range, the linear approximation used for $f_{\rm aniso}(\boldsymbol{k})$ is no longer valid.

We commence with an examination of the explicit form for the QLS $\ket{\Psi_{\mathrm{QLS}}}$, which can be obtained by solving its eigenequation $H_{\mathrm{eff}}\ket{\Psi_{\mathrm{BS}}} = E_{\mathrm{BS}}\ket{\Psi_{\mathrm{BS}}}$, where the effective Hamiltonian under the single-sublattice dissipation configuration is
\begin{align}\label{S129}
H_{{\rm eff}}=&\,\Delta_{e}\sigma^{\dagger}\sigma+\text{g}(\sigma^{\dagger}a_{n_{e},n_{e}}+\sigma a^{\dagger}_{n_{e},n_{e}})-\frac{\kappa_a}{2}i\sum_{n_{x},n_{y}}a_{n_{x},n_{y}}^{\dagger}a_{n_{x},n_{y}}+J\sum_{n_{x},n_{y}}(\beta a_{n_{x},n_{y}}^{\dagger}b_{n_{x},n_{y}}+a_{n_{x},n_{y}}^{\dagger}b_{n_{x}+1,n_{y}}\nonumber\\
&\,+a_{n_{x},n_{y}}^{\dagger}b_{n_{x},n_{y}+1}
+\beta b_{n_{x},n_{y}}^{\dagger}a_{n_{x},n_{y}}+b_{n_{x}+1,n_{y}}^{\dagger}a_{n_{x},n_{y}}+b_{n_{x},n_{y}+1}^{\dagger}a_{n_{x},n_{y}}),
\end{align}
and the general ansatz for the bound state in the single-excitation manifold is given by
\begin{align}\label{S130}
\ket{\Psi_{\mathrm{BS}}}=\left[C_{e}\sigma^{\dagger}+\sum_{n_{x},n_{y}}\left(C^{\A}_{n_{x},n_{y}}a_{n_{x},n_{y}}^{\dagger}+C^{\B}_{n_{x},n_{y}}b_{n_{x},n_{y}}^{\dagger}\right)\right]\ket{g,\mathrm{vac}}.
\end{align}
Substituting Eqs.\,(\ref{S129}) and (\ref{S130}) into the Schr\"{o}dinger equation, $H_{\text{eff}}\ket{\Psi_{\mathrm{BS}}} = E_{\mathrm{BS}}\ket{\Psi_{\mathrm{BS}}}$, yields the following coupled equations
\begin{align}
\Delta_{e}C_{e}+\text{g}C^{\A}_{n_{e},n_{e}}=& \,E_{\mathrm{BS}} C_{e},\label{S131}\\
J(\beta C^{\B}_{n_{x},n_{y}}+C^{\B}_{n_{x}+1,n_{y}}+C^{\B}_{n_{x},n_{y}+1})-\frac{\kappa_a}{2}iC^{\A}_{n_{x},n_{y}}+\text{g}C_{e}\delta_{n_{x},0}\delta_{n_{y},0}=& \,E_{\mathrm{BS}} C^{\B}_{n_{x},n_{y}},\label{S132}\\
J(\beta C^{\A}_{n_{x},n_{y}}+C^{\A}_{n_{x}-1,n_{y}}+C^{\A}_{n_{x},n_{y}-1})=& \,E_{\mathrm{BS}} C^{\A}_{n_{x},n_{y}}.\label{S133}
\end{align}
Setting $\Delta_e = E_{\mathrm{BS}} = 0$ forces $C^{\A}_{n_e, n_e} = 0$ via Eq.\,(\ref{S131}), thereby causing the entire photon field on sublattice A to vanish, $C^{\A}_{n_x, n_y} = 0$, in accordance with the homogeneous difference equation Eq.\,(\ref{S133}). The amplitudes $C^{\B}_{n_{x}, n_{y}}$ are then determined by solving the sourced difference equation $J(\beta C^{\B}_{n_{x},n_{y}} + C^{\B}_{n_{x}+1,n_{y}} + C^{\B}_{n_{x},n_{y}+1}) = -\text{g}C_{e}\delta_{n_{x},0}\delta_{n_{y},0}$, which in the thermodynamic limit reads
\begin{align}\label{S134}
C^{\B}_{n_{x}, n_{y}}=-\frac{\text{g}C_{e}}{J}\int_{-\pi}^{\pi}\frac{\dd{k_{x}}}{2\pi}\int_{-\pi}^{\pi}\frac{\dd{k_{y}}}{2\pi}\frac{e^{i(k_{x}n_{x}+k_{y}n_{y})}}{\beta+e^{ik_{x}}+e^{ik_{y}}}.
\end{align}
 In order to compute the residue integral in Eq.\,(\ref{S134}), we first integrate over the variable $k_{y}$. Whether the corresponding pole $z_{0} = -(\beta + e^{ik_{x}})$ lies inside or outside the unit circle is determined exclusively by the anisotropic parameter $\beta$. Specifically, when $|\beta|>2$, the pole $z_{0}$ always lies outside the unit circle. For $\beta \in (0,2)$, the pole lies inside the unit circle when $k_{x} \in \left(-\pi, -\arccos\left(-\frac{\beta_{1}}{2}\right)\right) \cup \left(\arccos\left(-\frac{\beta}{2}\right),\pi\right)$, and outside for all other intervals of $k_{x}$. For $\beta \in (-2,0)$, the situation is reversed: the pole lies outside the unit circle when $k_{x} \in \left(-\pi, -\arccos\left(-\frac{\beta}{2}\right)\right) \cup \left(\arccos\left(-\frac{\beta}{2}\right),\pi\right)$, and inside for other intervals of $k_{x}$. Using contour integration, the integral in Eq.\,\eqref{S134} for the parameter range $\beta \in (0,2)$ can be evaluated analytically as follows
\begin{align}
C^{\B}_{n_{x}, n_{y}}=&-\frac{\text{g}C_{e}}{J}\int_{|k_{x}|\ge k_{c}}\frac{\dd{k}_{x}}{2\pi}e^{in_{x}k_{x}}(-\beta-e^{ik_{x}})^{n_{y}-1}\mathcal{H}(n_{y}-1)+\frac{\text{g}C_{e}}{J}
\int_{|k_{x}|<k_{c}}\frac{\dd{k}_{x}}{2\pi}e^{in_{x}k_{x}}(-\beta-e^{ik_{x}})^{n_{y}-1}\mathcal{H}(-n_{y})\nonumber\\
=&-\frac{\text{g}C_{e}}{J}\int_{k_{c}}^{\pi}\frac{\dd{k}_{x}}{2\pi}\left(e^{in_{x}k_{x}}S_{k_{x}}^{n_{y}-1}+\mathrm{c.c.}\right)\mathcal{H}(n_{y}-1)
+\frac{\text{g}C_{e}}{J}\int_{-k_{c}}^{k_{c}}\frac{\dd{k}_{x}}{2\pi}e^{in_{x}k_{x}}S_{k_{x}}^{n_{y}-1}\mathcal{H}(-n_{y})\nonumber\\
=&\sum_{j=0}^{n_{y}-1}C_{e}C_{n_{y}-1}^{j}[\mathcal{K}(-n_{x}-j)+\mathcal{K}(j+n_{x})]\mathcal{H}(n_{y}-1)
+\frac{\text{g}C_{e}}{J}\int_{-k_{c}}^{k_{c}}\frac{\dd{k}_{x}}{2\pi}e^{in_{x}k_{x}}S_{k_{x}}^{n_{y}-1}\mathcal{H}(-n_{y}),\label{S135}
\end{align}
where $S_{k_{x}} \equiv -\beta - e^{ik_{x}}$, and $\mathcal{K}(q)$ is defined as $-(-1)^{n_{y}-1}\text{g}(\pi-k_{c})/2\pi J$ for $q=0$ and $-(-1)^{n_{y}-1}\text{g}\left(e^{iq\pi} - e^{iqk_{c}}\right)/2\pi qJi$ otherwise. For $\beta>2$, the amplitudes $C^{\B}_{n_{x}, n_{y}}$ can be calculated as
\begin{align}
C^{\B}_{n_{x}, n_{y}}=&-\frac{\text{g}C_{e}}{J}\int_{-\pi}^{\pi}\frac{\dd{k}_{x}}{2\pi}(-1)^{n_{y}}(\beta+e^{ik_{x}})^{n_{y}-1}e^{in_{x}k_{x}}\Theta(-n_{y}+0^{+})\nonumber\\
=&-\frac{\text{g}C_{e}}{J}(-1)^{n_{y}}\Theta(-n_{y}+0^{+})\times\left\{ \begin{array}{cc}
C_{n_{y}-1}^{-n_{x}}\beta^{n_{x}+n_{y}-1}, & {\rm if}\,\,\,n_{x}\le0\\
0, & {\rm else}
\end{array}\right.,\label{S136}
\end{align}
where $C_{n_{y}-1}^{-n_{x}}$ is the generalized binomial coefficient.

We have thus fully determined the QLS $\ket{\Psi_{\mathrm{QLS}}}$ for a single QE coupled to an anisotropic bath, with its explicit expression given by $\left(R_{0}\sigma^{\dagger} + \sum_{n_{x},n_{y}}C^{\B}_{n_{x},n_{y}}b_{n_{x},n_{y}}^{\dagger}\right)\ket{g;\mathrm{vac}}$. Figures \ref{figS10}(c)-(f) depict the evolution of the spatial profiles $\left|C^{\B}_{n_{x},n_{y}}\right|$ for different values of the anisotropic parameter $\beta$ (see legend), as calculated from the analytical solutions in Eqs.\,(\ref{S135}-\ref{S136}). Based on these results, we summarize the following key findings: First, for anisotropy strengths in the range $\beta \in (1,2)$, the atom‑photon bound state shows a power‑law decay around the lattice site coupled to the QE, and its spatial extent broadens as $\beta$ increases. Second, when $\beta \ge 2$, the bound state exhibits directional behavior, becoming strongly localized along a specific lattice direction, as clearly described by Eq.\,\eqref{S136}. Third, at $\beta = 2$, the envelope of $\left|C^{\B}_{n{x},n_{y}}\right|$ extends over a significantly longer range compared to other values of $\beta$.

\subsection{Dissipation-free interactions between two quantum emitters coupled to an anisotropic photonic graphene}\label{IVC}
In this subsection, we systematically investigate and describe in detail the quantum dynamics of two QEs coupled to an anisotropic photonic graphene with single‑sublattice dissipation. Recall that in Subsection \ref{IIIC}, we have identified an atomic interaction in an isotropic photonic graphene which remains immune to dissipative environments, where the coherent exciton transfer between QEs is achieved through the combined participation of a QLS and a dark state. The main goal of this subsection is to examine how the inclusion of anisotropy modifies the QLS and, consequently, the corresponding dissipation‑free atomic interactions.

Below we describe the dynamics in which a donor QE, initially excited and coupled to a specific site of the anisotropic lattice, transfers its excitation to an acceptor QE coupled to the same cavity. We continue to assume a direct coherent coupling of strength $\Omega$ between the QEs. Since the detailed computational procedures have been presented in Subsection \ref{IIIC}, we omit redundant details here and directly show the time evolution of the probability amplitudes for the donor and acceptor QEs, obtaining
\begin{align}
e_{1}(t)=&\,\langle s_{1}|\mathcal{P}U(t)\mathcal{P}|s_{1}\rangle=\frac{1}{2\pi i}\int_{\mathcal{C}}\frac{z-\Delta_{e}-\Sigma_{22}^{\A\A}(z)}{\mathscr{D}_{\text{A}}(z)}e^{-izt}\dd{z}\nonumber\\
=&\,\frac{1}{2\pi i}\int_{\mathcal{C}}\frac{z-\Delta_{e}-\Sigma_{22}^{\A\A}(z)}{[z-\Delta_{e}-\Sigma_{11}^{\A\A}(z)][z-\Delta_{e}-\Sigma_{22}^{\A\A}(z)]-[\Omega+\Sigma_{12}^{\A\A}(z)][\Omega+\Sigma_{21}^{\A\A}(z)]}e^{-izt}\dd{z}\nonumber\\
=&\,\frac{1}{2\pi i}\int_{\mathcal{C}}\frac{z-\Delta_{e}-\Sigma_{11}^{\A\A}(z)}{[z-\Delta_{e}-\Sigma_{11}^{\A\A}(z)]^{2}-[\Omega+\Sigma_{11}^{\A\A}(z)]^{2}}e^{-izt}\dd{z}\nonumber\\
=&\,\frac{1}{2}e^{2i\Omega t}+\frac{1}{2}[\frac{1}{1+2\text{g}^{2}\mathcal{G}(\beta,N)/J^{2}}],\label{S137}\\
e_{2}(t)=&\,\langle s_{2}|\mathcal{P}U(t)\mathcal{P}|s_{1}\rangle=\frac{1}{2\pi i}\int_{\mathcal{C}}\frac{\Omega+\Sigma_{12}^{\A\A}(z)}{\mathscr{D}_{\text{A}}(z)}e^{-izt}\dd{z}\nonumber\\
=&\,\frac{1}{2\pi i}\int_{\mathcal{C}}\frac{\Omega+\Sigma_{12}^{\A\A}(z)}{[z-\Delta_{e}-\Sigma_{11}^{\A\A}(z)][z-\Delta_{e}-\Sigma_{22}^{\A\A}(z)]-[\Omega+\Sigma_{12}^{\A\A}(z)][\Omega+\Sigma_{21}^{\A\A}(z)]}e^{-izt}\dd{z}\nonumber\\
=&\,\frac{1}{2\pi i}\int_{\mathcal{C}}\frac{\Omega+\Sigma_{11}^{\A\A}(z)}{[z-\Delta_{e}-\Sigma_{11}^{\A\A}(z)]^{2}-[\Omega+\Sigma_{11}^{\A\A}(z)]^{2}}e^{-izt}\dd{z}\nonumber\\
=&\,-\frac{1}{2}e^{2i\Omega t}+\frac{1}{2}[\frac{1}{1+2\text{g}^{2}\mathcal{G}(\beta,N)/J^{2}}],\label{S138}
\end{align}
where the self-energies $\Sigma_{mn}^{\alpha\beta}(z)$ are defined by Eqs.\,(\ref{S118}-\ref{S125}) and the residue $\mathcal{G}(\beta,N)$ is given by Eq.\,(\ref{S128}). Evidently, a decoherence‑free interaction, mediated by a modified QLS and a dark state, persists between the two QEs. While this mechanism resembles the isotropic case, the energy‑transfer efficiency now additionally depends on the anisotropy strength.

In Fig.\,\ref{figS11}(a), we plot the maximal steady-state excitation probability of the acceptor QE, $\max\left(\left|e_{2}(\infty)\right|^{2}\right)$, as a function of the light-matter coupling strength $\text{g}$, based on Eqs.\,(\ref{S137}) and (\ref{S138}). The solid and dashed lines in the simulation correspond to anisotropy parameters $\beta = 1.9$ and $\beta = 1.4$, respectively. The results imply the joint dependence of the probability both on the lattice size $N$ and the anisotropy parameter $\beta$. In previous discussions of isotropic scenario, we have pointed out that $\max\left(\left|e_{2}(\infty)\right|^{2}\right)$ consistently follows the trend that a larger lattice size leads to a smaller steady-state acceptor population at a fixed coupling strength. However, this trend is now broken with the introduction of anisotropy, as presented in Fig.\,\ref{figS11}(a). Furthermore, for a fixed lattice size $N = 64$, $\max\left(\left|e_{2}(\infty)\right|^{2}\right)$ exhibits a non-monotonic dependence on $\beta$. Specifically, it shows intermittent minima at certain specific values of $\beta$, followed by plateaus of high acceptor population over certain ranges of $\beta$, repeating periodically, as shown in Fig.\,\ref{figS11}(b). For illustration, in Fig.,\ref{figS11}(c) we select $\beta = 1.4$ to simulate the two‑QE dynamics, observing nearly perfect dissipation‑free energy transfer in the weak‑coupling regime ($g = 0.01J$). Therefore, achieving perfect excitation transfer between the two QEs in an anisotropic photonic graphene requires tuning the anisotropy parameter to a specific, well‑chosen range rather than an arbitrary one. The decoherence‑free interaction is mediated by the atomic dark state and an anisotropy-controlled QLS. The corresponding real‑space wavefunctions for these bound‑state are derived analogously to Subsection \ref{IIIC} and are omitted here for brevity.
\begin{figure}
  \centering
  \includegraphics[width=16.0cm]{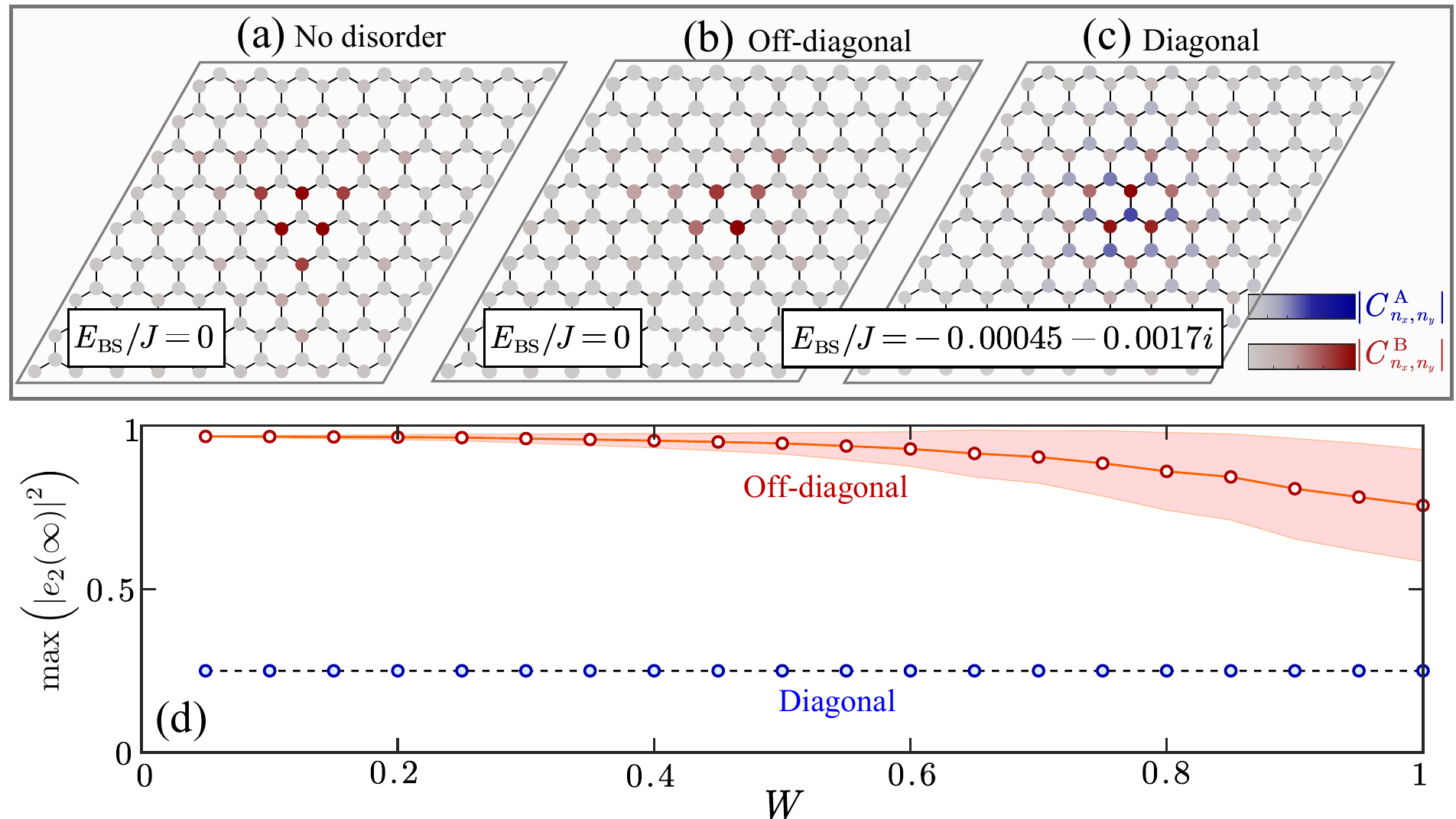}
  \caption{Panels (a-c) illustrate the properties of the bound states in the absence and presence of disorder. The probability amplitudes $\lvert C^{\text{A}}_{n_{x},n_{y}} \rvert$ and $\lvert C^{\text{B}}_{n_{x},n_{y}} \rvert$ are represented by blue and red color bars, respectively. Panel (a) corresponds to the spatial wavefunction profile without disorder, panel (b) to the case with off-diagonal disorder in inter-cavity couplings, and panel (c) to the case with diagonal disorder in cavity resonant frequencies. For both disorder types, the strength is set to $W = 0.5$. The energies of these bound states, e.g., $E_{\mathrm{BS}} = 0$, are shown below each panel. Panel (d) shows the maximum transferred population $\text{max}\left(\left|e_{2}(\infty)\right|^{2}\right)$ of the acceptor QE as a function of disorder strength $W$ for two types of disorder. The red (blue) dots represent the average over $10^3$ disorder realizations for off-diagonal (diagonal) disorder, and the shaded regions indicate the corresponding standard deviation. Throughout all panels, the implemented parameters are fixed at $\text{g} = 0.1J$, $\Delta_{e} = -\Omega = 0.1J$, $\kappa_a = 10J$, and $\kappa_b = 0$.
}\label{figS12}
\end{figure}
%
\section{Robustness of Dissipation-Free Interactions against Disorder}\label{V}
\renewcommand\theequation{S\arabic{equation}}
\makeatletter
\renewcommand{\thefigure}{S\@arabic\c@figure}
\makeatother
The inherent presence of disorder in practical systems significantly influences the efficiency of excitation transfer between QEs. This section investigates the consequences of disorder within a photonic graphene, focusing specifically on its manifestations and their impact on the modified QLS and therefore the decoherence-free interactions.

We focus on two primary types of disorder: one affecting the cavities' resonant frequencies (diagonal) and the other affecting the tunneling amplitudes between them (off-diagonal). The diagonal disorder introduces random diagonal terms into the bath's Hamiltonian, modifying it as $H_{\text{pg}}\to H_{\text{pg}}+\sum_{\boldsymbol{n}}(\epsilon_{a,\boldsymbol{n}}a_{\boldsymbol{n}}^{\dagger} a_{\boldsymbol{n}}^{}+\epsilon_{b,\boldsymbol{n}}b_{\boldsymbol{n}}^{\dagger} b_{\boldsymbol{n}}^{})$, thereby breaking the chiral symmetry of the original model. In contrast, the off-diagonal disorder introduces random off-diagonal terms via $H_{\text{pg}}\to H_{\text{pg}}+\sum_{\boldsymbol{n}}(\epsilon_{1,\boldsymbol{n}}a_{\boldsymbol{n}}^{\dagger} b_{\boldsymbol{n}}^{}+\epsilon_{2,\boldsymbol{n}}b_{\boldsymbol{n}}^{\dagger} a_{\boldsymbol{n}+\boldsymbol{v}_{1}}^{}+\epsilon_{3,\boldsymbol{n}}b_{\boldsymbol{n}}^{\dagger} a_{\boldsymbol{n}+\boldsymbol{v}_{2}}^{}+\text{H.c.})$, which preserves the chiral symmetry. The disorder parameters $\epsilon_{\nu,\boldsymbol{n}}/J\ (\nu=a,b,1,2,3)$ are sampled from a uniform distribution over $[-W, W]$ for each $\boldsymbol{n}$th unit cell, where $W$ denotes the disorder strength. Our analysis is restricted to the single-sublattice dissipation configuration and focuses on the maximum transferred polulation denoted by $\max\left(\left|e_{2}(\infty)\right|^{2}\right)$. As established in Sec.~\ref{IIIC} and Sec.~\ref{IVC}, nearly perfect excitation transfer from the donor QE to the acceptor QE can occur even with environmental dissipation and anisotropy, respectively, mediated primarily by the dark state and the QLS. Note that the dark state remains completely unaffected by either disorder type because it is inherently decoupled from the environment. Therefore, any disorder-induced changes in the energy‑transfer efficiency arise solely from modifications of the QLS.

In the first row of Fig.\,\ref{figS12}, we compare the spatial profiles of typical bound states for three cases: without disorder, with diagonal disorder, and with off-diagonal disorder. The A and B sublattices are colored blue and red, respectively, according to the amplitude of the photonic wave function $\left|C_{\boldsymbol{n}}^{\text{A/B}}(\infty)\right|$. The QLS appears only in the absence of disorder or with off-diagonal disorder, whereas it is absent under diagonal disorder. For symmetry-preserving cases, i.e., without disorder or with off-diagonal disorder, as shown in Figs.\,\ref{figS12}(a) and (b), the bound states exhibit a spatial profile localized around the cavity that is coupled to the QEs, with components exclusively on sublattice B. Compared to the clean system, the bound-state energy remains at zero $E_{\mathrm{BS}} = 0$ in the presence of off‑diagonal disorder, confirming that its existence is protected by chiral symmetry, even though the amplitude of its sublattice‑B component shows slight variations. Moreover, under symmetry-breaking case when the diagonal disorder is introduced [see Fig.\,\ref{figS12}(c)], the bound state acquires non-zero photonic components on both sublattices. Its energy now becomes complex and acquires a nonzero imaginary part, indicating that chiral‑symmetry breaking disrupts the QLS. In Fig.\,\ref{figS12}(d), we plot the functional dependence of the maximum transferred population $\max\left(\left|e_{2}(\infty)\right|^{2}\right)$ on the varied strength of the two kinds of disorder. On the one hand, we demonstrate that the preservation of chiral symmetry guarantees a high energy-transfer efficiency even in disordered settings, facilitated by the synergistic action of the atomic dark state and the QLS. That is to say, the decoherence-free interactions exhibit strong robustness to off-diagonal disorder. Importantly, as the strength $W$ of off‑diagonal disorder increases, the average value of $\max\left(\left|e_{2}(\infty)\right|^{2}\right)$ follows a power-law decay, indicated by the orange line in Fig.\,\ref{figS12}(d). In the other hand, once chiral symmetry is broken and the QLS disappears, the maximum excitation transfer is strictly reduced to 1/4 for any strength of diagonal  disorder.

\begin{figure}
  \centering
  \includegraphics[width=18.0cm]{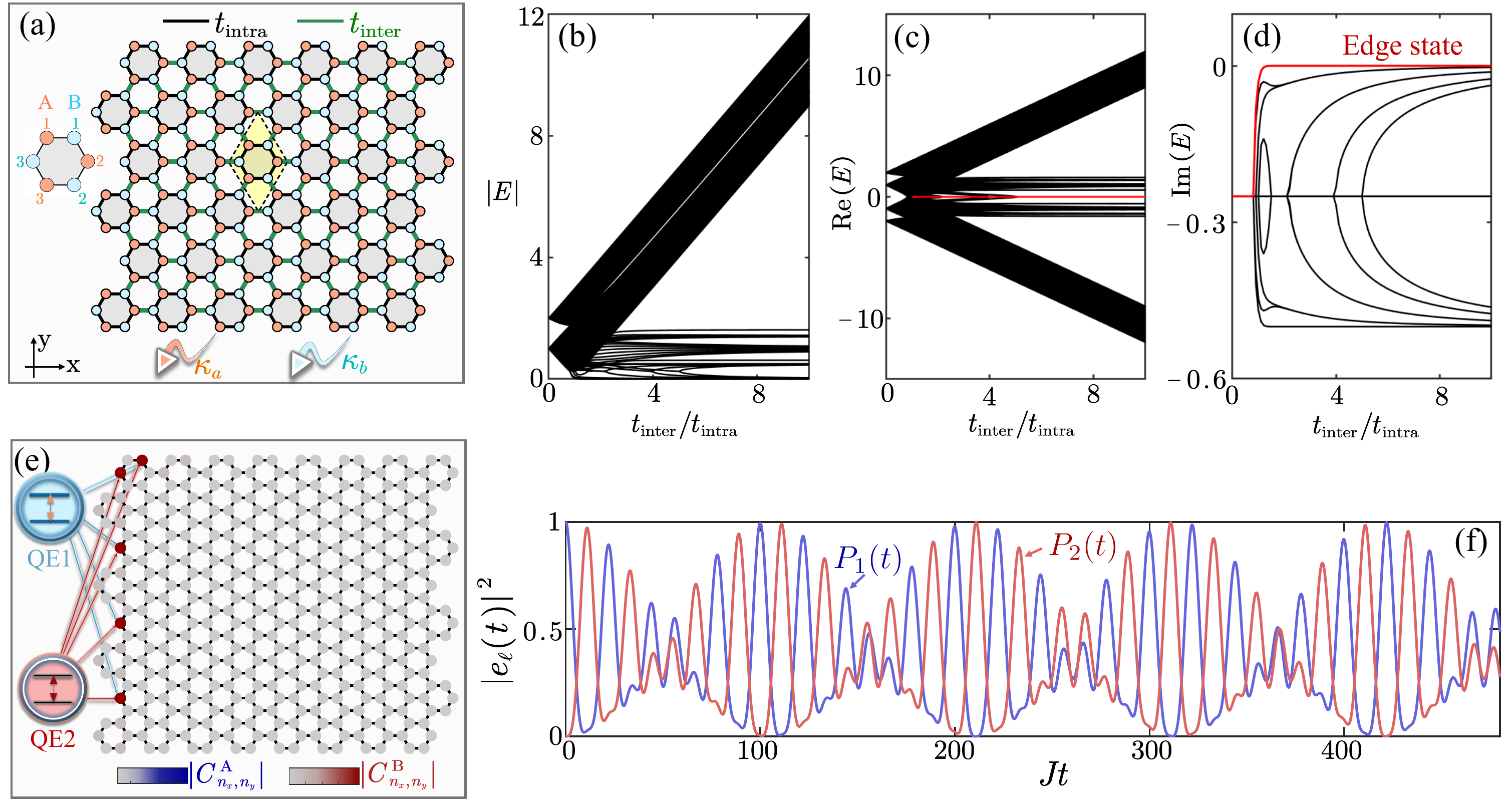}
  \caption{(a) Schematic of the $6\times 6$ photonic graphene that is characterized by two types of hopping parameters:  intracell hopping $t_{\text{intra}}$ and intercell hopping $t_{\text{inter}}$ similar to the Su-Schrieffer-Heeger model. Each unit cell (outlined by a gray hexagon) contains six lattice sites. The indices of the A and B sublattices within a unit cell are labeled in the left. Sublattice A and B sites are represented by orange and blue dots, respectively, and experience external dissipation with rates $\kappa_a$ and $\kappa_b$. We call the edge along the y direction zigzag, while that along the x direction armchair. For a lattice size $L=10$ with dissipations $\kappa_{a}=J$ and $\kappa_{b}=0$, panels (b-d) display the spectral magnitude $|E|$, real component $\text{real}(E)$, and imaginary component $\text{imag}(E)$ versus the coupling ratio $t_{\mathrm{inter}}/t_{\mathrm{intra}}$. The red curves highlighted in (c) and (d) correspond to the edge state. In panel (e), we display the spatial profiles of $\lvert C^{\mathrm{A}}_{n{x},n_{y}} \rvert$ and $\lvert C^{\mathrm{B}}_{n{x},n_{y}} \rvert$ for this edge state, with lattice size $L=8$ and ratio $t_{\mathrm{inter}}/t_{\mathrm{intra}}=15$. The amplitudes are color‑coded in blue and red, respectively. Panel (f) plots the time evolution of $P_{1}(t) = |e_{1}(t)|^{2}$ and $P_{2}(t) = |e_{2}(t)|^{2}$ for a pair of giant atoms that are coupled to this edge state, where the performed parameters are $L=8, \Delta_{e} = 0$, $\Omega = 0.02J$, $t_{\mathrm{inter}} = 1.5J$, $t_{\mathrm{intra}} = 0.1J$, $\kappa_a = J$, and $\text{g} = 0.2J$.
}\label{figS13}
\end{figure}
%

\section{Dissipation-Free Interactions in Topological Platforms with Mixed Boundaries}\label{VI}
\renewcommand\theequation{S\arabic{equation}}
\makeatletter
\renewcommand{\thefigure}{S\@arabic\c@figure}
\makeatother
In the preceding sections, we have explored efficient energy transfer between QEs that remains immune to single-sublattice dissipation, within both isotropic and anisotropic photonic graphene. This exotic transfer dynamics is mediated exclusively by the coexistence of atomic dark states and QLS. In this section, we investigate an alternative dissipation-immune mechanism by leveraging edge states or corner states in a modified photonic graphene with specific boundary conditions to achieve this goal. To maintain generality, we continue to regard sublattice A as the dissipative sublattice and sublattice B as the non-dissipative one, i.e., $\kappa_{a}\neq 0,\kappa_{b}=0$.

We begin by investigating a protocol that utilizes edge states, whose band properties are controlled by precise engineering of the graphene geometry. Inheriting a structural analogy to the Su-Schrieffer-Heeger model, the modified photonic graphene considered here features a Kekulé‑type hopping texture, as shown in Fig.\,\ref{figS13}(a)\,\cite{SMPhysRevLett.122.086804,SMPhysRevLett.125.255502}. A unit cell now contains six lattice sites [gray hexagon in Fig.\,\ref{figS13}(a)]. The intra-cell coupling strength, denoted by $t_{\mathrm{intra}}$ (black solid lines), is uniform among all sites within the unit cell, while the inter-cell coupling is represented by $t_{\mathrm{inter}}$ (green solid lines). Note that we consider a $L \times L$ photonic graphene with mixed boundary conditions: armchair edge along the $x$-axis and zigzag edge along the $y$-axis, where the sublattices A and B are represented by orange and blue dots, respectively. Figures \ref{figS13}(b)–(d) show the spectrum of the honeycomb lattice plotted against the hopping ratio $t_{\mathrm{inter}}/t_{\mathrm{intra}}$, displaying the magnitude $|E|$, real part $\Re(E)$, and imaginary part $\Im(E)$. The real part of the energy spectrum, $\Re(E)$, is symmetric about the zero energy, while its imaginary part, $\Im(E)$, is symmetric with respect to $-\kappa/4$. More importantly, when $t_{\mathrm{inter}} > t_{\mathrm{intra}}$, this model supports a dissipation-free edge state at zero energy, characterized by a strictly vanishing imaginary part, as indicated by the red curve in Fig.\,\ref{figS13}(d). Furthermore, we find that this edge state is localized only at a few lattice sites belonging to the dissipationless sublattice B, as illustrated in Fig.\,\ref{figS13}(e).

\begin{figure}
  \centering
  \includegraphics[width=16.0cm]{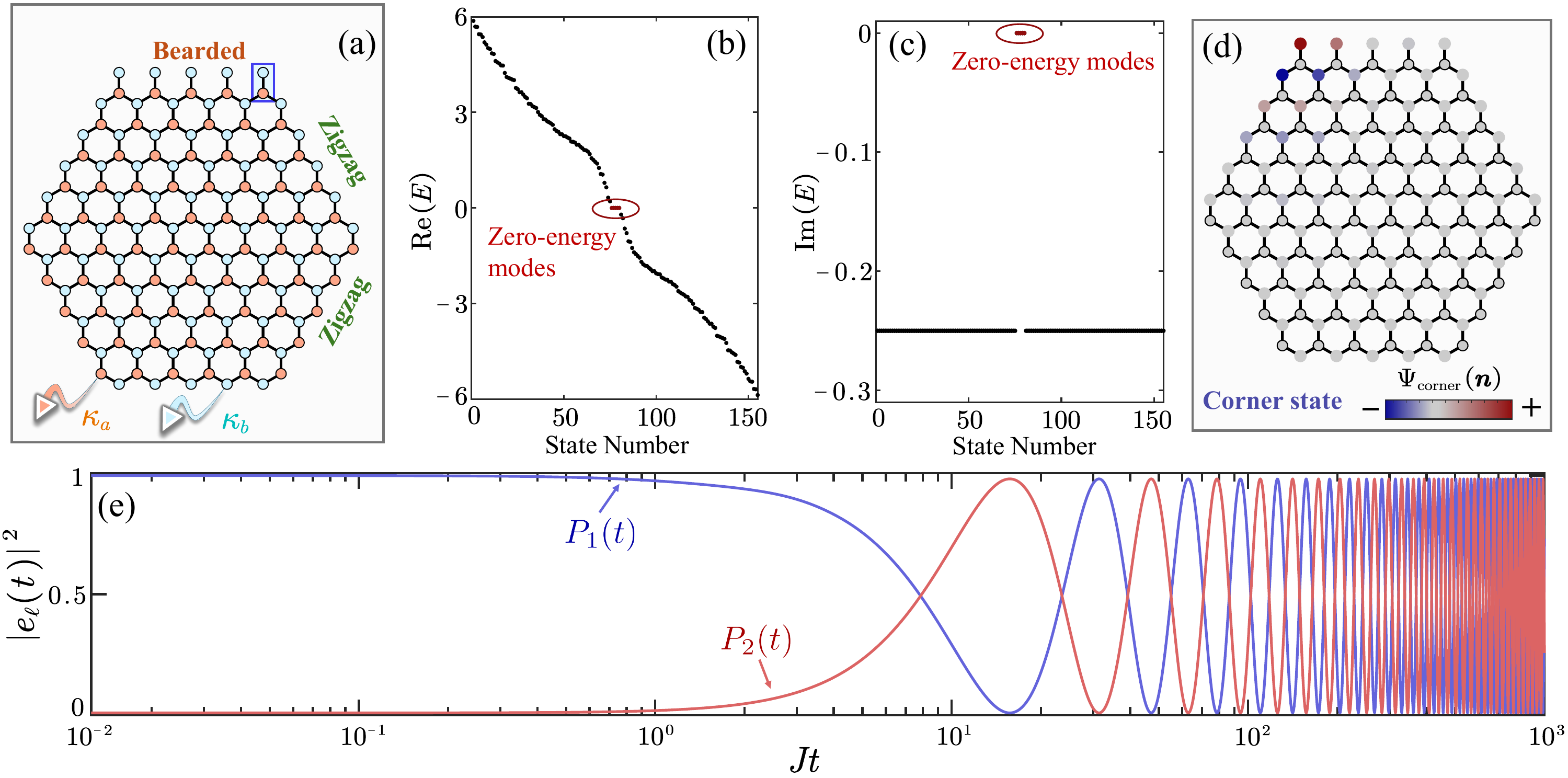}
  \caption{(a) Schematic of the photonic graphene with zigzag-bearded boundaries, where the boundary type of each edge is illustrated. Each unit cell contains two lattice sites, enclosed in a blue rectangular box. In this lattice, the couplings between all nearest-neighbor sites, including both intracell hopping and intercell hopping, are characterized by a uniform strength $J$. The sites of sublattice A and B are marked by orange and blue dots, respectively, and are subject to external dissipation with strengths $\kappa_a$ and $\kappa_b$.  For a finite lattice in panel (a) with dissipations $\kappa_{a}=J$ and $\kappa_{b}=0$, panels (b) and (c) present the corresponding eigenvalue spectrum, including the real $\Re(E)$ and imaginary $\Im(E)$ parts. The red dots highlighted in (b-c) correspond to the zero-energy states. (d) Calculated spatial profile of the corner state $\Psi _{\mathrm{corner}}\left( \boldsymbol{n} \right)$, color-coded by the sign of the wavefunction. Panel (e) plots the time evolution of $P_{1}(t) = |e_{1}(t)|^{2}$ and $P_{2}(t) = |e_{2}(t)|^{2}$ for a pair of QEs coupled to a same cavity belonging to sublattice A, with the implemented parameters $\Delta_{e} = 0.1J$, $\Omega = -0.1J$, $\kappa_a = J$, and $\text{g} = 0.1J$.
}\label{figS14}
\end{figure}

A key prerequisite for these edge states is their complete localization on the dissipation-free sublattice B. Their field distribution is confined to a finite set of cavity modes at spatial positions $\boldsymbol{n}_{1},\boldsymbol{n}_{2}, \ldots, \boldsymbol{n}_{M}$, where the integer $M$ quantifies the total number of cavity modes that are predominantly occupied by the edge state. The decoherence‑free interaction is physically realized by coupling both QEs simultaneously to the edge mode of the photonic graphene. In order to enable a single QE to couple to multiple spatial locations simultaneously, the proposed scheme necessitates the use of superconducting giant artificial atoms\,\cite{SMAndersson2015,SMKannan2020}, which inherently exhibit spatial extensibility. The dynamics of the two giant atoms coupled to the edge state is described by the following effective Hamiltonian:
\begin{align}\label{S139}
H_{{\rm eff}}=&\,\Delta_{e}(\sigma_{1}^{\dagger}\sigma_{1}+\sigma_{2}^{\dagger}\sigma_{2})+\left[(\sigma_{1}^{\dagger}+\sigma_{2}^{\dagger})\sum\limits_{m=1}^{M}\text{g}_{m}a_{\boldsymbol{n}_{m}}+\text{H.c.}\right] -\frac{\kappa_a}{2}i\sum_{n_{x},n_{y}}(a_{n_{x},n_{y}}^{(1),\dagger}a^{(1)}_{n_{x},n_{y}}+a_{n_{x},n_{y}}^{(2),\dagger}a^{(2)}_{n_{x},n_{y}}+a_{n_{x},n_{y}}^{(3),\dagger}a^{(3)}_{n_{x},n_{y}})\nonumber\\
&+t_{\mathrm{intra}}\sum_{n_{x},n_{y}}(a_{n_{x},n_{y}}^{(1),\dagger}b^{(1)}_{n_{x},n_{y}}+a_{n_{x},n_{y}}^{(1),\dagger}b^{(2)}_{n_{x},n_{y}}+a_{n_{x},n_{y}}^{(2),\dagger}b^{(2)}_{n_{x},n_{y}}+
a_{n_{x},n_{y}}^{(2),\dagger}b^{(3)}_{n_{x},n_{y}}+a_{n_{x},n_{y}}^{(3),\dagger}b^{(3)}_{n_{x},n_{y}}+a_{n_{x},n_{y}}^{(3),\dagger}b^{(1)}_{n_{x},n_{y}}+\text{H.c.})\nonumber\\
&+t_{\mathrm{inter}}\sum_{n_{x}\in\text{odd},n_{y}}(a_{n_{x},n_{y}}^{(2),\dagger}b^{(3)}_{n_{x}+1,n_{y}}+b_{n_{x},n_{y}}^{(2),\dagger}a^{(1)}_{n_{x}+1,n_{y}-1}+a_{n_{x},n_{y}}^{(3),\dagger}b^{(1)}_{n_{x},n_{y}-1}+
b_{n_{x},n_{y}}^{(3),\dagger}a^{(2)}_{n_{x}-1,n_{y}}+a_{n_{x},n_{y}}^{(1),\dagger}b^{(2)}_{n_{x},n_{y}+1}\nonumber\\
&+b_{n_{x},n_{y}}^{(1),\dagger}a^{(3)}_{n_{x}+1,n_{y}+1}+\text{H.c.})
+t_{\mathrm{inter}}\sum_{n_{x}\in\text{even},n_{y}}(a_{n_{x},n_{y}}^{(2),\dagger}b^{(3)}_{n_{x}+1,n_{y}}+b_{n_{x},n_{y}}^{(2),\dagger}a^{(1)}_{n_{x},n_{y}-1}+a_{n_{x},n_{y}}^{(3),\dagger}b^{(1)}_{n_{x}-1,n_{y}-1}\nonumber\\
&+b_{n_{x},n_{y}}^{(3),\dagger}a^{(2)}_{n_{x}-1,n_{y}}+a_{n_{x},n_{y}}^{(1),\dagger}b^{(2)}_{n_{x}-1,n_{y}+1}
+b_{n_{x},n_{y}}^{(1),\dagger}a^{(3)}_{n_{x},n_{y}+1}+\text{H.c.}),
\end{align}
where $a_{n_{x},n_{y}}^{(p),\dagger}\equiv( a_{n_{x},n_{y}}^{(p)})^{\dagger}$ and $b_{n_{x},n_{y}}^{(p),\dagger}\equiv( b_{n_{x},n_{y}}^{(p)})^{\dagger}$ are the creation operators of the cavity modes on sites A and B located at $(n_{x},n_{y})$, respectively, with $p$ indexing the bosonic modes whthin the unit cell [see left of  Fig.\,\ref{figS13} (a)]. The nonlocal light-matter coupling strengths are chosen as $\text{g}_{m} = \text{g} C^{\B}_{\boldsymbol{n}}$, where $C^{\B}_{\boldsymbol{n}}$ is the amplitude of the edge state, thus ensuring that the giant atoms couple selectively to it.

By applying the numerical method demonstrated in Sec.\,\ref{IIIC},  we simulate the time evolution of the excitation probabilities $P_{1,2}(t)=\left|e_{1,2}(t)\right|^{2}$ for the two giant atoms coupled to this modified photonic graphene. The initial excitation is placed in the donor QE. Note that a direct coupling of strength $\Omega$ persists between the QEs, thereby ensuring the existence of a dark state. As expected, the edge state and the dark state jointly enable the purely coherent and dissipation-immune interactions between the QEs, as illustrated in Fig.\,\ref{figS13}(f). Unlike the point-like coupling in the previous scheme, where a point-like QE interacts with only a single cavity mode, the giant atoms considered in this section feature a distributed coupling structure that enables simultaneous interaction with multiple cavity modes, thereby introducing complex interference effects. As a result, the population transfer dynamics of the QEs deviates from a simple coherent oscillation and exhibits a modulated envelope.

Motivated by the fundamental insight that the corner and edge geometry of a photonic graphene critically determines its band properties, we explore an alternative lattice configuration for engineering decoherence‑free interactions. In Fig.\,\ref{figS14}(a), we consider a photonic graphene with hybrid boundary conditions, where one edge is of the bearded type and the remaining five edges are of the zigzag type\,\cite{SMPhysRevResearch.3.023121,SMACSXie}. In this setup, each unit cell [see blue box in Fig.\,\ref{figS14}(a)] contains one A site and one B site, subject to dissipation with strengths $\kappa_a\neq 0$ and $\kappa_b=0$, respectively. It is expected that this lattice structure with specific boundary geometry may support topological quantum states that are immune to dissipation.

We then numerically compute the eigenspectrum of this finite photonic graphene, with the real and imaginary parts of the eigenvalues displayed in Figs.\,\ref{figS14}(b) and (c), respectively. From these spectra, we identify five zero‑energy (or dissipation‑free) eigenstates, marked by the red circle, which are localized exclusively on sublattice B. We find this geometric configuration supports the formation of topological corner states. It should be noted, however, that the emerging ``gapless" corner states hybridize with edge states, making it impossible to directly extract a pure corner state from these zero-energy eigenstates. To address this, we employ an inner product method to isolate the target corner state, with the result presented in Fig.\,\ref{figS14}(d). Interestingly, the corner state exhibits an out-of-phase relation along the zigzag boundary and an in-phase relation along the bearded boundary, as clearly demonstrated from the color-coded sign of the corner state $\Psi _{\mathrm{corner}}\left( \boldsymbol{n} \right)$. We then couple two directly coupled QEs simultaneously to a cavity in sublattice A, which is not necessarily located at corners or edges. By simulating the corresponding dynamic of atomic excited-state populations in Fig.\,\ref{figS14}(e),  it is found that the excitation transfer between the two QEs exhibits is immune to single-sublattice dissipation. Unlike the previously studied Kekulé‑type graphene lattice, which supports only a single dissipation‑free eigenstate in the quadrupole phase, the present model hosts multiple dissipation‑free modes. Furthermore, in this model the interatomic population‑transfer efficiency depends strongly on the light‑matter coupling strength $\text{g}$.

\end{document}